\documentclass[iop]{emulateapj}		% Nice 2-col format w/out space jumps

\usepackage{apjfonts, natbib}
\usepackage{graphicx}

\newcommand{\tnm}[1]{\tablenotemark{#1}}	% For several tables
\newcommand{\tnt}[1]{\tablenotetext{#1}}	% ditto
\newcommand{\pnn}{\phn\phn}			% ditto
\newcommand{\bvri}{\mbox{$BV\!RI$}}		% BVRI
\newcommand{\av}{\mbox{$A_V$}}			% A_V
\newcommand{\avmw}{\mbox{$A_{V,\,{\rm MW}}$}}	% A_V,MW (Milky Way extinction)
\newcommand{\avint}{\mbox{$A_{V,\,{\rm i}}$}}	% A_V,i  (internal extinction)
\newcommand{\mv}{\mbox{$M_V$}}			% M_V (absolute magn in V)
\newcommand{\mi}{\mbox{$M_I$}}			% M_I (absolute magn in I)
\newcommand{\mk}{\mbox{$M_K$}}			% M_K (abs magn in K_s band)
\newcommand{\mktwo}{\mbox{$M_{K,2}$}}		% M_K (ditto for Nuc_2)
\newcommand{\vtot}{\mbox{$V_{\rm T}$}}		% V_T
\newcommand{\vtotzero}{\mbox{$V_{\rm T,0}$}}	% V_T,0
\newcommand{\itotzero}{\mbox{$I_{\rm T,0}$}}	% I_T,0
\newcommand{\ks}{\mbox{$K_{\rm s}$}}		% K_s
\newcommand{\kstwo}{\mbox{$K_{\rm s,2}$}}	% K_s of Nuc_2
\newcommand{\vi}{\mbox{$V\!-\!I$}}		% V-I
\newcommand{\vitot}{\mbox{$(V\!-\!I)_{\rm T}$}}	% (V-I)_T
\newcommand{\vitotzero}{\mbox{$(V\!-\!I)_{\rm T,0}$}}
\newcommand{\vihst}{\mbox{$(V_{555}\!-\!I_{814})$}}
\newcommand{\arcone}{Arc$\;$1}			% Arc 1
\newcommand{\arctwo}{Arc$\;$2}			% Arc 2
\newcommand{\chandra}{{\em Chandra}}		% Chandra X-ray Observatory
\newcommand{\coonezero}{\mbox{{\rm CO\,($J$~= 1--0)}}}
\newcommand{\czhel}{\mbox{$cz_{\rm hel}$}}	% Heliocentric radial velocity
\newcommand{\czhelone}{\mbox{$cz_{\rm hel,1}$}}	% Hel. radial vel. of Nuc_1
\newcommand{\czheltwo}{\mbox{$cz_{\rm hel,2}$}}	% Hel. radial vel. of Nuc_2

\newcommand{\czlg}{\mbox{$cz_{_{\rm LG}}$}}	% Vel. rel to Local Group [sic]
\newcommand{\dproj}{\mbox{$d_{\rm proj}$}}	% Projected radial distance
\newcommand{\dvlos}{\mbox{$\Delta v_{_{\rm LOS}}$}}
\newcommand{\ellip}{\mbox{$\epsilon$}}		% Ellipticity
\newcommand{\ergscm}{\mbox{{\rm erg s}$^{-1}$ {\rm cm}$^{-2}$}}
\newcommand{\ergscma}{\mbox{{\rm erg s}$^{-1}$ {\rm cm}$^{-2}$} \AA$^{-1}$}
\newcommand{\ergsec}{\mbox{{\rm erg s}$^{-1}$}}	% Luminosity units
\newcommand{\fhi}{\mbox{$F_{\rm H\,I}$}}	% Integrated H~I flux
\newcommand{\hi}{\ion{H}{1}}			% Symbol for neutral hydrogen
\newcommand{\hii}{\ion{H}{2}}			% Symbol for H~II region

\newcommand{\hst}{{\em HST}}			% Abbrev. for Hubble Space Tel.
\newcommand{\jykms}{\mbox{Jy~km~s$^{-1}$}}	% Jansky kilometers per second
\newcommand{\kms}{\mbox{km~s$^{-1}$}}		% kilometers per second

\newcommand{\lineratzero}{\mbox{$F_0/F_0({\rm H}\beta)$}}
\newcommand{\lineratzeromw}{\mbox{$F_{0,{\rm MW}}/F_{0,{\rm MW}}({\rm H}\beta)$}}
\newcommand{\logmbh}{\mbox{$\log(\mbh/\msun)$}}	% Log {M_SMBH/M_sun}

\newcommand{\lbzero}{\mbox{$L_{B,0}$}}		% Intrinsic blue luminosity
\newcommand{\lk}{\mbox{$L_K$}}			% K/K_s-band luminosity
\newcommand{\lksun}{\mbox{$L_{K,\odot}$}}	% K/K_s-band lumin. of Sun
\newcommand{\lktwo}{\mbox{$L_{K,2}$}}		% K/K_s-band lumin. of Nuc_2
\newcommand{\loiii}{\mbox{$L_{\rm [O III]}$}}	% Luminosity in [O III]_5007
\newcommand{\loiiitwo}{\mbox{$L_{\rm [O III],2}$}}
\newcommand{\lv}{\mbox{$L_V$}}			% V-band luminosity
\newcommand{\lvsun}{\mbox{$L_{V,\odot}$}}	% V-band luminosity of Sun
\newcommand{\lx}{\mbox{$L_{\rm X}$}}		% Intrinsic X-ray luminosity
\newcommand{\lxone}{\mbox{$L_{\rm X,1}$}}	% X-ray luminosity of Nuc_1
\newcommand{\magarcs}{mag arcsec$^{-2}$}	% magnitudes/arcsec^2
\newcommand{\magarcstab}{mag/$\square$\arcsec}  % magnitudes/arcsec^2 f/Tables
\newcommand{\mdyn}{\mbox{$M_{\rm dyn}$}}	% Dynamical mass
\newcommand{\mdyntwo}{\mbox{$M_{\rm dyn,2}$}}	% Dynamical mass of Nuc_2
\newcommand{\mtot}{\mbox{$M_{\rm tot}$}}	% Total mass
\newcommand{\mbh}{\mbox{$M_{\bullet}$}}		% Mass of black hole
\newcommand{\mbhmw}{\mbox{$M_{\bullet,{\rm MW}}$}}
\newcommand{\mbhtwo}{\mbox{$M_{\bullet,2}$}}	% Mass of black hole in Nuc_2
\newcommand{\mbulge}{\mbox{$M_{\rm \star,bulge}$}}
\newcommand{\mhi}{\mbox{$M_{\rm H\,I}$}}	% Mass of H~I
\newcommand{\mhtwo}{\mbox{$M_{\rm H_2}$}}	% Mass of H_2 (molecular gas)
\newcommand{\mhihtwo}{\mbox{$M_{\rm H\,I\,+\,H_2}$}}
\newcommand{\mstar}{\mbox{$M_{\star}$}}		% Stellar mass
\newcommand{\mhost}{\mbox{$M_{\rm \star,host}$}}
\newcommand{\mseven}{\mbox{$M_{\rm \star,N7727}$}}
\newcommand{\mstartwo}{\mbox{$M_{\star,2}$}}	% Stellar mass of Nuc_2
\newcommand{\msun}{\mbox{$M_{\odot}$}}		% Solar mass
\newcommand{\msunpc}{\mbox{$M_{\odot}\>$pc$^{-2}$}}
\newcommand{\mueffzero}{\mbox{$\mu_{\rm e,0}$}}	% Effective SB corr'd f/MW ext
\newcommand{\muzerozero}{\mbox{$\mu_{0,0}$}}	% App. central SB corrected
\newcommand{\n}{NGC~}				% NGC = New General Catalogue
\newcommand{\nucone}{Nucleus$\;$1}		% Nucleus 1
\newcommand{\nuctwo}{Nucleus$\;$2}		% Nucleus 2
\newcommand{\rc}{\mbox{$r_{\rm c}$}}		% Core radius
\newcommand{\rcone}{\mbox{$r_{\rm c,1}$}}	% Core radius of Nuc_1
\newcommand{\rctwo}{\mbox{$r_{\rm c,2}$}}	% Core radius of Nuc_2
\newcommand{\reff}{\mbox{$r_{\rm e}$}}		% Effective radius
\newcommand{\reffone}{\mbox{$r_{\rm e,1}$}}	% Eff. radius of Nuc_1
\newcommand{\refftwo}{\mbox{$r_{\rm e,2}$}}	% Eff. radius of Nuc_2
\newcommand{\rhalf}{\mbox{$r_{\rm h}$}}		% Half-mass radius
\newcommand{\rtid}{\mbox{$r_{\rm t}$}}		% Tidal (cutoff) radius
\newcommand{\sapec}{Sa\,pec}			% Sa pec (morphological type)
\newcommand{\sighi}{\mbox{$\Sigma_{\rm H\,I}$}}	% Surface density of H~I
\newcommand{\sigvel}{\mbox{$\sigma_v$}}         % Velocity dispersion
\newcommand{\sigveleff}{\mbox{$\sigma_{v,{\rm e}}$}}
\newcommand{\sigvelone}{\mbox{$\sigma_{v,1}$}}  % Velocity disp of Nuc_1
\newcommand{\sigveltwo}{\mbox{$\sigma_{v,2}$}}  % Velocity disp of Nuc_2

\newcommand{\zsun}{\mbox{$Z_{\odot}$}}		% Solar metallicity
\newcommand{\halp}{\mbox{${\rm H}\alpha$}}	% H-alpha
\newcommand{\hbet}{\mbox{${\rm H}\beta$}}	% H-beta
\newcommand{\hgam}{\mbox{${\rm H}\gamma$}}	% H-gamma
\newcommand{\hdel}{\mbox{${\rm H}\delta$}}	% H-delta
\newcommand{\heps}{\mbox{${\rm H}\epsilon$}}	% H-epsilon
\newcommand{\mgi}{\mbox{\ion{Mg}{1}}}		% Mg I (triplet)
\newcommand{\nai}{\mbox{\ion{Na}{1}}}		% Na I (D doublet)
\newcommand{\nii}{\mbox{[\ion{N}{2}]}}		% Forbidden [N II]

\newcommand{\niithr}{\mbox{[\ion{N}{2}] $\lambda$6583}}
\newcommand{\oi}{\mbox{[\ion{O}{1}]}}		% Forbidden [O I]
\newcommand{\oizer}{\mbox{[\ion{O}{1}] $\lambda$6300}}
\newcommand{\oii}{\mbox{[\ion{O}{2}]}}		% Forbidden [O II]

\newcommand{\oiisev}{\mbox{[\ion{O}{2}] $\lambda$3727}}
\newcommand{\oiii}{\mbox{[\ion{O}{3}]}}		% Forbidden [O III]

\newcommand{\oiiisev}{\mbox{[\ion{O}{3}] $\lambda$5007}}
\newcommand{\oiiiboth}{\mbox{[\ion{O}{3}] $\lambda\lambda$4959, 5007}}
\newcommand{\sii}{\mbox{[\ion{S}{2}]}}		% Forbidden [S II]
\newcommand{\siiboth}{\mbox{[\ion{S}{2}] $\lambda\lambda$6716, 6731}}
\slugcomment{Accepted by {\em The Astrophysical Journal}, 2017 December 25}
\shorttitle{Second Nucleus of \n7727}
\shortauthors{Schweizer et al.}

\begin{document}

\title{The Second Nucleus of \n7727: Direct Evidence for the Formation and
       Evolution of \\
       an Ultracompact Dwarf Galaxy\altaffilmark{1}}

\author{
Fran\c cois Schweizer\altaffilmark{2},
Patrick Seitzer\altaffilmark{3},
Bradley C.\ Whitmore\altaffilmark{4},
Daniel D.\ Kelson\altaffilmark{2}, \\
and Edward V.\ Villanueva\altaffilmark{2}
}
%\setcounter{footnote}{4}			% Necessary f/normal style, but
						% not f/emulateapj

\altaffiltext{1}{Based in part on data gathered with the 6.5 m Magellan
Telescopes located at Las Campanas Observatory, Chile.}
\altaffiltext{2}{The Observatories of the Carnegie Institution for Science,
   813 Santa Barbara St., Pasadena, CA 91101, USA;
   schweizer@carnegiescience.edu}
\altaffiltext{3}{Department of Astronomy, University of Michigan,
   1085 S.\ University Ave., Ann Arbor, MI 48109, USA}
\altaffiltext{4}{Space Telescope Science Institute, 3700 San Martin Drive,
Baltimore, MD 21218, USA}

% Notice that each of these authors has alternate affiliations, which
% are identified by the \altaffilmark after each name.  The actual alternate
% affiliation information is typeset in footnotes at the bottom of the
% first page, and the text itself is specified in \altaffiltext commands.
% There is a separate \altaffiltext for each alternate affiliation
% indicated above.

% The abstract environment prints out the receipt and acceptance dates
% if they are relevant for the journal style.  For the aasms style, they
% will print out as horizontal rules for the editorial staff to type
% on, so long as the author does not include \received and \accepted
% commands.  This should not be done, since \received and \accepted dates
% are not known to the author.

\begin{abstract}
We present new observations of the late-stage merger galaxy \n7727,
including \hst/WFPC2 images and long-slit spectra obtained with the Clay 
telescope.
\n7727 is relatively luminous (\mv=$-$21.7) and features two unequal tidal
tails, various bluish arcs and star clusters, and {\em two bright nuclei}\, 
480~pc apart in projection.
These two nuclei have nearly identical redshifts, yet are strikingly
different.
The primary nucleus, hereafter \nucone, fits smoothly into the central
luminosity profile of the galaxy and appears---at various
wavelengths---``red and dead.''
In contrast, \nuctwo\ is very compact, has a tidal radius of 103 pc, and
exhibits three signs of recent activity:
a post-starburst spectrum, an \oiii\ emission line, and a central X-ray
point source.
Its emission-line ratios place it among Seyfert nuclei.
A comparison of \nuctwo\ (\mv=$-$15.5) with ultracompact dwarf galaxies
(UCDs) suggests that it may be the best case yet for a massive UCD having
formed through tidal stripping of a gas-rich disk galaxy.
Evidence for this comes from its extended star-formation history, long blue
tidal stream, and elevated dynamical-to-stellar-mass ratio.
While the majority of its stars formed $\ga$10 Gyr ago, $\sim$1/3 formed
during starbursts in the past 2 Gyr.
Its weak AGN activity is likely driven by a black hole of mass
$3\times 10^{6{\rm -}8}\,\msun$.
We estimate that the former companion's initial mass was less than half
that of then-\n7727, implying a minor merger.
By now this former companion has been largely shredded, leaving behind
\nuctwo\ as a freshly minted UCD that probably moves on a highly eccentric
orbit.
\end{abstract}

% The different journals have different requirements for keywords.  The
% keywords.apj file, found on aas.org in the pubs/aastex-misc directory, 
% contains a list of keywords used with the ApJ and Letters.  These are 
% usually assigned by the editor, but authors may include them in their 
% manuscripts if they wish. 

\keywords{galaxies: active --- galaxies: dwarf --- galaxies: formation ---
galaxies: individual (NGC 7727) --- galaxies: interactions ---
galaxies: nuclei}

\section{INTRODUCTION}
\label{sec1}

The hierarchical assembly of galaxies \citep{white78} is now widely
accepted and has become a paradigm of $\Lambda$ cold dark matter
cosmology \citep[e.g.,][]{frwh12,prim17,peeb17,rodr17}.
In parallel, studies of galaxy interactions and mergers have slowly gained
prominence as part of the search for primary drivers of galaxy growth and
evolution, both observationally
\citep[e.g.,][]{zwic53,arp66,lt78,schw86,schw00,kbs98,bell06}
and theoretically \citep[e.g.,][]{tt72,bh92,bh96,hopk06,hopk13}.
Major mergers (i.e., with mass ratios $m/M\ga 1/3$) seem to be especially
involved in the formation of early-type galaxies and may contribute, in ways
still poorly understood, to the quenching of star formation
\citep[e.g.,][]{fabe07,hain15}.
Because in the local universe they often involve a pair of full-grown disk
galaxies and generate spectacular phenomena, such as long tidal tails,
central starbursts, and active galactic nuclei (AGNs), they have been studied
extensively \citep[e.g.,][]{sand96,hopk08}.
The crucial role they play in transforming the morphological types of
some galaxies guarantees them continued attention in the future from
observers and numerical modelers alike.

In contrast, minor mergers (with mass ratios $m/M\la 1/3$) have been studied
less and are generally less well understood than major mergers for a variety
of reasons.
Their effects on galaxies are often less spectacular, whence they can be
more difficult to discover and observe.
Since they involve galaxies of distinctly different mass, the dynamical
friction they experience tends to be less effective than in major mergers,
leading to longer accretion times with a strong dependence on the impact
parameter.
As a result, some minor mergers can lead to prolonged tidal stripping of
the intruder in the halo of the main galaxy, with little or no delivery
of intruder mass to the center.
The stellar tidal streams that such prolonged minor mergers leave behind
in the halos of all types of galaxies can be spectacular, once galaxies and
their outskirts get imaged to very faint levels of surface brightness.
Disk galaxies wrapped in faint tidal streams, such as \n5907 \citep{mart08}
and M63 (= \n5055, \citealt{chon11}), are being discovered in growing numbers
\citep{mart10,misk11,carl16}.
Intermediate between them and classical major-merger remnants are early-type
galaxies that feature some long filaments, whether tidal streams or parts of
former tidal tails, and disturbed-looking bodies, often with ripples and
shells \citep[e.g.,][]{arp66,schw80,mali83,tal09,duc15}.
\n7727, the \sapec\ galaxy that is the subject of the present paper,
belongs in this intermediate category.

The discovery of ultracompact dwarf galaxies (UCDs) in the Fornax Cluster
\citep{hilk99,drin00} may have originally appeared unrelated to the kind
of major and minor mergers just described.
Soon, however, some UCDs were linked to nuclei of tidally stripped dwarf
galaxies \citep[e.g.,][]{bekk01,hase05,pfef13,liu15a} and, possibly, of even
larger and more metal-rich galaxies \citep[e.g.,][]{brod11,norr14,sand15}.
Recent findings that some UCDs had extended star-formation histories
\citep{norr15}, are sometimes associated with tidal stellar streams
\citep{vogg16}, and can even harbor supermassive black holes (SMBHs)
unexpectedly massive for their
luminosity \citep{mies13,seth14,ahn17} have strengthened suspicions that
some of them may be the long-lived nuclear remnants of tidally disrupted
companion galaxies of considerable mass.
Our observations of the intriguing ``second nucleus'' of \n7727 presented
here add what may be the strongest evidence yet that this process is still
occurring, even in the present-day local universe.

The study of \n7727 and its complex structure has an interesting history.
Probably because of its bright inner region and unresolved ``amorphous
spiral arms'' \citep{arp66}, now known as likely tidal tails,
\citet{hubb26} classified it as a normal spiral of type ``Sa.''
Both G.\ de Vaucouleurs and A.\ Sandage stuck to this morphological type,
refining it to ``SAB(s)a pec'' and ``Sa pec,'' respectively, based on
plates of superior resolution \citep{deva64,sand81}.
\citet{voro59} was the first to point out \n7727's peculiar structure,
placing the galaxy among ``Fused Systems'' in his {\em Atlas and
Catalogue of Interacting Galaxies}\, and noting that it might be the
result of an interaction between an elliptical and a spiral galaxy.
Following up on this tip, \citet{arp66} photographed the galaxy with the
Hale 5-m telescope and chose to include it as Number 222 in his
{\em Atlas of Peculiar Galaxies.}

The second nucleus of \n7727, hereafter ``\nuctwo'' for short, was
mistaken by many astronomers to be a Milky Way foreground star or was
outright overlooked, including by Arp himself.\footnote{
Arp noted on the envelope of his plate PH-4002-A, a 25-min visual exposure
of \n7727 taken with the Hale 5-m telescope: ``Star in interior seems to
be sitting on dark nebula --- investigate!''  This foreground star lies
$17\arcsec$ NNW of the primary nucleus, which---like \nuctwo---is not visible
owing to overexposure of the central region; it is this plate that is reproduced
in Arp's {\em Atlas}.  However, \nuctwo\ is strikingly visible on plate
PH-3963-A, a 60-min exposure of \n7727 through an \halp\ interference filter
also taken by Arp, yet there is no note on the envelope of this plate, nor
are there any markings on the plate itself.}
After inspecting old Mount Wilson plates, \citet{deva64} noted it as
``There is a star or bright knot superimposed.''
The first to recognize \nuctwo\ as highly peculiar was A.\ Sandage, who
compared \n7727 to the classical merger remnant \n7252
\citep{toom77,schw82} and then commented:
``The merger hypothesis for \n7727 is strengthened by the presence of
two nuclei near the center of the image seen on short-exposure Mount
Wilson 100-inch plates taken by Duncan in 1925 and 1938 and by Hubble
in 1946.  One of these nuclei is at the center of the bulge plus inner
disk of the parent galaxy.  The other unresolved bright nucleus is well
separated from the primary nucleus by $3\arcsec$. However, at $0\farcs8$
resolution for the seeing disk, it is not possible to decide if the
secondary nucleus is a superposed star or is the nucleus of a proposed,
nearly merged previous companion'' \citep{sand94}.

Sandage's description is right on the mark, as two of us (F.S. and B.W.)
realized after obtaining images of \n7727 with the {\em Hubble Space
Telescope} (\hst) for the purpose of studying the blue star-cluster
population \citep{crab94,tran14} of this merger candidate
\citep{tt72,schw86}.
It is these \hst\, images and our ground-based follow-up observations that
form the basis of the present study focused on \nuctwo.

\n7727, also known as Arp 222, VV 67, and MCG$-$02-60-006, is the dominant
member of a small group of three galaxies, the other two members being
\n7723 and \n7724 (Group 1522 in \citealt{croo07}).
Its primary nucleus, hereafter ``\nucone'' for short, is located at
   $\alpha_{{\rm J}2000} =  23^{\rm h}39^{\rm m}53\fs80$,
   $\delta_{{\rm J}2000} = -12\degr17\arcmin34\farcs0$
(Section \ref{sec31}) and has a recession velocity relative to the
Local Group of $\czlg = +2000\pm 8$ \kms\ \citep{kara96}, which places the
galaxy at a distance of $D = 27.4$ Mpc for $H_0 = 73$ \kms\ Mpc$^{-1}$
\citep{free10}.
At that distance, adopted throughout the present paper, $1\arcsec = 132.8$ pc.
The corresponding true distance modulus is $(m-M)_0 = 32.19$.
The Milky Way foreground extinction is relatively small, with values in the
literature ranging between $\avmw = 0.075$ (\citealt{rc3}) and 0.113
\citep{schl98}.
We adopt the latest revised value of $\avmw = 0.098$ \citep{schl11}, with
which the absolute visual magnitude of \n7727 becomes $M_V = -21.69$
(Section \ref{sec33}).
Finally, with $\mk = -24.51$ \citep{jarr03}, \n7727 has an estimated stellar
mass of $\mstar\approx 1.4\times 10^{11}\,\msun$ (Section \ref{sec421}).

In the following, Section \ref{sec2} describes our observations and
reductions, including imaging and spectroscopy of \n7727 and its two nuclei.
Section \ref{sec3} presents results concerning the optical structure of
\n7727; the nuclei's positions at optical, X-ray, and radio wavelengths;
their radial velocities, velocity dispersions, stellar populations, star
formation histories, and dynamical and stellar masses; and the interstellar
medium (ISM) of \n7727.
Section \ref{sec4} then discusses \nuctwo\ as a smoking gun for UCD formation
and evolution, the apparent absence of AGN activity in \nucone\ and presence
of it in \nuctwo, and \n7727 as an interesting example of the aftermath of
ingesting a gas-rich companion.
Finally, Section \ref{sec5} summarizes our main results and conclusions.
Two appendices add some details about image-masking techniques used and
emission lines measured in the nuclei.

%%%%%%%%%%%%%%%%%%%%%%%%%%%%%%%   Fig. 01   %%%%%%%%%%%%%%%%%%%%%%%%%%%%%%%%
\begin{figure*}
  \begin{center}
    \includegraphics[width=17.8cm]{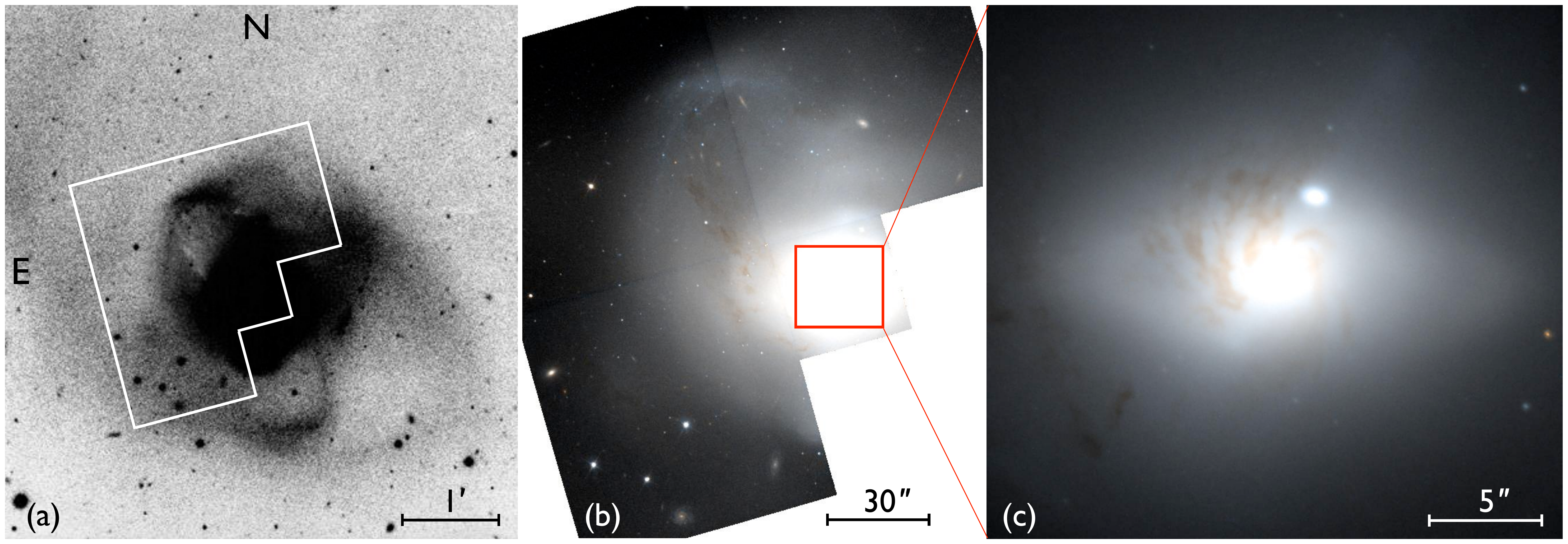}
    \caption{
Ground-based and \hst/WFPC2 images of \n7727 and its second nucleus.
(a) Hale 5-m telescope photograph reproduced from the {\em Carnegie Atlas of
    Galaxies} \citep{sand94}, with the footprint of WFPC2 observation marked
    in white.
(b) Portion of \hst/WFPC2 images in $V$ and $I$, combined into a color image
    by the HLA; the red box marks the central section of the galaxy shown
    enlarged in the next panel.
(c) Portion of PC images in $V$ and $I$, again combined into a color image.
Notice the spirally dust lanes in the bright central region of \n7727 and
the bluish, slightly elongated \nuctwo, located $3\farcs64$ NNW of the
primary nucleus.
North is up and east to the left in all three panels, and each panel features
its own scale bar to the lower right.
    \label{fig01}}
  \end{center}
\end{figure*}
%%%%%%%%%%%%%%%%%%%%%%%%%%%%%%%%%%%%%%%%%%%%%%%%%%%%%%%%%%%%%%%%%%%%%%%%%%%%

%% ##############################  Table 1  #################################
\begin{deluxetable*}{lccccrcccc}
\tablecolumns{10}
\tablewidth{0pt}
%\tabletypesize{\scriptsize}
\tablecaption{Log of Observations of \n7727\label{tab01}}
\tablehead{
  \colhead{Date}         &
  \colhead{Telescope}    &
  \colhead{Instrument\tnm{a}} &
  \colhead{CCD}          &
  \colhead{Filter}       &
  \colhead{P.A.}         &
  \colhead{Total\ }      &
  \colhead{Wavelength}   &
  \colhead{Seeing}       &
  \colhead{Notes\ \ }
                           \\
  \colhead{}             &
  \colhead{}             &
  \colhead{}             &
  \colhead{Detector}     &
  \colhead{}             &
  \colhead{}             &
  \colhead{\,Expos.}     &
  \colhead{Coverage}     &
  \colhead{}             &
                           \\
  \colhead{}             &
  \colhead{}             &
  \colhead{}             &
  \colhead{}             &
  \colhead{}             &
  \colhead{(deg)}        &
  \colhead{(s)}          &
  \colhead{(\AA)}        &
  \colhead{(arcsec)}     &
  \colhead{}
}
\startdata
1999 May 10 &   \hst        & WFPC2 & Loral   & F555W  & 0.0\tnm{b}& 2693& 4500--6100& \nodata& From HLA \\
            &               &       &         & F814W  & 0.0\tnm{b}& 3260& 7100--9600& \nodata& From HLA \\
2008 Aug 23 &  du Pont 2.5 m& DC    & Tek \#5 &$R_{\rm KC}$&0.0& 1200& 5670--7180& 1.2&      \nodata	 \\
2009 Aug 21 &  Clay 6.5 m   & LDSS-3& STA0500A& \nodata&  151.9& 1380& 3850--6590& 0.7&      Slit $1\farcs1\times 240\arcsec$ \\
2009 Aug 23 &  Clay 6.5 m   & MagE  & E2V 42-20&\nodata&  151.9& 2160& 3300--8250& 0.7--0.8& Slit $0\farcs7\times \phn10\arcsec$
\enddata
\tnt{a}{DC: Direct camera;\ \ LDSS-3: Low-Dispersion Survey Spectrograph 3; MagE: Magellan Echellette Spectrograph.}
\tnt{b}{Position angle of HLA-reprocessed images, not of original WFPC2 frames.}
\end{deluxetable*}
%% ###########################  End of Table 1  #############################

\section{OBSERVATIONS AND REDUCTIONS}
\label{sec2}

Our observations of \n7727 described below include images obtained with
the {\em Hubble Space Telescope} (\hst\,) and the du Pont 2.5-m telescope
at Las Campanas Observatory, as well as spectra taken with the
Low-Dispersion Survey Spectrograph (LDSS-3) and the Magellan Echellette
(MagE) spectrograph, both on the Clay 6.5-m telescope at Las Campanas.
Table~\ref{tab01} presents a log of these observations.

\subsection{Imaging}
\label{sec21}

\subsubsection{HST/WFPC2 Images}
\label{sec211}

We observed \n7727 with the Wide Field and Planetary Camera 2 (WFPC2)
aboard \hst\, on 1999 May 10 as part of Program GO-7468 (PI: Schweizer).
Five exposures each were taken through filters F555W and F814W (hereafter
$V$ and $I$, respectively), one short (60~s) and four long and dithered,
for a total exposure time of 2693~s in $V$ and 3260~s in $I$.
Although the various frames were processed in a standard manner at the
time \citep[e.g.,][]{tran03}, the new measurements presented below
(Sections \ref{sec31}, \ref{sec33}) were made from reprocessed $V$ and $I$\,
images downloaded from the Hubble Legacy Archive (hereafter HLA;
\citealt{whit07}) in 2014 December.
The reprocessing, done in 2009 and 2010, included new drizzling that
significantly improved the geometric rectification and the accuracy of
the coordinate system \citep{jenk10}.
This is a crucial advantage when comparing the optical coordinates derived
for the two nuclei of the galaxy with coordinates derived from \chandra\,
X-ray and ALMA radio observations, as we do in Section \ref{sec31} below.

Figure~\ref{fig01} shows a photomontage of one ground-based and two
\hst/WFPC2 images of \n7727 and its two nuclei.
The ground-based image (Figure~\ref{fig01}(a)) is based on a photograph of
this galaxy obtained by A.\ Sandage with the Hale 5-m telescope at prime
focus (Kodak IIIa-J plate, no filter, 75-min exposure) and here reproduced
from the {\em Carnegie Atlas of Galaxies} \citep[][Vol.\ I, Panel 83]{sand94}.
On this image the central part of the galaxy's tidally disturbed body
appears ``burnt out'' owing to overexposure.
The two WFPC2 images (Figures~\ref{fig01}(b) and (c)) are color renditions
prepared by the HLA through a combination of the $V$ and $I$ exposures
obtained with \hst.
Figure~\ref{fig01}(b) shows a large portion of the mosaicked WFPC2 image,
reproduced at a contrast that emphasizes faint structures and dust lanes
in the outer body at the expense of details in the central region, which
again appears burnt out.
Finally, Figure~\ref{fig01}(c) displays a $25\farcs0\times 23\farcs1$ segment
of the central body of \n7727, imaged by the PC chip of WFPC2 and centered
on \nucone.
The contrast has been chosen to show both the bright regions surrounding
this nucleus and the slightly elongated \nuctwo\ located about $3\farcs6$
north--northwest of it at P.A.\,= 332\degr.
Notice the {\em spiral-shaped dust lanes} emerging from the bright central
region to the northeast, the extended fainter body's {\em spindle shape}
which may indicate that it is a stellar disk, and the surprisingly steep
brightness drop-off of \nuctwo.
These features and their photometric properties are further discussed
in Sections~\ref{sec32} and \ref{sec33} below.

\subsubsection{Ground-Based Images}
\label{sec212}

Direct, mostly $R$-band images of \n7727 were obtained with the CCD camera of
the du Pont 2.5-m telescope on 2008 August 23 (Table~\ref{tab01}).
Conditions were photometric, with a seeing of $\sim${}$1\farcs2$ (FWHM).
The $R$-band images were taken through a standard $R\,(\rm Harris)$ filter
designed to match the Kron-Cousins photometric system.
The camera was equipped with the chip Tek~5 (2048$\,\times\,$2048 pixels),
which yielded a scale of $0\farcs2602$ pixel$^{-1}$ in $R$ and a field of
view (FOV) of $8\farcm8\times 8\farcm8$.
The CCD frames were flat-fielded, co-added, and reduced in standard manner
with IRAF.\footnote{
The Image Reduction and Analysis Facility (IRAF) is distributed by the
National Optical Astronomy Observatories (NOAO), which are operated by
the Association of Universities for Research in Astronomy (AURA), Inc.,
under a cooperative agreement with the National Science Foundation.}

A final $R$-band image of \n7727 was produced in two steps.
A full-FOV image was first computed as the median of the four best-seeing
frames of 180~s exposure each, for a total of 720~s of exposure.
Since the very center of \n7727 was mildly saturated on each frame, the
central region was then excised and replaced with a properly scaled median
image computed from five 60~s exposures that were clearly unsaturated
even at the nucleus.
This composite $R$-band image of \n7727 is presented and discussed in
Section \ref{sec32} below.

\subsection{Spectroscopy}
\label{sec22}

Long-slit spectra of \n7727 were obtained both with LDSS-3 (see
\citealt{alli94} for LDSS-2) at the Clay 6.5-m telescope during the night
of 2009 August 21/22 and with the MagE spectrograph \citep{marsh08} at the
same telescope two nights later (Table~\ref{tab01}).

\subsubsection{LDSS-3 Spectra}
\label{sec221}

For the observations with LDSS-3, the $1\farcs1\times 240\arcsec$ slit
of the spectrograph was placed at a position angle of P.A.\,= $151\fdg9$\ 
across the two nuclei of \n7727.
One exposure of 120~s and three exposures of 420~s each were obtained with
the ``VHP Blue" grism (1090 g mm$^{-1}$), yielding a wavelength coverage of
3850--6590~\AA\ and---for uniform illumination---a spectral resolution of
$R\approx 1300$ (4.0~\AA\ at $\lambda$5200) after full reduction.
The spatial scale along the slit was $0\farcs1890$ pixel$^{-1}$, with the
spatial resolution set by the $\sim${}$0\farcs7$ seeing during the
observations.

\n7727 was observed in the morning hours under somewhat cirrussy conditions,
following a prolonged telescope closure due to clouds.
To permit at least a relative flux calibration, two standard stars (Feige~110
and LTT~1020) were also observed, one before and the other after the
galaxy,  using a $1\farcs5$ wide slit at parallactic angle.
The subsequent reduction of the various spectra included COSMOS-pipeline
processing to flat-field, wavelength-calibrate, and rectify the spectra
frame by frame \citep{oeml17}.
For \n7727, the four frames of 120s and $3\times 420$~s exposure were then
co-added and cleaned of cosmic rays with the IRAF task {\em imcombine}.
Finally, the resulting rectified two-dimensional (2D) spectrum was
flux-calibrated and sky-subtracted in preparation for the various object
extractions and measurements to follow (see Section~\ref{sec341}).
Because of the cirrussy conditions, we estimate that the {\em absolute\,}
flux calibration of this spectrum is uncertain by about $\pm 30$\%, while
the relative flux calibration in wavelength is reliable at the few percent
level.

\subsubsection{MagE Spectrum}
\label{sec222}

For our 2009 August 23 observations of \n7727 with the MagE spectrograph,
the $0\farcs7\times 10\arcsec$ slit was oriented at P.A.\,=  $151\fdg9$
again and placed across both nuclei.
The main goal of taking this spectrum was to measure the radial-velocity
difference between the two nuclei and their individual velocity dispersions.
The total exposure of 36 minutes was broken into six 6-minute subexposures,
during which the airmass increased from 1.30 to 1.49 and the seeing
slightly deteriorated from $0\farcs7$ to $0\farcs8$.
A separate sky exposure, also of 6-minute duration, was then obtained
with the slit offset to a patch of blank sky 2\arcmin\ N of \nucone.
To calibrate fluxes, several standard stars were observed with a
$1\farcs0\times 10\arcsec$ slit at parallactic angle.
We also observed two Lick-index standards of spectral types G8\,III and
A5\,V to help measure the systemic velocity of \n7727 and permit a simple
assessment of the old- and young-star light contributions to the spectra of
the two nuclei.

The subsequent reduction of the various MagE spectra included pipeline
processing to flat-field and co-add frames, rectify spectral orders,
calibrate wavelengths, and subtract the sky spectrum \citep{kels00,kels03}.
Although MagE covers the wavelength range of 3100~\AA\ -- 1.0~$\mu$m in
orders 20--6 \citep{marsh08}, the two most ultraviolet orders yielded
no signal for the galaxy, and the two most infrared orders could not be
reliably processed because of scattered-light problems.
Hence, the final spectra, extracted from orders 18--8, cover the
wavelength range 3300--8250~\AA\ with a resolution of $R \approx 5800$.
The spatial scale along the slit varies slightly from order to order
and is $0\farcs275$ pixel$^{-1}$ at the \mgi\ triplet ($\lambda\approx
5200$~\AA).
In a final step, one-dimensional (1D) spectra of the two nuclei of \n7727
were extracted from each 2D order spectrum and were spliced together in
wavelength to form a single 1D spectrum for each nucleus, as described in
more detail in Section \ref{sec34}.

\section{RESULTS}
\label{sec3}

The following subsections first present our astrometry of \n7727's two
nuclei at optical, X-ray, and radio wavelengths, then describe the general
optical structure of the galaxy and the detailed photometric structure
of both nuclei, and go on to present the nuclei's velocities, velocity
dispersions, stellar populations, star-formation histories, and masses.
The final subsection briefly describes some interesting properties of the
galaxy's sparse ISM.

\subsection{Positions of the Two Nuclei}
\label{sec31}

We have measured new optical positions for the two nuclei of \n7727
from the \hst/WFPC2 images obtained in the $V$ and $I$ passbands
(Section \ref{sec211}) and have compared them with the positions of peaks
observed in X-ray and radio images of the galaxy, as described below.
Table~\ref{tab02} presents all measured coordinates.

%% ###############################  Table 2  ################################
\begin{deluxetable*}{llccccccccc}
\tablecolumns{11}
\tablewidth{0pt}
\tablecaption{Astrometric Positions of the Two Nuclei of \n7727\label{tab02}}
\tablehead{
  \colhead{Telescope/\pnn\pnn}       &
  \colhead{Source Image\pnn}         &
  \colhead{}                         &
  \multicolumn{2}{c}{\nucone}        &
  \colhead{}                         &
  \multicolumn{2}{c}{\nuctwo}        &
  \colhead{}                         &
  \colhead{$s$\tnm{a}}               &
  \colhead{P.A.\tnm{b}}              \\[1pt]
  \cline{4-5}  \cline{7-8}           \\[-7pt]
  \colhead{\pnn\ Instrument}         &
  \colhead{}                         &
  \colhead{}                         &
  \colhead{$\alpha_{\rm J2000}$}     &
  \colhead{$\delta_{\rm J2000}$}     &
  \colhead{}                         &
  \colhead{$\alpha_{\rm J2000}$}     &
  \colhead{$\delta_{\rm J2000}$}     &
  \colhead{}                         &
  \colhead{(\arcsec)}                &
  \colhead{(\degr)}            
}
\startdata
\hst/WFPC2	&  PC $V$ (F555W)	&& 23:39:53.796 & $-$12:17:34.03       && 23:39:53.679 & $-$12:17:30.83    &&  3.63   &   331.8  \\
		&  PC $I$\, (F814W)	&& 23:39:53.796 & $-$12:17:34.04       && 23:39:53.679 & $-$12:17:30.83    &&  3.64   &   331.9  \\
\chandra/ACIS	&  Center-of-field	&& 23:39:53.804 & $-$12:17:34.48\tnm{c}&& 23:39:53.687 & $-$12:17:31.23\tnm{d}&& 3.67 &   332.2  \\
ALMA\tnm{e}	&  3 mm continuum\tnm{f}&& 23:39:53.801 & $-$12:17:34.67       &&  \nodata     &    \nodata        && \nodata &  \nodata \\
		&  CO(1-0) line\tnm{f}	&& 23:39:53.838 & $-$12:17:34.42       &&  \nodata     &    \nodata        && \nodata &  \nodata
\enddata
\tnt{a}{Separation $s$\, between \nucone\ and \nuctwo\ (in arcseconds).}
\tnt{b}{Position angle of \nuctwo\ relative to \nucone.}
\tnt{c}{Center position of diffuse, extended X-ray emission.}
\tnt{d}{Position of bright X-ray point source named ``Source 5'' by Brassington et al.\ (2007), who treated its
   position as being that of the main optical nucleus, which it clearly is not.  Instead, Source 5 is here measured to
   coincide with \nuctwo\ to within 0.42\arcsec, which is well within the combined errors of
   \hst\, and \chandra\ positions.}
\tnt{e}{ALMA observations from 2011 November made in the Cycle 0 compact configuration (Ueda et al.\ 2014).}
\tnt{f}{The radio positions were measured from Figure 1h in Ueda et
   al.\ (2014), where high-accuracy coordinate grid marks are provided.}
\end{deluxetable*}
%% ###########################  End of Table 2  #############################

The centroids of the {\em optical}\, nuclear light distributions were
measured with the IRAF task {\em imexam} from the Planetary Camera (PC)
frames (scale of $0\farcs050$ per drizzled pixel) and, as a check, also
from the WFPC2 mosaics delivered by the HLA.
The measurements all agreed with each other to within $\pm 0\farcs01$
in each coordinate (R.A., Decl.), much better than the reported typical
{\em absolute} positional accuracy of $\sim${}$0\farcs3$ per coordinate for
\hst\, images reprocessed by the HLA.
Table~\ref{tab02} lists the PC coordinates of the two nuclei in $V$ and $I$.
Note that the near-identity of the nuclear positions in the two passbands
suggests that dust extinction near the nuclei is either
relatively small or rather uniform across each nucleus.

In order to address the question whether either nucleus of \n7727 might
show any signs of AGN activity, we also measured the
positions of peaks observed in the {\em X-ray count} distribution of
this galaxy.
As \citet{bras07} found from a 19 ks exposure taken with \chandra's
Advanced CCD Imaging Spectrometer (ACIS), the galaxy's X-ray emission
consists of a diffuse photon distribution stemming from the hot ISM and
a number of point sources, of which the brightest, {\em Source~5,}\, lies
close to the center of the diffuse emission.
In their Table~3 \citeauthor{bras07} assumed that the center of the diffuse
emission coincides with Source~5 and gave it the same coordinates.
However, their Figure~14 (right panel) clearly shows a small offset between
Source~5 and the peak of the diffuse distribution.

To determine new astrometric positions for the distinct centers of the
diffuse X-ray emission and Source~5 we downloaded the reprocessed X-ray
images of \n7727 from the \chandra\ Archive.
The original exposure (ID: Observation 2045) was obtained on
2001 December 18 with ACIS-S (PI: A.\ M.\ Read) and was reprocessed on
2012 October 1 by B.\ Sundheim.
To measure peak positions, we used the so-called ``Center image'' (ACIS chip
S3), which contains count numbers in 1025x1024 square pixels of size
$0\farcs492$ summed over a net exposure time of 19.01 ks.
The main result of our astrometry is that the centroid of the diffuse
X-ray emission lies at
   $\alpha_{{\rm J}2000} =  23^{\rm h}39^{\rm m}53\fs804$,
   $\delta_{{\rm J}2000} = -12\degr17\arcmin34\farcs48$
and coincides with the optical position of \nucone\ to within
$0\farcs46$, while X-ray Source~5 lies at
   $\alpha_{{\rm J}2000} =  23^{\rm h}39^{\rm m}53\fs687$,
   $\delta_{{\rm J}2000} = -12\degr17\arcmin31\farcs23$
and coincides with the optical position of \nuctwo\ to within $0\farcs42$
(see Table~\ref{tab02}).
We conclude that the hot ISM of \n7727 is clearly centered on \nucone,
while Source~5 closely coincides with \nuctwo.

Two arguments strongly support this conclusion.
First, the coordinate differences between the measured X-ray and optical
positions are nearly identical for the ISM centroid versus \nucone\
($\Delta\alpha = +0\farcs12$, $\Delta\delta = -0\farcs44$) and the
Source~5 versus \nuctwo\
($\Delta\alpha = +0\farcs12$, $\Delta\delta = -0\farcs40$), clearly
suggesting a systematic shift of $\sim${}$0\farcs44$ between the
\chandra-based and \hst-based coordinate systems.
Such a shift falls well within the range of the combined absolute-coordinate
uncertainties for \chandra\ and \hst, being $\la\,${}$0\farcs30$ radial errors 
for 68\% of all point sources measured from ACIS-S images and typically
$\sim${}$0\farcs3$ errors per coordinate for WFPC2 images from the HLA.

Second, as Table~\ref{tab02} shows, the separation $s$ and position angle
between the two measured X-ray centroids,
$(s, {\rm P.A.}) = (3\farcs67, 332\fdg2)$, are nearly identical to those
between the two optical nuclei, $(3\farcs64, 331\fdg9)$.
Hence, there can be little doubt about the spatial congruence of the true
(i.e., physical) X-ray and optical peaks.
Specifically, this implies that X-ray Source~5 probably is coincident with,
or lies very close ($\la0\farcs1$) to, the center of \nuctwo, a point
discussed further in Section~\ref{sec422}.

Finally, we also measured precise positions for the {\em radio peaks} of
the 3~mm continuum and \coonezero\ line emission observed with ALMA during
Cycle~0 by \citet{ueda14}.
These authors found a rotating central molecular-gas disk and remarked
that the 3~mm continuum peak is associated with the nucleus and coincides
with the peak seen in the CO integrated-intensity map.
Since they seemed unaware of the presence of two optical nuclei in \n7727
and did not give precise positions for the 3~mm continuum and CO emission
peaks, we determined these positions ourselves from the contour map shown
in the middle panel of their Figure~1h.
This panel measures $29\farcs7\times 29\farcs7$ and is framed with precise
coordinate grid marks.
Table~\ref{tab02} lists the coordinates
$(\alpha_{{\rm J}2000}, \delta_{{\rm J}2000})$
measured separately for the 3~mm continuum peak and the \coonezero\
integrated-intensity peak.
From the beam size and observational signal-to-noise ratios we estimate
the $1\sigma$ positional uncertainties to be $\sim${}$0\farcs3$ in
each coordinate for the 3~mm continuum peak and $\sim${}$0\farcs4$ for
the CO emission-line peak.
As a comparison of the positions of the radio peaks with those of the
optical and X-ray peaks given in the table shows, both the 3~mm continuum
peak and the CO emission-line peak coincide clearly with \nucone\ (to
within the combined errors) and not with \nuctwo\
or X-ray Source 5.
Hence, the small central molecular-gas disk found by \citet{ueda14} is
indeed associated with the main nucleus of \n7727.

\subsection{Optical Structure of \n7727}
\label{sec32}

The optical structure of \n7727 is complex and has long been understood to
point to a relatively recent gravitational interaction between two galaxies
(\citealt{voro59}; \citealt{tt72}; \citealt{schw86}, esp.\ Fig.~4;
\citealt{sand94}).
Presciently, \citet{arp66} placed this galaxy (Arp 222) in his
morphological category named ``Galaxies [with] Amorphous Spiral Arms,'' 
which also contains the two merger remnants \n3921 (Arp 224) and \n7252
(Arp 226).
The main difference between \n7727 and these two remnants is that it
features only one {\em obvious}\, tidal tail, while they clearly feature
{\em two}\, such tails each.
In the following, we first use ground-based images to describe \n7727's
rich inner and outer fine structure, then use \hst/PC images to illustrate
the dust lanes and two nuclei of the central region at high resolution,
and finally present evidence for a tidal stream emanating from \nuctwo.

\subsubsection{Inner and Outer Fine Structure}
\label{sec321}

%%%%%%%%%%%%%%%%%%%%%%%%%%%%%%%   Fig. 02   %%%%%%%%%%%%%%%%%%%%%%%%%%%%%%%%
\begin{figure*}
  \begin{center}
    \includegraphics[width=16.8cm]{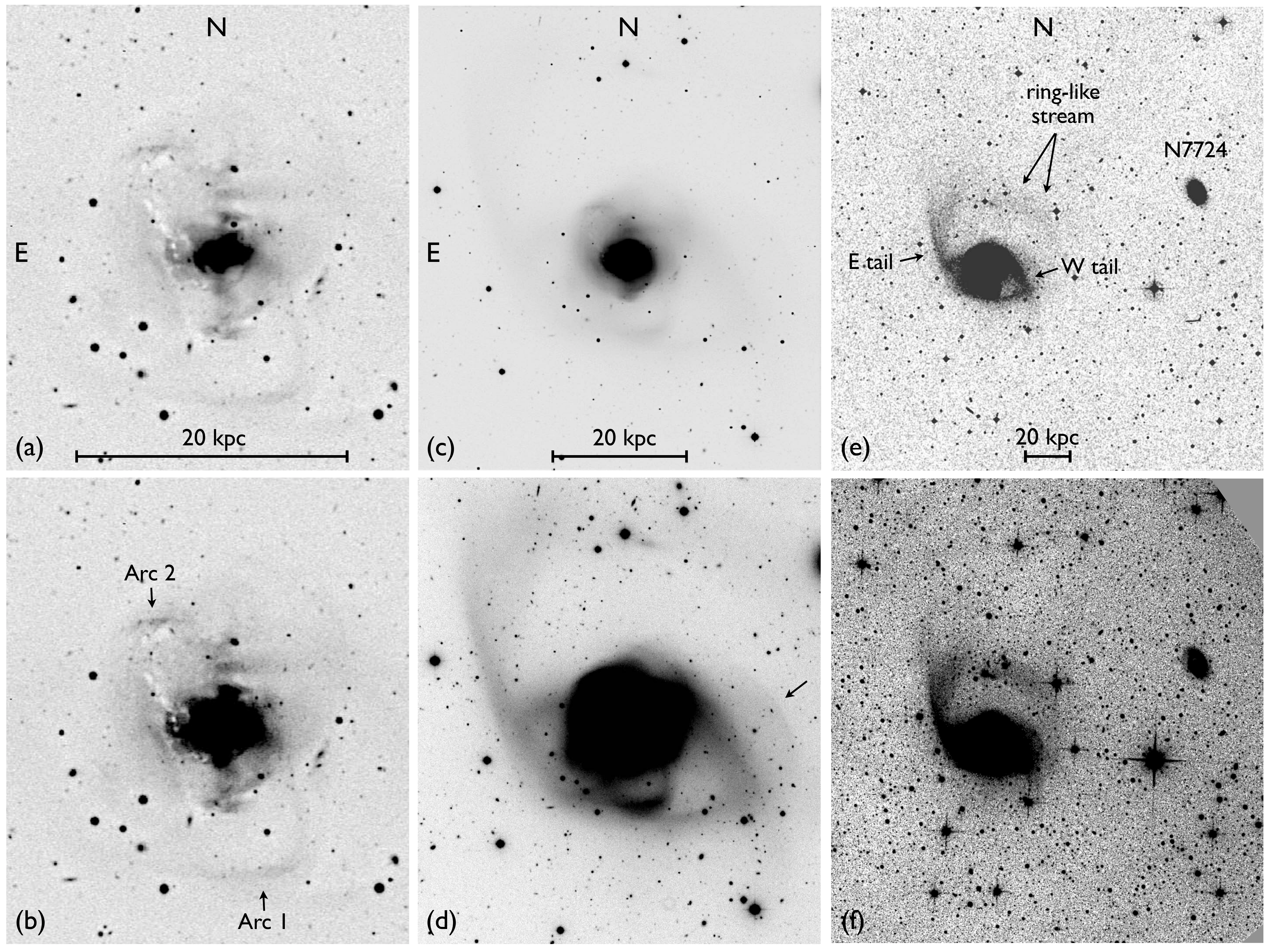}
    \caption{
Six displays of ground-based images illustrating the rich inner and outer
fine structure of \n7727.
(a) and (b) Partially masked $R$-band images revealing some of the complex
inner structure of the galaxy, including luminous arcs, ripples, and dust
patches and lanes.
\nuctwo\ is invisible, being buried in the black central area.
(c) and (d) Direct $R$-band image displayed at low and high contrast and
showing the connection between the inner structure, outer loops, and E tail.
Notice the faint outer ripple marked by the arrow.
(e) and (f) Medium- and high-contrast displays of \n7727 and its companion
\n7724 to the west.
Note the faint outer {\em ring-like stream}\, seemingly connecting
the N end of the E tail with the westernmost ripple.
North is up and east to the left in all panels.
The scale bars in the top panels measure 20 kpc ($150\farcs6$) each and
apply to the panels below them as well.
Panels (a)--(d) are based on a 12-min $R$-band exposure taken with
the CCD camera at the du Pont telescope; panel (e) is made from a Digitized
Sky Survey scan of a IIIa-J plate of 1~hr exposure taken with the UK Schmidt
telescope; and panel (f) reproduces part of a 6~hr luminance exposure obtained
by W.\ Probst with a CCD camera on his 0.25~m Newtonian reflector near Gurk,
Austria, and kindly made available.
    \label{fig02}}
  \end{center}
\end{figure*}
%%%%%%%%%%%%%%%%%%%%%%%%%%%%%%%%%%%%%%%%%%%%%%%%%%%%%%%%%%%%%%%%%%%%%%%%%%%%

Figure~\ref{fig02} illustrates the inner and outer fine structure of \n7727
with two masked images (panels (a) and (b)) and four direct images
(panels (c)--(f)).
The scales of the left, middle, and right panels differ and are indicated
by 20~kpc ($150\farcs6$) long bars in the top panels.

The inner fine structure, best seen in Figures~\ref{fig02}(a) and
(b),\footnote{
Details about the digital masking process used to produce these two
panels are presented in Appendix~\ref{appa}.}
appears chaotic, with various luminous protrusions, ripples, and
{\em loop-like arcs}\, interwoven with dust lanes and patches.
There are two main such arcs: ``\arcone'' extending $\sim\,$11~kpc to the
SSW and possibly connecting to the E tail (see esp.\ Figures~\ref{fig02}(b)
and (d)), while ``\arctwo'' appears as a nearly 4~kpc long feature lying
about 9~kpc NE of the center of the galaxy.
Both arcs are blue in color, making them especially well visible on
Sandage's blue-band photograph reproduced in Figure~\ref{fig01}(a).
Their blue color also shows nicely in a $(B\!-\!R)$-index map of \n7727
\citep{lho11} and is indicative of young to intermediate-age stellar
populations in these arcs.
\arctwo, seen faintly in Figures~\ref{fig02}(a)--(c) and much better in
Figures~\ref{fig01}(a) and (b), coincides with an abrupt, probably tidal
cutoff in the galaxy's surface brightness, seen prominently in
Figure~\ref{fig02}(d).
Notice in Figures~\ref{fig02}(a) and (b) that while many {\em dust lanes}
and {\em patches} appear chaotic, some do align with luminous matter, like
the dust lane $\sim\,$5.4 kpc SW of the center.
These aligned dust lanes may well mark gas layers seen nearly edge-on and
embedded in sheets of luminous matter, often called ``ripples'' or
``shells'' \citep{schw80,schw86,mali83,quin84}.

Perhaps the most obvious outer structure of \n7727 is its prominent tidal
``E~tail,'' which measures at least 60~kpc in projected length (Figures
\ref{fig02}(d)--(f)).
Given the main body's many ripples and arcs, most presumably consisting of
former-disk material, the question is whether there is a {\em second}\, tidal
 tail that might stem from a second disk galaxy involved in the interaction.
A diffuse candidate for such a second tail can be seen in
Figures~\ref{fig01}(a), \ref{fig02}(c), and \ref{fig02}(d), curving away
from the main body first northward, then westward, and finally southward
in direction of the bright star near the SW corners of these two panels.
If this apparent curved stream of diffuse light indeed is a coherent structure
and tidal tail, as we believe, then it forms a roughly symmetric counterpart
to the E tail, although clearly less extended than the latter.
This ``W~tail'' and the more prominent E tail then indicate that {\em two}\,
disk galaxies, likely of unequal mass, were involved in forming the current
near-remnant system.

Notice also a faint, $\sim$\,16~kpc ($\sim$\,120$\arcsec$) long ripple about
22~kpc west of \n7727's center (near the W edge of Figure~\ref{fig02}(d),
marked by the arrow), yet another sign of dynamically cold disk material
involved in this messy merger.

%%%%%%%%%%%%%%%%%%%%%%%%%%%%%%%   Fig. 03   %%%%%%%%%%%%%%%%%%%%%%%%%%%%%%%%
\begin{figure*}
  \begin{center}
    \includegraphics[width=17.8cm]{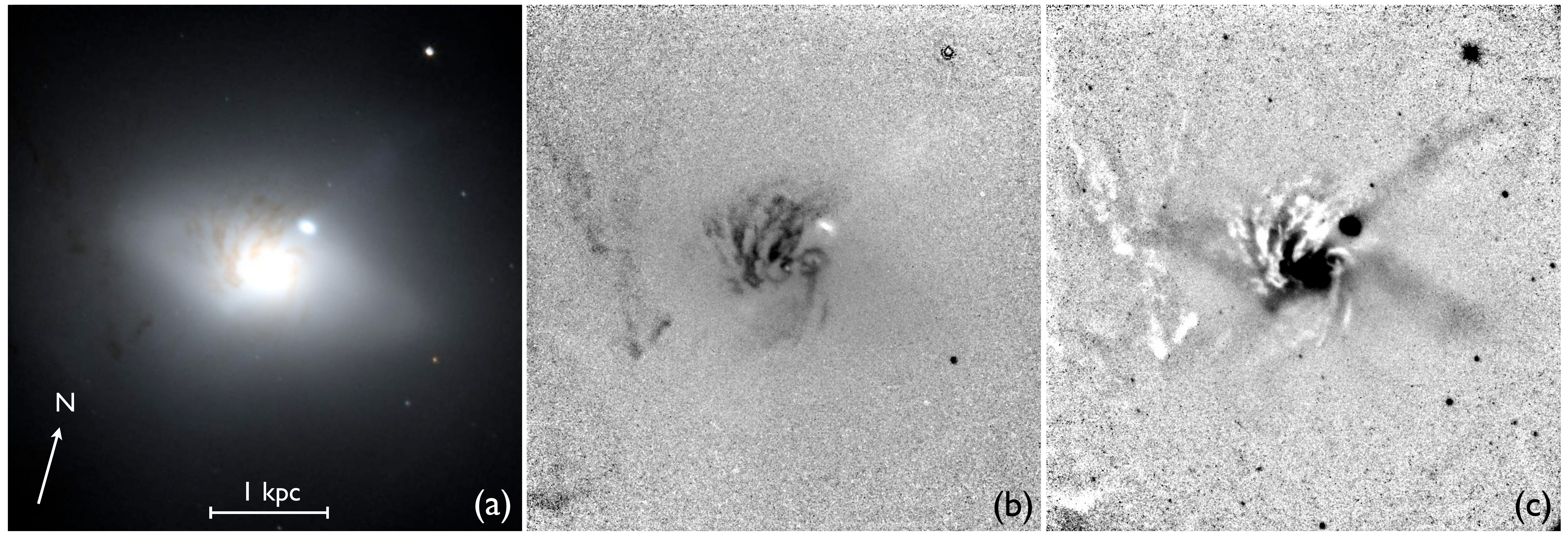}
    \caption{
Structure of central dust lanes in \n7727 imaged with the PC chip of
\hst/WFPC2.
The field of view shown in each panel measures $32\farcs8\times 33\farcs6$
($4.36\times 4.46$ kpc), and the arrow pointing north is 5\arcsec\ long.
(a) The color image downloaded from the HLA shows the two nuclei plus,
    faintly, the innermost part of the central dust lanes.
(b) The ratio of $I/V$ counts (displayed linearly) shows the reddish dust
    lanes as dark filaments and the bluish \nuctwo\ as a whitish oval.
(c) An unsharply masked $V$ image shows the extinction caused by the dust
    lanes (whitish areas) and emphasizes \nuctwo\ and a bluish tidal stream
    extending from it to the NNW.
For details, see text.
    \label{fig03}}
  \end{center}
\end{figure*}
%%%%%%%%%%%%%%%%%%%%%%%%%%%%%%%%%%%%%%%%%%%%%%%%%%%%%%%%%%%%%%%%%%%%%%%%%%%%

The most surprising outer structure is the faint {\em ring-like stream}\, of
luminous matter that seems to connect the N end of the E tail with this
outer ripple and the westernmost part of the galaxy (Figures~\ref{fig02}(e)
and (f)).
Interestingly, this diffuse stream {\em may}\, continue into a faint giant
southern arc passing slightly south of the much brighter \arcone\ (see
Figure~\ref{fig02}(d)).
{\em If}\, the E tail and this ring-like structure were to form a large
physical ring (of major-axis diameter $\sim$\,50 kpc), then this ring might
indicate that the dense center of one disk galaxy plunged through the disk
of another galaxy at some time early during the interaction, forming a
classical ring galaxy (\citealt{lynd76}; \citealt{toom78}, esp.\ Fig.~5).
In principle, \n7724---the companion galaxy located 97~kpc WNW of \n7727
(see Figures~\ref{fig02}(e) and (f))---could have been the plunging perturber,
yet we see no trace of any luminous connection between the two systems, and
\n7724 itself looks like a rather normal, unperturbed barred spiral galaxy.
For this reason, and also because \n7724 is gas-rich and lies far off the
minor axis of \n7727's ring, we deem it very unlikely to have been involved
in creating the apparent ring-like stream.

Alternatively, this stream might itself be some tidal tail, perhaps
indicative of a minor third galaxy having been involved in giving \n7727
its present messy outer structure.

\subsubsection{Central Dust Lanes}
\label{sec322}

The center region of \n7727 is laced with prominent dust lanes that make it
difficult to recognize any luminous fine structure.

Figure~\ref{fig03} illustrates the structure of these central dust lanes with
three derivatives of the $V$ and $I$\, frames obtained with the PC chip
of \hst/WFPC2 (Section \ref{sec211}).
The color image downloaded from the HLA (Figure~\ref{fig03}(a)) shows not
only the galaxy's two nuclei, but also what appears to be an inner lens or
bar of $\sim$\,2.7~kpc ($\sim$\,20\arcsec) length.
Emerging from the bright center toward the NE are spiral-shaped dust lanes
that are more clearly visible in an image displaying the ratio $I/V$ of
counts (or flux), shown in Figure~\ref{fig03}(b).
The fact that these reddish central dust lanes appear strong (i.e., dark) in
the half-plane NE of \nucone\ and weak in the SW half-plane suggests that
they trace an {\em inclined disk of dust and gas}\, with a semimajor axis
extending from the NW to the SE and its NE side protruding from the sky plane
through \nucone.
Assuming that the spirally dust filaments are trailing, this inferred
disk geometry predicts that the SE side of the gas disk should be receding.
Velocity maps of this disk made in CO(1--0) with ALMA confirm this
morphology-based prediction and yield a disk inclination of
$62\degr\pm 2\degr$, major-axis orientation at P.A.\,= $113\degr\pm 1\degr$,
and maximum rotation velocity of $\sim$\,150 \kms\ 
\citep[][esp.\ Fig.\ 1h and Table 6]{ueda14}.

In an unsharply masked version of the PC $V$ image shown in
Figure~\ref{fig03}(c),\footnote{
For details about the masking process, see Appendix~\ref{appa}.}
the dust lanes appear as whitish areas of diminished brightness because
they absorb light originating from luminous matter behind them.
In two of the strongest dust patches, located 320 pc ($2\farcs4$) N and
290 pc ($2\farcs2$) NNE of \nucone, the measured extinction reaches values
of $A_V = 0.42$ mag and $A_I = 0.22$ mag.
These {\em apparent} extinctions in $V$ and $I$\, represent only lower
limits to the true extinction caused by the dust since starlight originating
in \n7727 between the dust patches and us tends to ``fill in'' some of
the observed extinction.

Finally, notice the more chaotic dust lanes near the E edge of the
$4.36\times 4.46$ kpc FOV displayed in Figure~\ref{fig03}.
These lanes are part of an extensive system of obscuring dust that extends
toward the NE at least as far as the luminous arc lying 9 kpc from \nucone\
(Figure~\ref{fig01} and Section \ref{sec321}).

\subsubsection{The Blue Tidal Stream of \nuctwo}
\label{sec323}

Immediately beyond \n7727's central, $r\la 5\arcsec$ region messed up by
dust lanes there is one luminous fine structure that stands out---despite
its faintness---when one inspects the PC color image of the galaxy made by
the HLA on a computer screen.
It is a $\sim$\,11\arcsec\ (1.5 kpc) long ``plume'' or ``streamer'' of
bluish light extending from \nuctwo\ to the NNW.
This faint bluish plume is difficult to discern on any printed version of
the above color image (Figure~\ref{fig03}(a)), but can easily be seen on an
unsharply masked version of the PC $V$ image, as shown in
Figure~\ref{fig03}(c).

%%%%%%%%%%%%%%%%%%%%%%%%%%%%%%%   Fig. 04   %%%%%%%%%%%%%%%%%%%%%%%%%%%%%%%%
\begin{figure}
  \begin{center}
    \includegraphics[width=8.2cm]{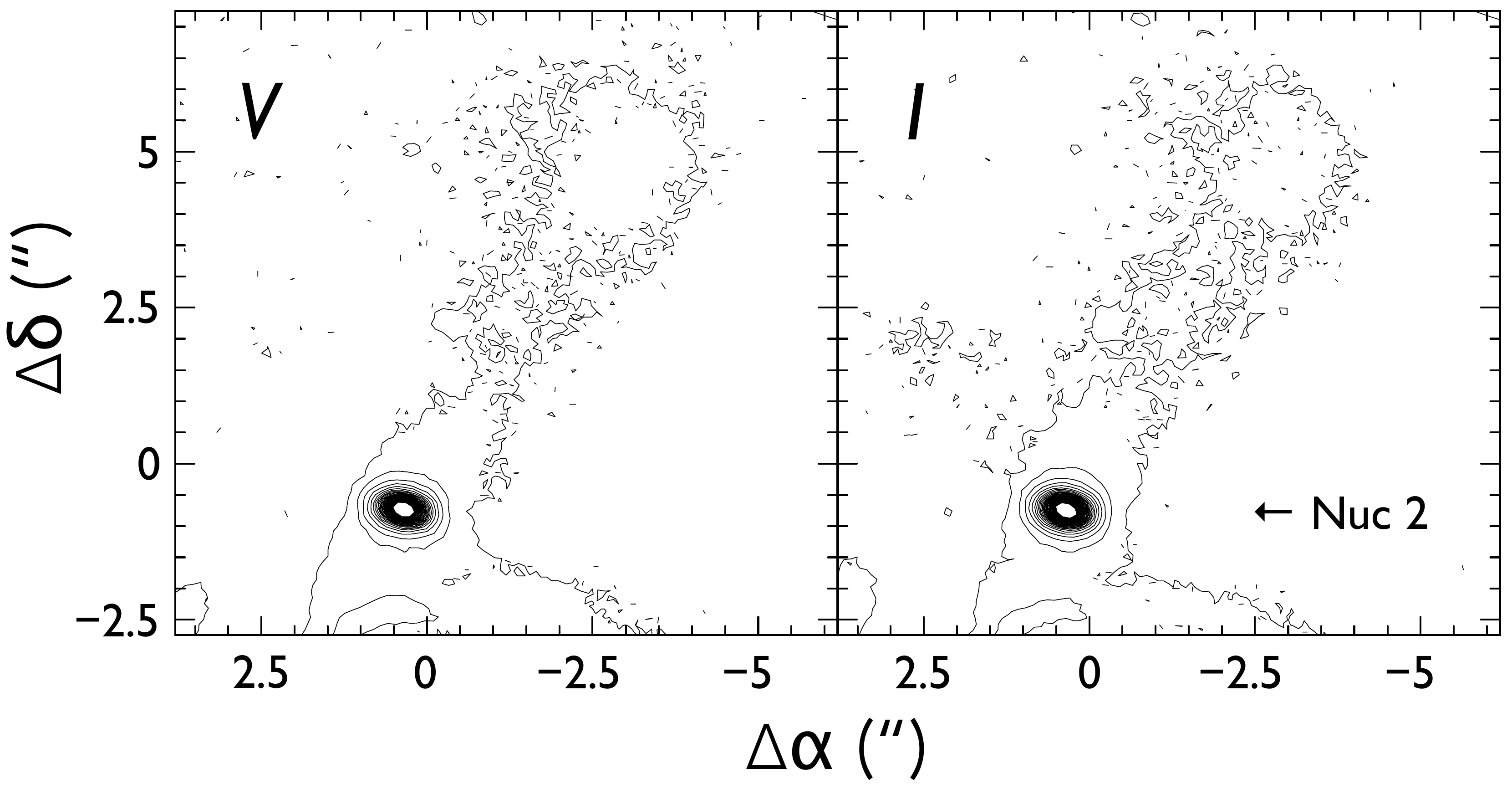}
    \caption{
Isophotal contours of \nuctwo\ and its blue tidal stream in (left) $V$ band
and (right) $I$ band, obtained from the \hst\ PC images after subtraction
of a smooth model light distribution for \n7727 (see text for details).
The outermost isophote outlining the tidal stream in $V$ ($I$) marks a
surface-brightness level of $\mu(V)~= 21.65$ mag arcsec$^{-2}$ ($\mu(I)~=
20.74$ mag arcsec$^{-2}$) in excess of the model, while the second-faintest
isophote---nearly circular around \nuctwo---marks $\mu(V)~= 19.32$
($\mu(I)~= 18.42$) in the same units.
    \label{fig04}}
  \end{center}
\end{figure}
%%%%%%%%%%%%%%%%%%%%%%%%%%%%%%%%%%%%%%%%%%%%%%%%%%%%%%%%%%%%%%%%%%%%%%%%%%%%

Given that unsharp masking of images of complex objects can occasionally
produce strange artifacts, we convinced ourselves of the reality of this
bluish plume by producing various isophotal plots of carefully
{\em model}-masked $V$ and $I$ images of \n7727 (see Section \ref{sec33} below
for details about the computation of the model light distributions).
Figure~\ref{fig04} shows two resulting contour plots, in $V$ and $I$, of
\nuctwo\ and the region NW of it.
In both cases the lowest contour was set to 0.040 electrons per second and
pixel, which corresponds to surface brightnesses of 21.65 \magarcs\ in
$V$ and 20.74 \magarcs\ in $I$.
Note that these surface brightnesses pertain to the {\em residual light}
of \n7727 after subtraction of model light distributions computed separately
for the $V$ and $I$ images (Section \ref{sec33}).
Since the luminous plume extending from \nuctwo\ to the NNW looks very
similar in both $V$ and $I$, it is unquestionably real.
From here on and for reasons to be detailed shortly, we will call it the
{\em Blue Tidal Stream (BTS)}.

Before deriving its photometric properties, we point out two things:
(1) the contour plots show the BTS only over its brightest $7\farcs5$
(1.0 kpc) in length, compared to its full $\sim$\,11\arcsec\ length seen
in Figure~\ref{fig03}(c);
(2) the BTS may have a counterpart to the SE of \nuctwo, as suggested by
what {\em may} be a faint bluish streak across the central area of
\n7727---glimpsed upon visual inspection of the HLA's color image---and by
a rectangular-shaped dark-gray ``slab'' of extra light seen protruding from
the central burnt-out area to the SE in Figure~\ref{fig03}(c).
However, because of the complex dust lanes and higher surface brightness
in that area, we have been unable to convince ourselves of its reality and
to produce evidential contour plots.
Hence, in the following we describe only the clearly established NNW part
of the BTS, but keep in mind that there may be an SSE part to it as well.

We estimate the integrated apparent $V$ and $I$ magnitudes of the BTS
from its isophotal contours shown in Figure~\ref{fig04} as follows.
Integrating flux over the entire area of the BTS contained within the
contour of lowest surface brightness from \nuctwo\ to the BTS's NW end
while excluding any flux from \nuctwo\ itself (i.e., flux within the
second and higher contours), and assuming that the surface brightness
within this area is constant at the level of the lowest contour, yields
apparent magnitudes of $V \leq 19.08$ and $I \leq 18.28$ for the BTS.
These values clearly represent upper (i.e., faint) limits to the
{\em total}\, magnitudes since (1) the surface brightness within the
lowest contour exceeds the contour value and (2) the BTS extends
$\sim$\,47\% beyond its contour in the NW direction (11\arcsec\ vs.
$7\farcs5$).
After applying estimated corrections for these two missing contributions
to the flux and also correcting for Milky Way foreground extinctions of
$A_V = 0.097$ ($A_I = 0.053$), we find integrated true magnitudes for the
BTS of $\vtotzero = 18.68\pm 0.15$ and $\itotzero = 17.93\pm 0.15$.
The corresponding absolute magnitudes are $\mv = -13.51$ and $\mi =
-14.26$.
Note that when compared to those of \nuctwo\ (Section \ref{sec33}, esp.\
Table~\ref{tab03}), these absolute magnitudes for the BTS indicate that its
integrated $V$ ($I$) luminosity corresponds to about 16\% (14\%) of the
luminosity of \nuctwo.

Perhaps the most interesting photometric property of the BTS is its blue
integrated color, $\vitotzero = 0.75\pm 0.21$, which is 0.18 mag bluer
than that of \nuctwo, $\vitotzero = 0.93\pm 0.03$, but only by 
$0.85\,\sigma$ of the combined errors.
In contrast, the \vitotzero\ color of the BTS is 0.48 mag bluer than that
of \nucone, $\vitotzero = 1.23\pm 0.03$, a difference significant
at the $2.3\,\sigma$ level.
Thus, the blue color of the BTS strongly links the stream to \nuctwo,
its point of apparent origin.
Taken together, the apparent physical linkage and the common blue
color make a strong case for this stream consisting of stars being
tidally stripped from \nuctwo, which explains our choice of the
name Blue Tidal Stream.

\subsection{Photometric Structure of the Two Nuclei}
\label{sec33}

Detailed photometric analyses of the main body of \n7727 and \nucone\
have been published \citep{lho11,laue05} and will not be repeated here.
Instead, we concentrate on a photometric comparison of the galaxy's
two nuclei.

To compare the surface-brightness profile of \nuctwo\ with that
of \nucone, we performed $V$ and $I$ photometry from the PC
frames obtained with \hst/WFPC2 (Section \ref{sec211}).
We used the task {\em ellipse} of the {\em isophote} package in IRAF/STSDAS
\citep{busk96} to fit elliptical isophotes via the method developed by
\citet{jedr87}, itself based on earlier work by \citet{cart78} and
\citet{laue85}.
Because of the complex system of dust lanes (Section \ref{sec32}) we did the
primary ellipse fitting in the $I$ band, where the extinction is only 60\%
of that in the $V$ band \citep{schl98}.
In addition, we also used the iterative sigma-clipping option of the
task {\em ellipse} to reduce the influence of dust lanes when fitting
elliptical isophotes in $I$ and deriving the mean surface brightness
along them.
For the photometry in $V$ we used the $I$ ellipses and the same
sigma-clipping algorithm to make the $V$ and $I$ surface-brightness
measurements as congruent as possible.
Unlike \citet{laue05}, we did not attempt to deconvolve the PC images
for the effects of the point-spread function (PSF), again because of
the considerable uncertainties that would be introduced by the central
dust lanes.

In a first step, the photometric profile of \nucone\ and the surrounding
body was derived with the region of \nuctwo\ excluded from the fit.
In a second step, a model light distribution of the main body based on
this profile and computed with the STSDAS task {\em bmodel}\, was
subtracted from the $V$ and $I$ images to produce images of
%the second nucleus
\nuctwo\ alone (plus surrounding brightness residuals).
%The photometric profiles of the second nucleus in $V$ and $I$ were then
The photometric profiles of \nuctwo\ in $V$ and $I$ were then
again obtained with the task {\em ellipse} out to a radius of
$0\farcs80$ (= 16 PC pixels), where both profiles drop off sharply and
beyond which residuals make measurements unreliable.

%%%%%%%%%%%%%%%%%%%%%%%%%%%%%%%   Fig. 05   %%%%%%%%%%%%%%%%%%%%%%%%%%%%%%%%
\begin{figure}
  \begin{center}
    \includegraphics[width=8.0cm]{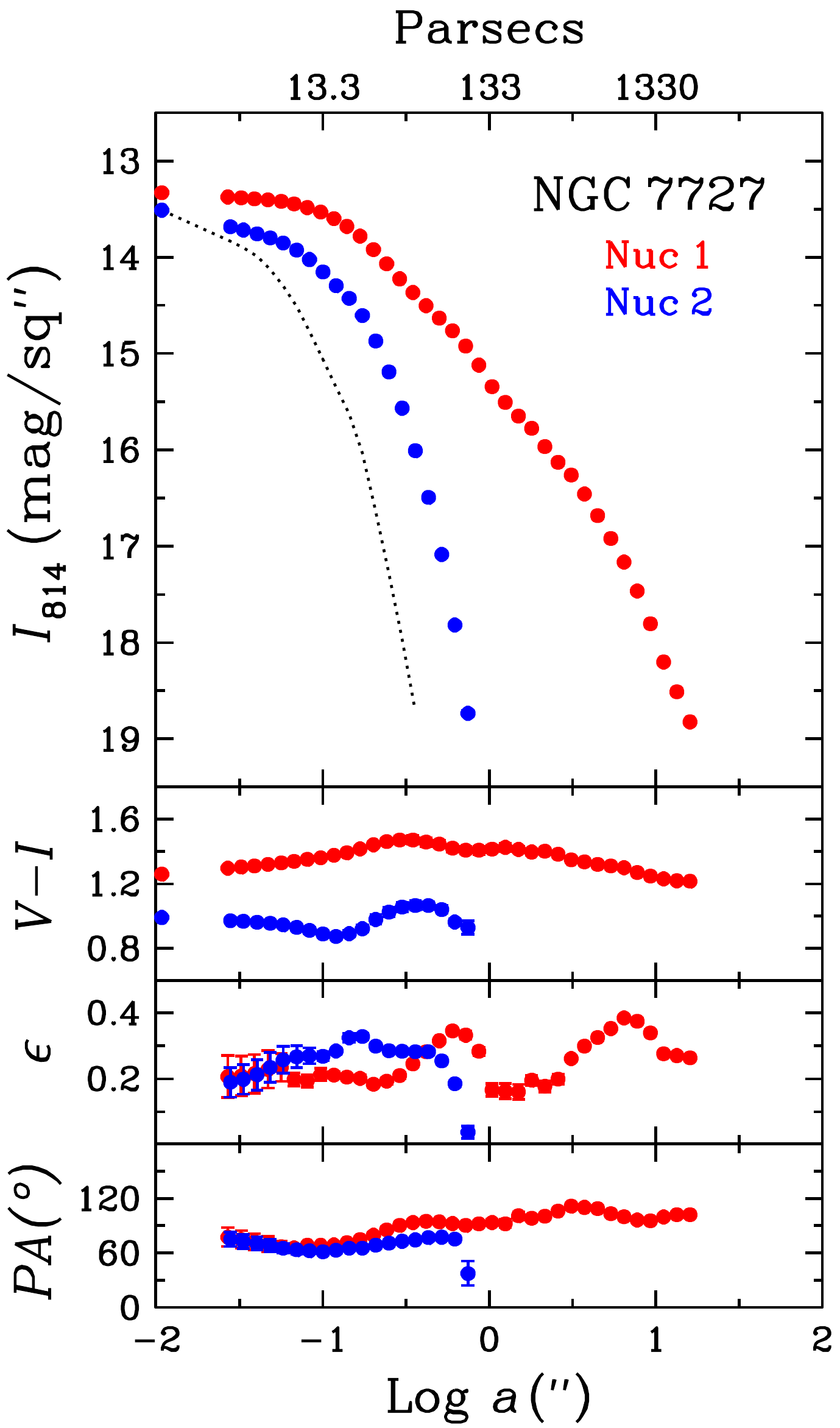}
    \caption{
Surface photometry of \nucone\ and the surrounding body (``Nuc 1'') and
\nuctwo, derived from the PC images taken with \hst/WFPC2.
The top panel shows the measured $I_{814}$ surface-brightness profiles of
both nuclei plotted vs the semimajor axis $a$ (red and blue data points
for Nuc~1 and Nuc~2, respectively), plus the PSF (black dotted line) derived
from a nearby star.
Error bars for the profile data points are all smaller than the data points.
The second panel shows the measured \vihst\ color profiles, while the third
and fourth panels show the ellipticities and position angles, respectively,
of the best-fit elliptical isophotes.
The scale at the top of the panels gives parsecs.
Note the sharp drop-off in the surface brightness of \nuctwo, which explains
why on ground-based photographic plates many observers have mistaken it for
a foreground star; for details, see text.
    \label{fig05}}
  \end{center}
\end{figure}
%%%%%%%%%%%%%%%%%%%%%%%%%%%%%%%%%%%%%%%%%%%%%%%%%%%%%%%%%%%%%%%%%%%%%%%%%%%%

Figure~\ref{fig05} shows the resulting $I$\, surface-brightness, \vi\ color,
ellipticity \ellip, and position-angle profiles for \nucone\ and the
surrounding body (red data points) and for \nuctwo\ (blue data points),
plotted as functions of the semimajor axis $a$ (in arcseconds) of the
best-fitting $I$ ellipses.
For comparison, the PSF derived from an $I\approx 17.3$\ star located
17$\arcsec$ NNW of \nucone\ is also plotted (black dotted line).
We note parenthetically that the $V$ profile of \nucone\ and the surrounding
body, not shown in Figure~\ref{fig05}, agrees well with the $V$ profile
derived by \citet{laue05} over the range $0\farcs10\la a \la 10\arcsec$, where
PSF smearing does not seriously affect our profiles of the main body.
Figure~\ref{fig05} illustrates several results of interest.

First, the $I$ surface-brightness profile of \nuctwo\ drops off very sharply
when compared to that of \nucone, yet not as steeply as the stellar PSF;
hence, \nuctwo\ is clearly resolved, as further discussed below.
Second, \nuctwo\ is about 0.45 mag bluer in \vi\,\ than \nucone, with a mean
color index of $\langle\vi\rangle_{\rm Nuc\,2}= 0.97$ within a radius of
$r=0\farcs80$ compared to
$\langle\vi\rangle_{\rm Nuc\,1}= 1.42$ within the same radius.
Third, given \nuctwo's low velocity dispersion when compared to that of
\nucone\ (Section \ref{sec34}), its measured central surface brightness
is surprisingly high.
In the $I$ band, it is only 18\% fainter than the measured surface brightness
of \nucone, while in the $V$ band the center of \nuctwo\ outshines that of
\nucone\
 by $\sim$9\%.
These relative central surface brightnesses could, of course, change after
proper deconvolution of the PC $V$ and $I$ images, but the striking result
remains that {\em the two nuclei of \n7727 have comparable optical central
surface brightnesses.}

Fourth, note the rapidly diminishing ellipticity of \nuctwo\ in the
outermost part of its measured profile ($a\approx 0\farcs5$ to $0\farcs8$).
As Figure~\ref{fig04} illustrates, this circularization of the isophotes is
real and marks the transition region between the bright inner part of \nuctwo,
which appears elliptical with $\ellip\approx 0.3$ and semimajor axis
oriented at P.A.$\:\approx 70\degr\pm 5\degr$, and the BTS, which itself is
oriented toward P.A.$\:\approx 330\degr\pm 5\degr$.
The $I$ isophotes appear circular ($\ellip = 0$) at\,\ $\rtid\!= 0\farcs78 \pm
0\farcs01 = 103 \pm 2$~pc, which we adopt as the {\em tidal cutoff radius}
%of the second nucleus.
of \nuctwo.

The surface-brightness profiles shown in Figure~\ref{fig05} further allow us
to derive photometric structural parameters for both nuclei and global
photometric parameters for \nuctwo.
For \nucone, which does not stand out as a separate structure
or star cluster \citep{laue05}, and the surrounding main body, we adopt
the corresponding global parameters obtained by \citet{lho11} from their
wide-field \bvri\ photometry of \n7727.
Table~\ref{tab03} presents the collated measurements, from which we again derive
several results.

%% ###############################  Table 3  ################################
\begin{deluxetable*}{lccc}
\tablecolumns{4}
%\tablewidth{0pt}
\tablewidth{12.5truecm}
%\tabletypesize{\footnotesize}		% Too small!
%\tabletypesize{\small}
\tablecaption{Photometric Parameters of \n7727 and \nuctwo\label{tab03}}
\tablehead{
\colhead{Parameter\phm{AAA}}    &
\colhead{Symbol}                &
\colhead{NGC 7727 (Nucleus 1)}  &
\colhead{Nucleus 2}
}
\startdata
Apparent total $V$ magnitude	&	\vtot &	    10.60 mag\tnm{a} &	16.80 mag 	   \\
Apparent total color index	&	\vitot &  \phn1.27 mag\tnm{a} &	\phn0.97 mag	   \\
Total $V$ magnitude\tnm{b}	&	\vtotzero &	10.50 mag    &	16.70 mag	   \\
Total color index\tnm{b}	&	\vitotzero & \phn1.23 mag    &	\phn0.93 mag	   \\
Absolute $V$ magnitude\tnm{c}	&	$M_V$ &	     $-$21.69 mag\phs &	$-$15.49 mag\phs   \\
Absolute $I$ magnitude\tnm{c}	&	$M_I$ &	     $-$22.92 mag\phs &	$-$16.42 mag\phs   \\
Effective radius in $I$ passband\tnm{d} & \reff & $28\arcsec=3.7$ kpc & $0\farcs22=29$ pc  \\
Surface brightness at \reff\tnm{b} &  $\mueffzero(I)$ & 19.33 \magarcstab\tnm{a} &
									14.92 \magarcstab  \\
Central $V$ surface brightness\tnm{e} &	$\muzerozero(V)$ &
						   14.49 \magarcstab &	14.40 \magarcstab  \\
Central $I$ surface brightness\tnm{e} &	$\muzerozero(I)$ &
						   13.28 \magarcstab &	13.46 \magarcstab  \\
Core radius in $I$ passband\tnm{d} &	\rc &      $0\farcs24=32$ pc & $0\farcs106=14$ pc  \\
Tidal cutoff radius in $I$ passband &	\rtid &	 	\nodata      &	$0\farcs78=103$ pc \\
Ellipticity at core radius &		\ellip(\rc) &	 0.19        &	 0.28              \\
P.A. of semimajor axis at \rc &		P.A.(\rc) &	$85\degr$    &	$63\degr$
\enddata
\tnt{a} {Values taken from Ho et al.\ (2011).}
\tnt{b} {Corrected for Milky Way foreground reddening ($A_V = 0.097$), but not for
         internal reddening.}
\tnt{c} {For the adopted distance of $D = 27.4$ Mpc.}
\tnt{d} {Corrected for PSF smearing (see text for standard deviations).}
\tnt{e} {Corrected for Milky Way foreground reddening, but not for internal reddening
         or PSF smearing.}
\end{deluxetable*}
%% ###########################  End of Table 3  #############################

With its total apparent magnitude of $\vtot = 16.80$, measured within \rtid,
\nuctwo\ appears 6.2~mag fainter than \n7727 itself ($\vtot = 10.60$).
Hence, \nuctwo\ contributes only $\sim$\,0.3\% to the total optical
luminosity of the galaxy despite its high central surface brightness.
With its absolute magnitude of $\mv = -15.49$, \nuctwo\ is about 2.5$\times$
less luminous than M32 ($\mv = -16.5$, \citealt{vdbe00}), the compact
elliptical (cE) companion of M31, and 4.3$\times$ less luminous than the
Small Magellanic Cloud ($\mv = -17.07$, \citealt{vdbe00}).
Yet it is clearly over 100$\times$ more luminous than any Milky Way
globular cluster, of which the most luminous is $\omega$~Cen
($\mv = -10.26$, \citealt{harr10}).
Hence, there can be little doubt that \nuctwo\ is of a galactic nature.

Yet, the effective (i.e., half-light) radius of \nuctwo\ measured in the
$I$ band and corrected for the width of the PSF,
$\refftwo = 0\farcs217 \pm 0\farcs015  = 29 \pm 2$~pc,
is very small, especially when compared with that of the main nucleus and
body of \n7727,
$\reffone = 28\farcs1 \pm 1\farcs5 = 3.7 \pm 0.2$~kpc
(measured from our ground-based $R$ image).
This small \refftwo\ and the absolute magnitude of $\mv = -15.5$ place
\nuctwo\ clearly in the domain of UCDs (see, e.g., \citealt{norr14}).
In agreement with this assessment, the $I$ surface brightness measured at
\reff\ is also much higher for \nuctwo, $\mueffzero(I) = 14.92$ \magarcs,
than for \n7727 (19.33 \magarcs).

Contrary to the above global photometric parameters, which differ sharply
between \nuctwo\ and the main nucleus and body of \n7727, the local
photometric parameters at the {\em centers}\, of \nuctwo\ and \nucone\ are
much more similar.
As already found above, the apparent central surface brightness in
$V$---corrected for Milky-Way foreground reddening but not for internal
reddening or PSF smearing---is $\muzerozero(V) = 14.40$ for \nuctwo, about
9\% higher than that of \nucone.
The corresponding central surface brightness in $I$ for \nuctwo\ is
$\muzerozero(I) = 13.46$ \magarcs, only $\sim$\,18\% less than that of
\nucone, which, however, probably suffers some internal extinction.
Finally, the apparent core radius in $I$ (i.e., the radius at half the
apparent peak surface brightness) of \nuctwo\ is $\rctwo = 0\farcs106 =
14$~pc (corrected for PSF width), not all that different from the apparent
core radius of \nucone, $\rcone = 0\farcs24 = 32$~pc.

In summary, \nucone\ and \nuctwo\ of \n7727 have comparable central
photometric properties, supporting the notion that \nuctwo\ is galactic in
nature.
However, \nucone\ appears embedded in---and indistinguishable from---the
surrounding body of the galaxy (Fig.~\ref{fig05}, and \citealt{laue05}),
while \nuctwo\ has a sharp tidal cutoff at $\rtid = 103$~pc and shares
some photometric properties with UCDs.

\subsection{Nuclear Spectra, Velocities, and Velocity Dispersions}
\label{sec34}

The spectra of \nucone\ and \nuctwo\ obtained with LDSS-3 and MagE (Section
\ref{sec22}) are of relatively short total exposure (21 min and 36 min,
respectively) and have, hence, limited signal-to-noise ratios, especially
in their UV--blue regions.
Nevertheless, they yield valuable information concerning the spectral
differences between the two nuclei, plus systemic velocities and velocity
dispersions for each nucleus.

\subsubsection{Comparison of the Nuclear Spectra}
\label{sec341}

\begin{figure}
  \begin{center}
    \includegraphics[width=8.5cm]{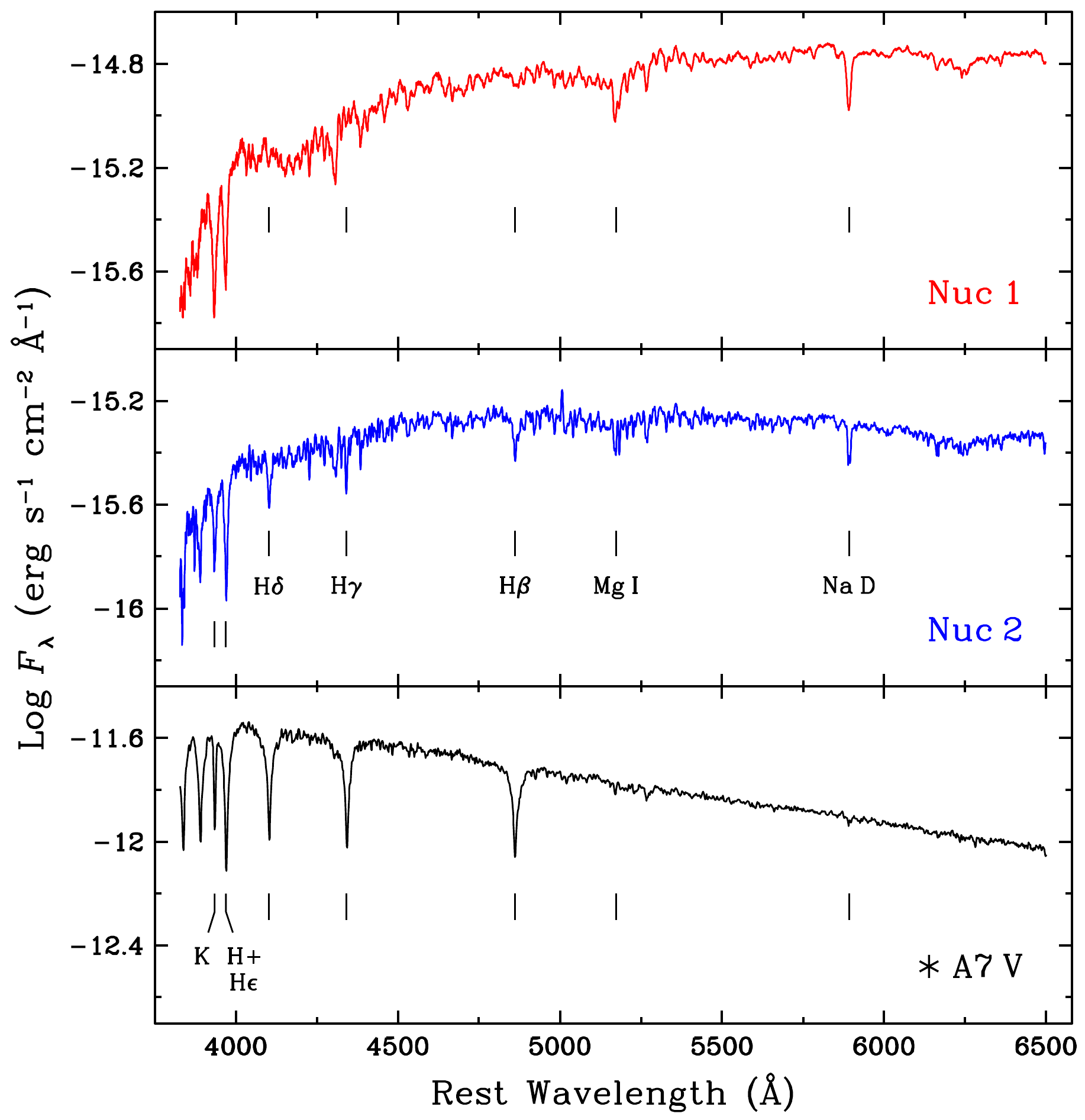}
    \caption{
UV--optical spectra of \nucone\ (top) and \nuctwo\ (middle) of \n7727,
obtained with LDSS-3 on the Clay 6.5~m telescope.
Fluxes are displayed logarithmically vs.\ rest wavelength.
The spectrum of HD\,111525 (bottom), an A7\,V star from the \citet{jaco84}
library, is shown for comparison.
Some of the main absorption features are marked by vertical bars and labeled
in the middle or bottom panels.
The spectrum of \nucone\ was extracted with an aperture of
$0\farcs8\times 1\farcs1$, while that of \nuctwo\ was extracted with one
of $1\farcs0\times 1\farcs1$ and slightly smoothed with a Gaussian
($\sigma = 0.7$ \AA) for better display.
Note that the spectral energy distribution of \nuctwo\ is significantly
bluer than that of \nucone\ and features Balmer absorption lines indicative
of an aging starburst.
    \label{fig06}}
  \end{center}
\end{figure}

Figure~\ref{fig06} shows the spectra of the two nuclei obtained with LDSS-3
and here displayed in the form of logarithmic flux versus rest wavelength.
Both spectra were extracted from the same rectified and sky-subtracted
2D spectrum described in Section \ref{sec221}.
The extraction apertures chosen for \nucone\ and \nuctwo\ were
$0\farcs80\times 1\farcs10$ ($106\times 146$ pc) and
$1\farcs00\times 1\farcs10$ ($133\times 146$ pc), respectively, and were
each centered on its nucleus.\footnote{
The extractions comprised four spatial pixels of the rectified 2D spectrum
(scale of $0\farcs200$ pixel$^{-1}$) for \nucone\ and five spatial pixels
for \nuctwo.}
For \nucone\ no background subtraction was necessary, since it is an integral
part of the galaxy and indistinguishable from its surroundings
(Section \ref{sec33}).
However, for \nuctwo\ the surrounding bright background of the main galaxy
had to be subtracted, which, because of its curvature and steep gradient,
poses a challenge.

We experimented with two background subtraction methods.
One was to make separate quadratic fits to the background on either side
of \nuctwo\ at each wavelength and subtract the fitted background that
falls into the aperture used for \nuctwo.
The second method was to extract a separate spectrum from a patch of the
galaxy located SE of the main nucleus, symmetrically opposite from \nuctwo\
and at the same level of surface brightness, and subtract it from the
``\nuctwo\ + local background'' spectrum.
For the LDSS-3 spectrum of \nuctwo\ displayed in Figure~\ref{fig06} we
chose this second method, since it yielded clearly more reliable continuum
fluxes, although at the price of slightly inferior spectral-line subtraction
because of the rotation of the main galaxy.
(For the higher-resolution MagE spectrum of \nuctwo\ discussed in
Section \ref{sec342} below, we chose the first method since good spectral-line
subtraction is imperative for measuring accurate velocities and velocity
dispersions.)

As Figure~\ref{fig06} shows, the spectra of the two nuclei of \n7727 are
strikingly different.
The spectrum of \nucone\ shows the usual absorption lines characteristic
of old stellar populations (e.g., \ion{Ca}{2} H\,+\,K, \ion{Mg}{1} b triplet,
and \nai\ D doublet) and thus resembles the nuclear spectra of ``red and
dead'' early-type galaxies.
In contrast, the spectrum of \nuctwo\ features a significantly bluer continuum
and shows---in addition to the tracers of old stellar populations---strong
Balmer absorption lines.
The lines \hbet, \hgam, \hdel, and \heps\ are easily visible in it, as a
comparison with the spectrum of the A7\,V star HD\,111525, taken from the
library of stellar spectra by \cite{jaco84}, shows.
Note that the next two Balmer lines, H8 and H9 (unmarked), are also visible
near the blue end of the displayed spectrum (corresponding to the two bluest
absorption lines seen in the A7\,V spectrum).

Hence, the spectrum of \nuctwo\ is clearly composite in nature and
reminiscent of the post-starburst spectra often observed in local merger
remnants \citep[e.g.,][]{schw78,schw82,zabl96,nort01} and sometimes dubbed
``E+A'' or ``K+A'' spectra \citep{dres83,fran93}.
We infer from this composite spectrum that \nuctwo\ probably experienced
a starburst in the recent past ($\la$\,2~Gyr), while \nucone\ apparently
did not.
Section \ref{sec35} below addresses this issue in more quantitative detail
via a spectral synthesis leading to approximate star-formation histories for
the two nuclei.

%%%%%%%%%%%%%%%%%%%%%%%%%%%%%%%   Fig. 06   %%%%%%%%%%%%%%%%%%%%%%%%%%%%%%%%
%%%%%%%%%%%%%%%%%%%%%%%%%%%%%%%%%%%%%%%%%%%%%%%%%%%%%%%%%%%%%%%%%%%%%%%%%%%%

\subsubsection{Systemic Velocities and Central Velocity Dispersions}
\label{sec342}

To measure the systemic velocities and central velocity dispersions of
the two nuclei, we extracted their 1D spectra from the MagE rectified and
sky-subtracted 2D spectrum (Section \ref{sec222}).
The extraction was somewhat complex and involved the first method alluded
to above.
For each of the 11 spectral orders used and each nucleus, we extracted
spectral strips 3 spatial pixels ($0\farcs82$) wide and centered each
on the brightness peak of its nucleus, using the IRAF task {\em apsum}.
Counts were summed at each wavelength to form a 1D order spectrum.
For \nucone\ we again did not subtract any background.
For \nuctwo, on the other hand, we fitted the background spectrum of the main
galaxy at each wavelength in the spatial direction with a quadratic function,
excluding a 10-pixel ($2\farcs75$) wide stretch centered on \nuctwo.
The fitted background falling into the 3-pixel aperture used for \nuctwo\
was then subtracted from the counts in this aperture to yield the 1D net
spectrum of \nuctwo.
In two final steps, each order spectrum was normalized to a continuum
level of 1, and the 11 order spectra were spliced into a single
1D spectrum for each nucleus.
This spectrum refers to a rectangular area of $0\farcs82\times 0\farcs70$
($109\times 93$ pc) centered on the nucleus.

%%%%%%%%%%%%%%%%%%%%%%%%%%%%%%%   Fig. 07   %%%%%%%%%%%%%%%%%%%%%%%%%%%%%%%%
\begin{figure*}
  \begin{center}
    \includegraphics[width=17.0cm]{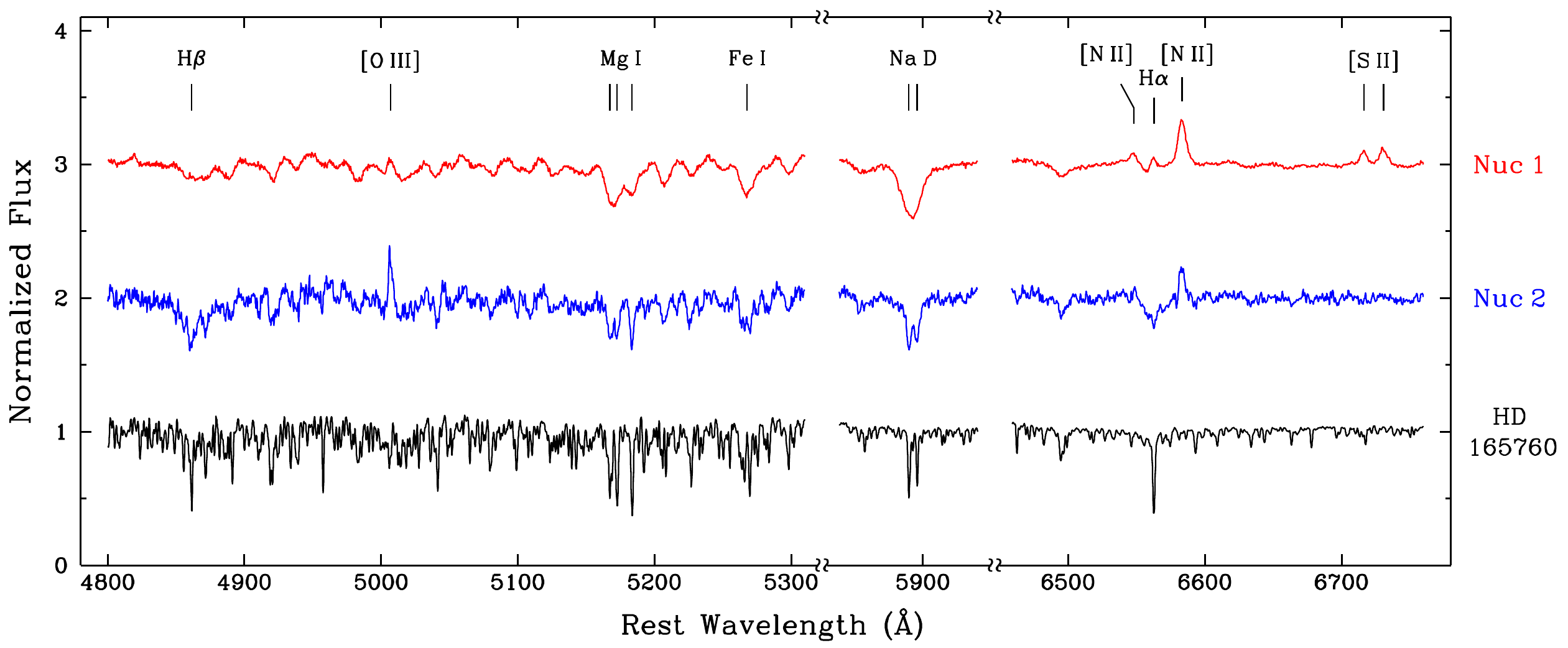}
    \caption{
Spectral details in three wavelength regions shown for the two nuclei of
\n7727 and a G8\,III star for comparison.
The spectra were obtained with MagE on the Clay 6.5 m telescope and are
plotted versus rest wavelength.
Each was extracted with an aperture of $0\farcs82\times 0\farcs70$
($109\times 93$ pc) centered on its respective nucleus.
Counts have been normalized to the continuum (set to 1), and the top
two spectra have been shifted by +2 and +1, respectively, relative to the
bottom spectrum for better display.
The spectral resolution is $R\approx 5800$.
Note the strong broadening of the stellar absorption lines in \nucone,
due to its central velocity dispersion of $\sigvelone = 190\pm 5$ \kms,
and the weaker---but still noticeable---broadening of the lines in \nuctwo\
($\sigveltwo = 79\pm 4$ \kms) when compared to those of the G8\,III star
HD\,165760.
    \label{fig07}}
  \end{center}
\end{figure*}
%%%%%%%%%%%%%%%%%%%%%%%%%%%%%%%%%%%%%%%%%%%%%%%%%%%%%%%%%%%%%%%%%%%%%%%%%%%%

Figure~\ref{fig07} shows segments of the resulting spectra of \nucone\
and 2 covering the rest wavelengths 4800--5310 \AA, 5830--5950 \AA,
and 6450--6780 \AA.
The spectrum of the G8\,III star HD\,165760, used in the subsequent
analysis, is included for comparison.
Note the broadening of spectral lines in the two nuclei when compared
to the same lines in the G8\,III star.
This broadening is especially easy to see for the \mgi\ triplet and the
\nai\ D doublet (both marked).

Using the spectra shown in Figure~\ref{fig07}, we measured the systemic
heliocentric velocities \czhel\ and central velocity dispersions \sigvel\
of each nucleus by cross-correlating its spectrum with the spectrum of
HD\,165760.
We used the technique developed by \citet[][hereafter TD79]{td79} and
encoded in the IRAF task {\em fxcor}.
We experimented with various spectral regions and used the ratio $r$,
defined as the height of the true correlation peak divided by the
average peak of the correlation function (TD79), to judge the quality of
the determined velocities and dispersions.
As is often found, a wavelength region centered on the \mgi\ triplet
($\lambda_0\approx 5175$) yielded measurements with the smallest errors,
although differences between results obtained with the adopted region,
$\lambda\lambda_{\rm obs} = 5100$--5450, and with various other regions
up to $\sim\,$1400 \AA\ wide were relatively minor.

The central {\em radial-velocity dispersions} measured in the
$109\times 93$ pc apertures were fully corrected for instrumental
broadening\footnote{
As explained in TD79, measured velocity dispersions have to be corrected
for the instrumental broadening of both the template star and the stars
in the galaxy.}
and are $\sigvelone = 190\pm 5$ \kms\ for \nucone\ and $\sigveltwo =
79\pm 4$ \kms\ for \nuctwo.
Note that the velocity dispersion of \nuctwo\ is surprisingly large.
Given the ratio $\sigveltwo/\sigvelone\approx 0.42$ (or its inverse,
$\sigvelone/\sigveltwo = 2.40\pm 0.14$) and the Faber--Jackson relation
$L\propto \sigvel^4$ \citep{fabe76}, one would expect \nuctwo\ to be
about 33$\times$ less luminous than the main galaxy, yet we measured it
to be 6.2 mag fainter in $V$ (Section \ref{sec33}), which means that it
is about 300$\times$ less luminous.
An interesting question, addressed in Section \ref{sec4} below, is how much
of this crass difference may be due to stellar population differences, the
presence of an abnormally massive SMBH, and/or tidal stripping.

The systemic {\em heliocentric radial velocities} measured for \nucone\
and 2 via cross-correlation with HD\,165760 are $\czhelone = 1865.3\pm 8.4$
\kms\ and $\czheltwo = 1881.4\pm 2.1$ \kms, respectively.
Our systemic velocity for \nucone\ and, hence, \n7727 agrees well with
that measured by \citet{roth06} from the \ion{Ca}{2} triplet in the near
IR,\,\ $\czhel = 1868\pm 2$ \kms.

The difference in line-of-sight (LOS) velocity between \nuctwo\ and the
main galaxy, $\dvlos\equiv \czheltwo - \czhelone = +16.1\pm 8.7$ \kms, is
of special interest for two reasons.
First, its small value shows that \nuctwo\ is kinematically closely
associated with \n7727, as we already guessed from its BTS (Section
\ref{sec323}).
And second, the quoted value places an essential constraint on acceptable
orbital parameters in any future model simulation of this interesting
late-stage merger.

The relatively large error in \dvlos, amounting to more than half the
measured value itself, is caused mainly by the $\pm\,$8.4 \kms\ error in
\czhelone, which is 4$\times$ larger than the error in \czheltwo.
Although this surprisingly large velocity error for \nucone\ may, in part,
be due to the 190 \kms\ velocity dispersion, it may also be caused by a
mismatch of spectral features between \nucone\ and the G8\,III template star.
A K0\,III or K2\,III star might have provided a better match, but
was not observed with MagE.
To test this possibility and obtain an improved value of \dvlos, we
cross-correlated the spectrum of \nucone\ with that of \nuctwo, thus
making a {\em direct}\, measurement of the velocity difference,
$\dvlos = +18.8\pm 3.1$ \kms.
We adopt this as our final value for \dvlos, and we also slightly adjust
the two measured nuclear recession velocities (in proportion to their
errors) to yield this directly measured difference.
Table~\ref{tab04} summarizes the results of our various cross-correlation
measurements just described.

%% ###############################  Table 4  ################################
\begin{deluxetable*}{lcccc}
\tablecolumns{5}
\tablewidth{0pt}
%%\tabletypesize{\footnotesize}		% Too small!
%\tabletypesize{\small}
\tablecaption{Velocities and Velocity Dispersions of \n7727 and \nuctwo\label{tab04}}
\tablehead{
\colhead{Parameter\phm{AAA}}    &
\colhead{Symbol}                &
\colhead{NGC 7727 (\nucone)}    &
\colhead{Nucleus 2}		&
\colhead{\dvlos(Nuc\,2 $-$ Nuc\,1)}
}
\startdata
Aperture size\tnm{a} &			\nodata &  $0\farcs82\times 0\farcs70$ &  $0\farcs82\times 0\farcs70$ &  \nodata \\
Measured heliocentric velocity\tnm{b} &	\czhel  &  $1865.3\pm 8.4$ &       $1881.4\pm 2.1$ &       $+16.1\pm 8.7$\tnm{c} \\
Adopted heliocentric velocity\tnm{d} &	\czhel  &  $1863.2\pm 8.7$ &       $1882.0\pm 2.2$ &       $+18.8\pm 3.1$\tnm{e} \\
Central velocity dispersion &		\sigvel &  $190\pm 5$ &            $79\pm 4$ &                    \nodata
\enddata
\tablecomments{All velocities and velocity dispersions are measured from the MagE spectrum and given in \kms.}
\tnt{a}{{L}onger dimension oriented at P.A. $= 151\farcs9$; the apertures project to $109\times 93$ pc.}
\tnt{b}{Measured by cross-correlation with the spectrum of HD\,165760 (G8\,III).}
\tnt{c}{Velocity difference between the two measured heliocentric velocities.}
\tnt{d}{Measured velocities adjusted to yield exactly the measured velocity difference; see text for details.}
\tnt{e}{Velocity difference measured directly by cross-correlation between the spectra of
        \nucone\ and 2.}
\end{deluxetable*}
%% ###########################  End of Table 4  #############################

The dynamical mass of \nuctwo\ derived from the measured velocity dispersion
is presented in Section \ref{sec36} below and further discussed in Section
\ref{sec422}.

\subsection{Stellar Populations and Star Formation Histories of the Two Nuclei}
\label{sec35}

As illustrated in Figure~\ref{fig06}, the two nuclei of \n7727 feature
markedly different spectra.
Whereas the spectrum of \nucone\ is characteristic of old stellar populations,
that of \nuctwo\ is of a more composite nature, suggesting the presence of
relatively young stars of spectral type A in addition to an old population.

To estimate the proportions of young and old stars in each nucleus, we
modeled the two flux-calibrated LDSS-3 spectra with the spectral-synthesis
code STARLIGHT (Version 4), developed and made publicly available by
\citet{cidf05,cidf07,cidf09}.
We used two libraries of model spectra provided with the package, both
based on \citet[][hereafter BC03]{bc03} models and a \citet{chab03}
initial mass function.
One library, called BC03.N, contains 45 model spectra for populations of
15 different ages and 3 metallicities, while the other library, BC03.S,
contains 150 model spectra for populations of 25 different ages and
6 metallicities.
We allowed for internal extinction with a \citet{card89} law and experimented
with various spectral-line maskings, finding that masking the three emission
lines \hbet\ and \oiiiboth\ improved the fits, while masking the Na D
absorption doublet made surprisingly little difference (with one exception
noted below).

The main difficulty we faced was a slightly incorrect shape of the spectral
continuum, especially a drooping of a few percent toward the red ends of
the two spectra.
This drooping was likely caused by the spectrograph slit deviating from
the vertical by $\sim\,$32$\degr$ during the observation, leading to some
light losses caused by differential atmospheric refraction.
Since by default the STARLIGHT code gives as much weight to continuum
fluxes as to spectral-line fluxes, the poor fits near the ends of the
spectra led to relatively large chi-squares per data point used
($\chi^2/N_{\rm u}$) and to unrealistic, negative values of internal
extinction.
We minimized the problem by restricting our spectral fits to the rest
wavelengths $\lambda_0 = 3850$--5950 \AA, which led to $\chi^2/N_{\rm u}$
values of $\sim\,$0.37 for fits to the spectrum of \nucone\ and
$\sim\,$0.80 for the more noisy spectrum of \nuctwo.
With this restricted wavelength range, we found the results from various
STARLIGHT fits remarkably robust against changes in spectral library,
extinction law, and initializing random number.

%%%%%%%%%%%%%%%%%%%%%%%%%%%%%%%   Fig. 08   %%%%%%%%%%%%%%%%%%%%%%%%%%%%%%%%
\begin{figure}
  \begin{center}
    \includegraphics[width=8.4cm]{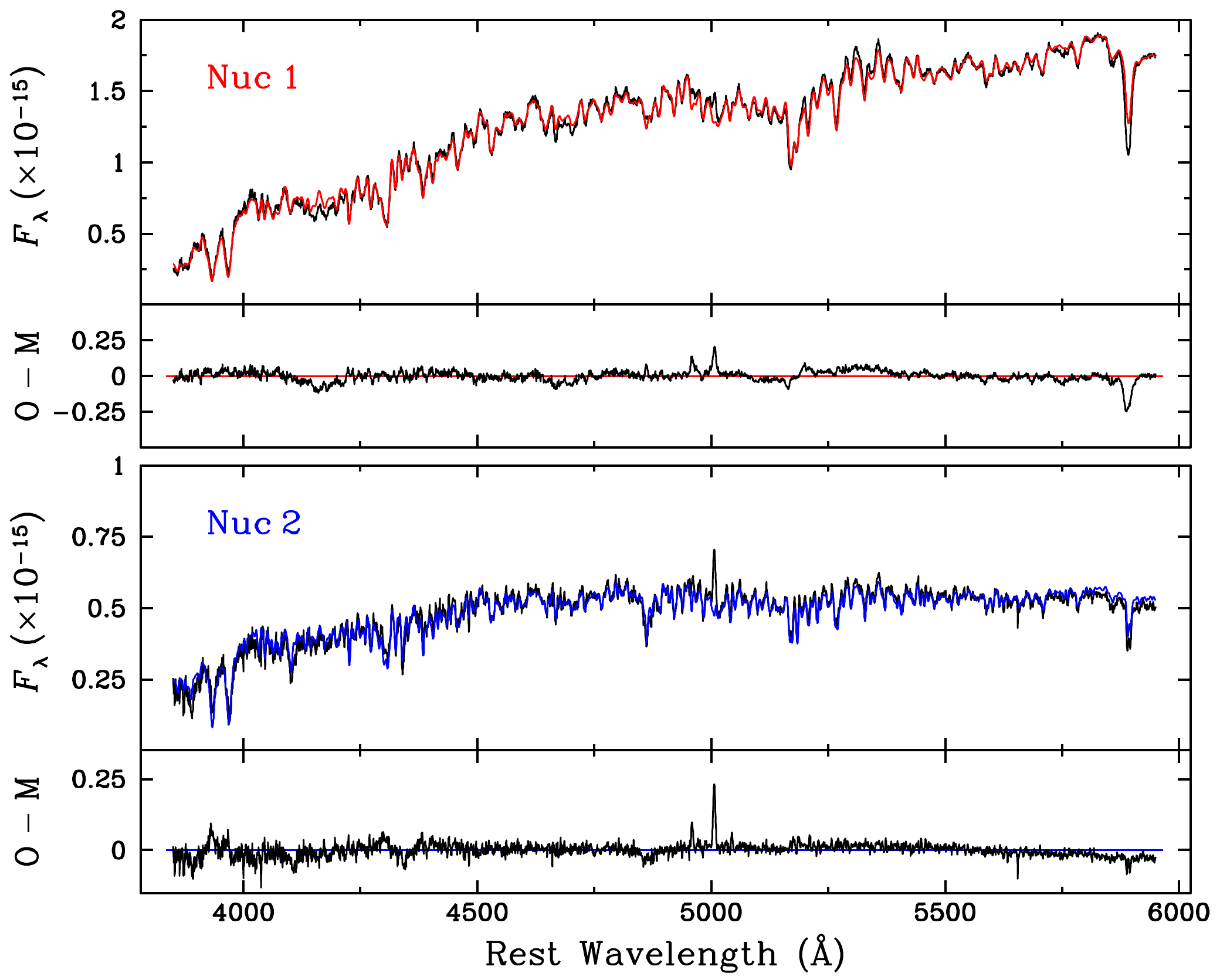}
    \caption{
Spectral-synthesis fits obtained with STARLIGHT for \nucone\ (top) and
\nuctwo\ (bottom) of \n7727.
The observed spectra are plotted in black vs rest wavelength, with fluxes
in units of $10^{-15}$ \ergscma, while model spectra synthesized and fit
by STARLIGHT are overplotted in color.
Note that the flux scale for \nuctwo\ is expanded by a factor of 2$\times$
compared to that for \nucone.
The residuals from each model fit (O $-$ M = observed minus model) are shown
in black underneath the fitted spectrum and on the same flux scale as the
spectrum.
Notice the emission lines \hbet\ and \oiiiboth\ visible in the residual
spectrum of \nucone\ and the two \oiii\ lines in that of \nuctwo.
For details about the fits, see text.
    \label{fig08}}
  \end{center}
\end{figure}
%%%%%%%%%%%%%%%%%%%%%%%%%%%%%%%%%%%%%%%%%%%%%%%%%%%%%%%%%%%%%%%%%%%%%%%%%%%%

Figure~\ref{fig08} shows the spectra of the two nuclei (black lines), each
superposed with one representative model fit (red and blue lines) picked
from the 10 final fits with different initializing random numbers.
These final fits were all made with the Na D line doublet {\em un}masked,
i.e., fully included in the fit, but differ only imperceptibly from those
with the doublet masked.
To facilitate judging the fits, the residuals ``O $-$ M'' (for Observed
minus Model spectrum) are shown below the fit for each nucleus.
Notice the \hbet\ and \oiiiboth\ emission lines in the residual spectrum
of \nucone\ and the same two \oiii\ emission lines in that of \nuctwo.
Of the absorption lines, the Na~D doublet is obviously insufficiently fit
by the model spectrum in \nucone, but better---though not completely---fit
in \nuctwo.
The poor fit for \nucone\ is likely due to a combination of interstellar
\nai\ absorption within the nuclear region of \n7727 and, perhaps, a
deficiency in giant stars with strong Na D absorption in the stellar model
library.
In contrast, the \ion{Ca}{2} H\,$+$\,K lines are well fit in \nucone, but
leave considerable residuals in \nuctwo.
Despite these flaws, both the $\chi^2/N_{\rm u}$ values quoted above and
the plotted spectral model fits appear overall satisfactory, given the
limited quality of our LDSS-3 spectra and, especially, the challenging
subtraction of the bright galaxy background from the spectrum extracted 
for \nuctwo\ (Section \ref{sec341}).

The STARLIGHT fits to the nuclear spectra yield interesting information
about the stellar populations and star-formation history of each nucleus.

For \nucone, all fits involve only the two oldest stellar populations
(11 and 13 Gyr) and the top two metallicities (\zsun\ and 2.5\,\zsun)
of the BC03.N model library that we used.
Hence, the stellar populations of \nucone\ are $\sim$100\% old, with
a formal light-weighted age of $12.9\pm 0.1$ Gyr and metallicity of
$2.0\pm 0.2\,\zsun$.
The small quoted error in age is derived from the 10 final STARLIGHT
fits with different random-number initializations and probably
underestimates the astrophysical uncertainties by a factor of $\sim\,$10--20.
But the main result remains that there is no sign of any young or
intermediate-age ($\la5$ Gyr) stellar population in the primary nucleus
of \n7727.
If present, any such young population would have to be heavily obscured.

The mix of stellar populations in \nuctwo\ is more newsworthy.
All STARLIGHT fits reveal the presence of a significant ``young'' population
of stars with a light-weighted mean age of $1.4\pm 0.1$ Gyr in
addition to the preponderant old stellar population.
The latter has itself a slightly younger light-weighted age,
$11.3\pm 0.3$ Gyr, than the old population of \nucone, as well as a
distinctly lower metallicity of $Z_{\rm old,2} \approx \zsun$.
Interestingly, all STARLIGHT fits for \nuctwo\ make use of only 4 of
the 15 age components in the BC03.N library: the ages of these four
components are 640 Myr and 1.43 Gyr for the young populations and 11 and
13 Gyr for the old populations.
Neither the intermediate-age 2.5 and 5 Gyr components in the library
nor any of the eight components with very young ages ($<\,$300 Myr) are
involved.
Hence, \nuctwo\ appears to have formed the majority of its stars in the
distant past ($\ga 10$ Gyr) and have then experienced one or perhaps two
starbursts during a period that occurred about 0.5--1.8 Gyr ago.
These starbursts were metal-rich ($Z \ga \zsun$) and contributed
significantly to the present stellar mass in \nuctwo: about $38\%\pm 4$\%
of the mass is in young stars, while $62\%\pm 4$\% of it is in old stars.
Although a minority in mass, the young stars contribute as much as
$\sim$\,80\% of the total flux at $\lambda_0 = 4020$ \AA\ (the default
normalization point used by the STARLIGHT code), while the mass-dominant
old stars contribute only $\sim$\,20\% of it.

These mass and light ratios between young and old stellar
populations in \nuctwo\ are rather sensitive to the treatment of the
Na D absorption doublet in the model fitting process.
If instead of including the doublet in the fit one masks it out, the mass
fraction in young stars increases from 38\% to 50\% (i.e., equal mass in
young and old stellar populations), and the violet-light ($\lambda$4020)
fraction contributed by the young population increases from 80\% to
nearly 87\% of the total.
This suggests that including the Na D doublet in the fit is important for
determining the fraction of light from late-type giant stars, even
though---if ISM absorption is not the culprit---such stars in the BC03.N and
BC03.S libraries seem to possess too weak Na doublets to fully reproduce
the observed doublet strength in \nuctwo\ as well as in \nucone.

The recent starburst episode in \nuctwo\ {\em may} have consisted of two
separate bursts of unequal strength.
STARLIGHT fits with both the BC03.N library and the BC03.S library suggest
that the main burst occurred about 1.2--1.5 Gyr ago, producing about
98\%--99\% of the present-day mass in the young stellar population.
A second, much weaker burst then occurred about 500--650 Myr ago,
producing the remaining 1\%--2\% of the mass in young stars
(corresponding to 0.3\%--0.7\% of the {\em total\,} stellar mass,
young and old).
Since with both libraries STARLIGHT made no use of an existing 905 Myr old
component (i.e., assigned zero mass to it), we believe that the separation
of the recent starburst into a main burst 1.2--1.5 Gyr ago, followed by
a weak second burst about 0.7--1.0 Gyr later, may be real.
However, new spectra of higher quality and more sophisticated population
syntheses will be needed to clarify this issue.

\subsection{Dynamical and Stellar Mass of \nuctwo}
\label{sec36}

The measurements and results presented in Sections \ref{sec33}--\ref{sec35}
above allow us to estimate both the dynamical mass and the stellar mass of
\nuctwo.

For a spherical stellar system in virial equilibrium, the {\em dynamical
mass}\, is $\mdyn = a \cdot \rhalf \cdot 3\sigvel^2 / G$, where \rhalf\ is
the half-mass radius, \sigvel\ the mean radial velocity dispersion, $G$ the
gravitational constant, and $a \approx 2.5$ a profile-dependent numerical
factor valid for most globular clusters in our Galaxy (\citealt{spit87},
pp.\ 11--12; \citealt{mara04}).
Approximating \rhalf\ with $4\reff/3$ \citep{spit87}, we get
$\mdyn = 10.0\cdot\reff\cdot\sigvel^2/G$, which---with the measured values of
$\refftwo = 29\pm 2$ pc and $\sigveltwo = 79\pm 4$ \kms\ (see Tables
\ref{tab03} and \ref{tab04})---yields the dynamical mass of \nuctwo,
$\mdyntwo = (4.2\pm 0.6)\times 10^8\,\msun$.
This mass is $\sim\,${}$10^2\times$ that of $\omega$~Cen, the most massive
globular cluster in the Milky Way \citep[e.g.,][]{meyl97}, and makes \nuctwo\
a close kin of the most massive UCD discovered so far in the Virgo Cluster,
M59-UCD3 ($\mdyn = 3.7\times 10^8\,\msun$, \citealt{liu15b}).

The above estimate of \mdyntwo\ might be questioned on grounds that
\nuctwo\ is the remnant of a former galaxy in the process of being stripped
and can, therefore, hardly be in virial equilibrium.
Yet, various arguments and simulations suggest that the mean velocity
dispersion of such a nucleus may change by only a few percent (see
\citealt{forb14} for a review), presumably especially if the nucleus
harbors an SMBH.
Here, we simply note that \nuctwo\ and M59-UCD3 have not only similar
dynamical masses (4.2 and $3.7\times 10^8\,\msun$, respectively), but also
similar effective radii ($29\pm 2$ pc and $25\pm 2$ pc, respectively)
and mean velocity dispersions ($79 \pm 4$ \kms\ and $78\pm 2$ \kms,
respectively).
Hence, the strength of tidal forces, likely several orders of magnitude
stronger in \nuctwo\ than in M59-UCD3, does not seem to make much difference,
as expected in galactic nuclei where self-gravitation dominates the dynamics
\citep[e.g.,][]{fabe73,pfef13}.

To estimate the {\em stellar mass}\, of \nuctwo, we used two methods, one
based on \ks-band photometry and the other on the STARLIGHT analysis of 
the LDSS-3 spectrum.

For the photometric estimate, we measured the integrated \ks\ magnitude of
\nuctwo\ relative to that of \n7727 from a \ks-band image obtained by
\citet{knap03} with the William Herschel 4.2 m telescope (WHT) on La Palma,
finding a magnitude difference of $\Delta\ks = 6.78\pm 0.15$ mag between
\nuctwo\ and the entire galaxy.\footnote{
The relatively large, $\sim$15\% error reflects mainly uncertainties in the
sky subtraction that are due to the limited, $4\farcm4\times 4\farcm4$
FOV of the image.}
Since the integrated magnitude of \n7727 is $\ks = 7.69\pm 0.03$
\citep{jarr03}, the total apparent magnitude of \nuctwo\ is
$\kstwo = 14.47\pm 0.15$, yielding an absolute magnitude of
$\mktwo = -17.73\pm 0.15$ at $D = 27.4$ Mpc.
This absolute magnitude corresponds to a \ks-band luminosity of
$\lktwo = 2.54_{-0.33}^{+0.37}\times 10^8\,\lksun$ and, via the mean
relation $\mstar/\lk = 0.10\,\sigveleff^{0.45}$ \citep{vdbo16}, to a stellar
mass for \nuctwo\  of $\mstartwo = 1.82_{-0.24}^{+0.27}\times 10^8\,\msun$.

For a second estimate, we used the stellar mass of
$(1.20\pm 0.02)\times 10^8\,\msun$ calculated by STARLIGHT in making its
best fit to the LDSS-3 spectrum of \nuctwo\ (Fig.~\ref{fig08}).
As described in Section \ref{sec35}, this mass is based on
stellar-population libraries and a \citet{chab03} initial mass function.
Two corrections need to be applied to it.
First, given the $\sim$\,$0\farcs7$ seeing during the LDSS-3 observation,
the $1\farcs0\times 1\farcs1$ aperture used to extract the spectrum of
\nuctwo\ missed $\sim$\,18\% of the integrated light.
And second, the $\pm 30$\% uncertainty in our flux calibration
(Section \ref{sec221}) translates into a similar uncertainty for \mstar.
With these two corrections applied, the population-synthesis-based estimate
of the stellar mass of \nuctwo\ becomes
$\mstartwo = (1.46\pm 0.44)\times 10^8\,\msun$.

Finally, we average the photometric and STARLIGHT-based values of \mstartwo\
with weights 2:1 and obtain, as our best estimate,
$\mstartwo = (1.7\pm 0.3)\times 10^8\,\msun$.

Note that the dynamical mass of \nuctwo\ exceeds this stellar mass
significantly, with the ratio between the two masses being
$\mdyntwo/\mstartwo = 2.47\pm 0.56$.
We will return to this interesting fact when we discuss the nature and
origin of \nuctwo\ (Section \ref{sec41}) and the likely mass of its SMBH
(Section \ref{sec422}).

\subsection{The Interstellar Medium of \n7727}
\label{sec37}

Given \n7727's system of prominent central dust lanes (Section \ref{sec322}),
it comes as somewhat of a surprise that the galaxy appears to be extremely
gas-poor.
In the following, we first review the evidence for a paucity of
interstellar gas in all its known forms (cold, hot, and warm) and then
present evidence that the central diffuse ionized gas counterrotates to
the stars.

\subsubsection{A Remarkable Paucity of Gas}
\label{sec371}

As already mentioned, the {\em cold gas} content of \n7727 appears to be
abnormally low.
In an \hi\ study of 40 Sa galaxies, \citet{bott80} found \n7727 to have
an \hi\ mass of only $\mhi = 3.6\times 10^8\,\msun$ (converted to the
$H_0 = 73$ distance scale, as all masses quoted here are).
They remarked that this galaxy ``appears highly discordant in $\log\mhi$
versus $\log\lbzero$ and $\log\sighi$ versus $\log\lbzero$ diagrams'' of
their sample, where \lbzero\ is a galaxy's intrinsic $B$-band luminosity and
\sighi\ is the mean \hi\ surface mass density within the optical $B = 25$
\magarcs\ isophote.
Note that \sighi\ is distance independent.
In both diagrams, \n7727 appears to have an order-of-magnitude lower
\hi\ content than other Sa galaxies of comparable luminosity.
Even though \citet{huch82} measured a somewhat higher mass of
$\mhi = (5.7\pm 1.8)\times 10^8\,\msun$, he---too---found an abnormally
low ratio of \hi\ mass to optical luminosity and of \hi\ mass to total
dynamical mass, $\mhi/\mtot \approx 0.002$.
%%%[CHECK: Does this need to be corrected for distance?  A: Yes, done;
%%%        i.e., changed fr/0.003 to 0.002 (exactly: 0.0028 -> 0.0019)]

Given that the \hi\ content of \n7727 seems to have been measured only
twice and long ago \citep{bott80,huch82}, we used data from the \hi\
Parkes All-Sky Survey (HIPASS; \citealt{barn01}) to check on the published
values.
From the publicly available HIPASS data
grid,\footnote{\url{http://www.atnf.csiro.au/research/multibeam/release/}}
we measured an integrated \hi\ flux of $\fhi = 1.4_{-0.4}^{+0.8}$ \jykms,
corresponding to a mass of $\mhi = 1.8_{-0.5}^{+1.0}\times 10^8\,\msun$
or about half the value measured by \citet{bott80}.
Our relatively large error bars reflect the fact that some \hi\ flux from
the companion galaxy \n7724 (type SBab), located at $12\farcm3$ projected
distance from \n7727, may contribute to the \hi\ flux measured for the latter.
This is because the beam size of the gridded HIPASS data has a FWHM
$= 15\farcm5$ and \n7724 contains about 20\% more \hi\ than \n7727 does.
Our low new value for \mhi\ agrees well with the value of
$\mhi = 1.86_{-0.50}^{+0.68}\times 10^8$ derived from homogenized \hi\ data
and currently (2017) given in the database HyperLEDA \citep{patu03}.

With the new, HIPASS-based value of \mhi\ we can compare the \hi\ content
of \n7727 to a sample of Sa, S0/a, and S0 galaxies observed at Arecibo
with high sensitivity \citep{eder91}.
The new value of $\sighi = 0.20_{-0.06}^{+0.11}\,\msunpc$ for \n7727 falls,
indeed, well below values for Sa galaxies, but is compatible with values
observed in S0 and S0/a galaxies, though it is low even there.
Specifically, the {\em logarithmic} value, $\log(\sighi/\msunpc) = -0.69$,
falls $4.2\,\sigma$ below the mean value of $0.57\pm 0.30$ (single-object
standard deviation) measured for Sa galaxies at Arecibo and completely
outside their observed range, but falls only $1.5\,\sigma$ below the mean
value of $0.12\pm 0.54$ for S0/a galaxies and $1.7\,\sigma$ below that of
$0.15\pm 0.50$ for S0 galaxies.
Overall, this comparison strongly supports the claim made by \citet{bott80}
that \n7727 has an abnormally low \hi\ content for a galaxy of its type
and luminosity.

This fact does not change when the {\em molecular} gas in the galaxy is
included.
Recent CO observations of \n7727 with ALMA yield a molecular-gas mass of
$\mhtwo = 1.05\times 10^8\,\msun$ \citep{ueda14}, confirming an earlier
estimate of $1.0\times 10^8\,\msun$ by \citet{crab94}.
Thus, the {\em total cold gas} content of this galaxy appears to be only
about $\mhihtwo = 2.8_{-0.5}^{+1.0}\times 10^8\,\msun$.

Interestingly, the {\em hot gas} ($T>10^5$\,K) content of \n7727 also
appears to be abnormally low when compared to that of eight other
well-known mergers and merger remnants \citep{bras07}.
After subtracting all X-ray point sources from the 19~ks exposure of \n7727
taken with \chandra/ACIS-S (see Section \ref{sec31}), these authors find
an intrinsic X-ray luminosity for the hot diffuse gas of
\lx(0.3--6.0 keV)\ $= (9.3\pm 0.5)\times 10^{39}\,\ergsec$ (adjusted to
the here-adopted distance).
This luminosity is a factor of $\sim$3 lower than that of the
hot gas in the optically similarly luminous merger remnant \n7252
($\lx = 2.7\times 10^{40}\,\ergsec$) and is the lowest value measured for
any of the nine mergers studied.
It is also at least one order of magnitude lower than values of \lx\
typically observed in giant ellipticals of similar optical luminosity
\citep{fabb89}.
According to \citet{bras07}, the spectrum of the hot gas in \n7727
is well fitted by a single-component MeKaL model with a global temperature
of $0.60_{-0.06}^{+0.07}$\,keV (corresponding to $T\approx 7\times 10^6$\,K),
a temperature similar to that observed in the other mergers studied.
The hot gas itself appears rather centrally concentrated, with its
detected emission extending only to about half the optical radius of the
galaxy's main body.
As noted in Section \ref{sec31}, this hot gas is centered on \nucone, the
primary optical nucleus of \n7727, rather than on \nuctwo.

Finally, \n7727 appears to also possess surprisingly little {\em warm
%% ($\sim$10$^4$\,K) gas}.
gas} ($\sim$10$^4$\,K).
The galaxy is devoid of any giant \hii\ regions, at least out to
$\sim\,$7~kpc ($0\farcm9$) radius in its main body. 
This can be seen from the nice continuum-subtracted \halp\ image obtained
of this region by \citet{knap06} with WHT on La Palma and shown in their
Figure~1.
The contrast with a similar \halp\ image of \n7723, the SBb galaxy with
$\sim\,$60\% the optical luminosity of \n7727 in the same group (Section
\ref{sec1}), is striking.
Taken with the same telescope, filter, and exposure time, the \halp\ image
of \n7723 shows about 170 distinct giant \hii\ regions, whereas that of
\n7727 shows none.
This severe lack of \hii\ regions in \n7727 does not, per se, argue against
the presence of cold or warm gas in the galaxy, but shows that---if such gas
%% is present---no massive OB stars have recently ($\la\,$10 Myr) formed in it.
is present---star formation including massive OB stars is not currently 
occurring in it.

A small amount of {\em diffuse}\, \halp\ emission near the center of \n7727
is suggested by the \citet{knap06} image and is confirmed spectroscopically
(Section \ref{sec372}).
However, as usual the mass of such rarefied \hii\ gas is insignificant
($< 10^6\,\msun$), even in comparison to the small total mass of
$2.8\times 10^8\,\msun$ in atomic and molecular hydrogen mentioned above.

In short, \n7727 appears to be unusually gas-poor, ranking low even when
compared to S0 galaxies.
This suggests that the galaxy may have experienced a {\em gas blowout}\, in
the past, possibly associated with the event that led to a starburst
in \nuctwo\ about 1.2--1.5 Gyr ago (Section \ref{sec35}), but possibly
also longer ago.

\subsubsection{A Counterrotating Gas Disk}
\label{sec372}

Although our long-slit spectrum of \n7727 obtained with LDSS-3 at
P.A.\,$= 152\degr$ across the two nuclei does not fall close to either
the major axis of the central body
(P.A.(stars)\,= $112\degr\pm 13\degr$, \citealt{lho11}) or the
minor axis, it yields interesting new information about the relative
kinematics of the ionized gas and the stars.

%%%%%%%%%%%%%%%%%%%%%%%%%%%%%%%   Fig. 09   %%%%%%%%%%%%%%%%%%%%%%%%%%%%%%%%
\begin{figure}
  \begin{center}
    \includegraphics[width=8.0cm]{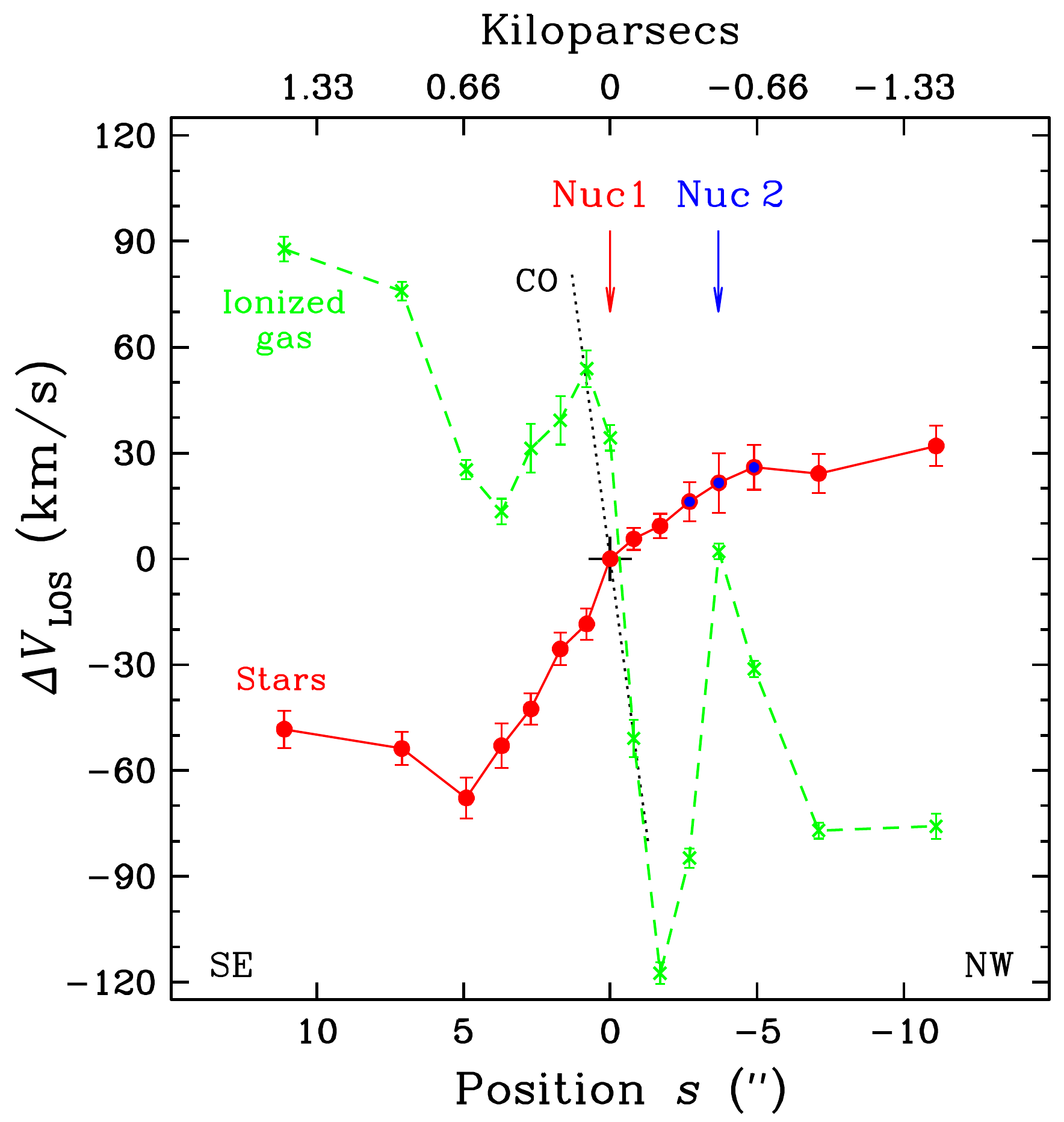}
    \caption{
Line-of-sight velocity profiles of stars and ionized gas in \n7727, measured
along a line crossing the two nuclei at P.A.\,= $151\fdg9$ from NW to SE.
Note that at that P.A. the ionized gas, measured from its \oiiisev\ emission
line, overall counterrotates to the stars.
However, it corotates with the molecular gas disk \citep{ueda14}, whose
central gradient is indicated by the dotted line labeled ``CO.''
Whereas the stars show roughly the velocity curve of a rotating disk, the
ionized gas exhibits major perturbations from a disk in pure rotation,
especially around the position of \nuctwo.
For further details, see text.
    \label{fig09}}
  \end{center}
\end{figure}
%%%%%%%%%%%%%%%%%%%%%%%%%%%%%%%%%%%%%%%%%%%%%%%%%%%%%%%%%%%%%%%%%%%%%%%%%%%%

Figure~\ref{fig09} shows radial velocities of the ionized gas and stars,
all relative to the systemic velocity of \n7727, plotted versus position
$s$ along the slit.
Since our LDSS-3 spectrum does not cover \halp, we measured the ionized-gas
velocities from the one easily visible emission line, \oiiisev, which on
our spectrum exposed for 23 minutes could be traced and measured out to
$\pm12\arcsec$ (1.6 kpc) from \nucone.
The stellar absorption-line velocities were measured by cross-correlating
spectra extracted at various positions along the slit with the spectrum of
\nucone, using the \mgi\ triplet region
($\lambda\lambda$5120--5540).\footnote{
The light from \nuctwo\ was not subtracted for these crosscorrelations.}
The striking result is that, overall, the ionized gas {\em counterrotates}\,
to the stars, at least along the cross section at P.A.\,$= 152\degr$.
Note that in the central region of $\sim\,$2$\arcsec$ radius the
ionized gas rotates at the same speed as does the molecular-gas disk,
whose velocity gradient measured from the CO(1--0) line by \citet{ueda14} is
indicated in Figure~\ref{fig09} by the dotted black line.
This strongly suggests that in this central region the ionized gas seen in 
\oiiisev\ emission forms part of the rotating gas disk of \n7727.

Yet, the velocity profile of the ionized gas clearly shows strong local
deviations from the overall rotation pattern.
We believe that the velocity ``dip'' extending from $s\approx +1\farcs5$
to $+6\farcs5$ (200--860 pc SE) may be caused by obscuring dust that limits
the visibility of the \oiiisev\ line to a foreground part of the gas disk,
while the major velocity ``spike'' near \nuctwo\ may be caused by ionized gas
associated with \nuctwo, rather than with the rotating gas disk of \n7727.
Clearly, deeper spectra obtained with an integral-field-unit (IFU)
spectrograph will be needed to further sort out the various possible velocity
components of the ionized gas (including any possible in- or outflows).

The counterrotation of the gas disk (both molecular and ionized) to the stars
in the central $\sim$\,3 kpc region of \n7727 (Figure~\ref{fig09}) is very
likely real and not just limited to a cross section along P.A.\,$= 152\degr$.
This is because the major axes of both the gas disk and the stellar light
distribution essentially coincide.
The major axis of the molecular-gas disk lies at
P.A.(gas)\,$= 113\degr\pm1\degr$ \citep{ueda14}, while that for the major
axis of the stellar distribution lies at P.A.(stars)\,$= 112\degr\pm 13\degr$
\citep{lho11}.
Hence, the slit P.A.\ of 152$\degr$ used for obtaining the LDSS-3 spectrum
intersected both the gas disk and the stellar body (or disk) at about the
same angle of $\alpha\approx 39\degr$--\,40$\degr$ against the major axes,
reducing the true rotation velocity of both structures by the same factor
of\,\ $\cos\alpha\approx 0.77$ for our measurements.
Of course, we cannot rule out the possibility that some merger-induced
chaotic motions are superposed on the central apparently disk-like rotation
of the stars or, especially, on stellar motions farther out in the galaxy.
Yet, counterrotation of gas to stars in galaxies has long been found to
correlate with past accretion or merger events
\citep[e.g.,][]{ward86,bert90,schw90,kann01}.
Hence, the situation in \n7727 clearly fits this well-established pattern.

\section{DISCUSSION}
\label{sec4}

The present discussion addresses various issues concerning the nature and
recent history of \n7727 and its two nuclei.
By comparing \nuctwo\ with four well-studied UCDs, we first make the case
that it itself is a UCD and a smoking gun for understanding UCD formation and
evolution.
We then describe our search for nuclear activities in \n7727, finding
\nucone\ to be a weak low-ionization nuclear emission-line region (LINER)
in an old stellar population, while \nuctwo\ shows evidence for low-level
AGN activity embedded in a post-starburst population.
We end this discussion by reviewing the evidence that in \n7727 we
witness the aftermath of the ingestion---over the past 2 Gyr---of a
gas-rich disk companion.

\subsection{\nuctwo: A Smoking Gun for UCD Formation and Evolution}
\label{sec41}

The second nucleus of \n7727 is a fascinating object.
As described above, it shares several properties with UCDs, but also
features some that are unique for a UCD so far:  it is very compact
($\reff = 29$ pc) and massive ($\mdyn = 4.2\times 10^8\,\msun$), has an
elevated ratio of $\mdyn/\mstar$ characteristic of massive UCDs, and
features a unique post-starburst spectrum and BTS.
In combination, these properties suggest that \nuctwo\ is the clearest case
yet of a {\em recently formed UCD}, besides also being on a par in mass
with the most massive UCD currently known, M59-UCD3 \citep{liu15b}.

Table \ref{tab05} compares \nuctwo\ with four well-studied massive UCDs,
with the objects listed in order of increasing luminosity and dynamical mass.
Note that in the visual passband \nuctwo\ is about twice as luminous as
M59-UCD3, in good part because of its major starburst $\sim$\,1.4~Gyr ago
(Section \ref{sec35}).
In that respect it is unique among the five objects, a testimony to its
recent formation as a UCD.
Because of the significant fraction ($\sim\;$38\%--50\%) of its stellar
mass in relatively young stars, there can be little doubt that its parent
galaxy contained copious amounts of gas.

%% ###############################  Table 5  ################################
\begin{deluxetable*}{lcccccccc}
\tablecolumns{9}
%\tablewidth{0pt}
\tablewidth{15.5truecm}
%%\tabletypesize{\small}		% Fills 1.05 pages!
%\tabletypesize{\footnotesize}		% "Better": small font, b/only 1 page
\tablecaption{Comparison of \nuctwo\ with Four Massive UCDs\label{tab05}}
\tablehead{
   \colhead{Parameter}	  &
   \colhead{}		  &
   \colhead{VUCD3\tnm{a}} &
   \colhead{M59cO}	  &
   \colhead{M60-UCD1}	  &
   \colhead{M59-UCD3}	  &
   \colhead{{\bf NGC\,7727:Nuc\,2}} &	    %% Editor: Please keep this boldfaced if possible.  Thank you. 
   \colhead{}		  &
   \colhead{Units}
}
\startdata
\mv		&&	$-$12.75	&   $-$13.26	    &	$-$14.2		&   $-$14.60	    &   $-$15.49	&&   mag      \\ 
\lv		&&	$1.1\times 10^7$&   $1.7\times 10^7$&	$4.1\times 10^7$&   $5.9\times 10^7$&	$1.3\times 10^8$&&   \lvsun   \\
Distance\tnm{b}	&&	16.5		&   16.5	    &	16.5		&   14.9	    &	27.4		&&   Mpc      \\
\dproj\tnm{c}	&&	14.4		&   10.2	    &	6.6		&   9.4		    &   0.48		&&   kpc      \\
\reff		&&	18		&   32		    &	24		&   $25\pm 2$	    &	$29\pm 2$	&&   pc       \\
\dvlos\tnm{d}	&&	$-573\pm 6$	&   $+259\pm 11$    &	$+182\pm 6$	&   $-12\pm 10$\tnm{e}&	$+19\pm 3$\phs&&   \kms     \\
\sigvel\tnm{f}	&&	33--53		&   29--40	    &	50--100$^+$     &   $78\pm 2$	    &	$79\pm 4$	&&   \kms     \\[1pt]
\mdyn\tnm{g}	&&	$6.6_{-0.8}^{+0.8}\times 10^7$ &
					    $8.3_{-1.2}^{+0.5}\times 10^7$  &
								$2.0_{-0.3}^{+0.3}\times 10^8$ &
										    $3.7_{-0.4}^{+0.4}\times 10^8$ &
													$4.2_{-0.6}^{+0.6}\times 10^8$ &&
															     \msun    \\[3pt]
\mstar		&&	$3.8_{-0.5}^{+0.5}\times 10^7$ &
					    $1.0_{-0.1}^{+0.1}\times 10^8$ &
								$1.2_{-0.4}^{+0.4}\times 10^8$ &
										    $1.8_{-0.3}^{+0.3}\times 10^8$ &
													$1.7_{-0.3}^{+0.3}\times 10^8$ &&
															     \msun    \\[1pt]
$\mdyn/\mstar$  &&	$1.7\pm 0.3$	&   $0.9\pm 0.2$    &	$1.7\pm 0.6$    &   $2.1\pm 0.4$    &   $2.5\pm 0.6$	&&   \nodata  \\[1pt]
\mbh\tnm{h}	&&	$4.4_{-0.9}^{+0.9}\times 10^6$ &
					    $5.8_{-0.9}^{+0.9}\times 10^6$ &
								$2.1_{-0.7}^{+1.4}\times 10^7$ &
										    \nodata	    &	$3\times 10^6$--$3\times 10^8$ &&
															     \msun    \\[1pt]
\lx\tnm{i}	&&	\nodata		&   \nodata	    &	$\leq 1.3\times 10^{38,}$\tnm{j} &
										    $3.0_{-1.6}^{+1.6}\times 10^{37}$ &
													$2.8_{-0.3}^{+0.3}\times 10^{39}$ &&
															     \ergsec  \\
Age\tnm{k}	&&	$9.6 + 11$	&   $5.5 + 11.5$    &	$\ga$\,10	&   $11\pm 3$	    &	$1.4 + 11.3$	&&   Gyr      \\
Tidal stream?	&&	no		&   no		    &   no		&   no (plume?)	    &	yes (BTS)	&&   \nodata  \\
\mhost		&&	$\sim$10$^9$	&   $\sim$10$^9$    & (0.4--1.1)$\times 10^{10}$& \nodata   & (0.6--6)$\times 10^{10}$&& \msun \\
References\tnm{l} &&	(1), (7)	&   (1), (7)	    &   (2), (3), (6)	&   (4), (5), (6)    &	this paper	&&   \nodata
\enddata
\tablecomments{Objects are listed in order of increasing $V$-band luminosity.}
\tnt{a}{Host galaxy is M87.}
\tnt{b}{Distances are from original sources; note different distances used for M59cO and M59-UCD3.}
\tnt{c}{Projected distance of object from center of its host galaxy.}
\tnt{d}{Velocity difference in sense ``UCD minus host galaxy,'' determined from most reliable 
	  radial velocity measurements for UCD and host galaxy in the literature.}
\tnt{e}{Value is based on velocity from Reference (5); it would be $-86\pm 21$ if based on velocity from  Reference (4).}
\tnt{f}{Velocity dispersion ranges are from published models, with higher value referring to object center, while single values with
		  standard deviations refer to integrated light.}
\tnt{g}{Total dynamical mass, uncorrected for effects of SMBH.}
\tnt{h}{Mass of SMBH.}
\tnt{i}{Intrinsic X-ray luminosity in energy range $\sim$\,0.3--8 keV for M60-UCD1 and M59-UCD3, and 0.3--6 keV for \n7727:Nuc\,2.}
\tnt{j}{X-ray luminosity varies.}
\tnt{k}{Ages expressed as sums represent results from two-population modeling.}
\tnt{l}{Key to references: (1) \citet{mies13}, (2) \citet{stra13}, (3) \citet{seth14}, (4) \citet{sand15},
		  (5) \citet{liu15b}, (6) \citet{hou16}, (7) \citet{ahn17}.}

\end{deluxetable*}
%% ###########################  End of Table 5  #############################

Table \ref{tab05} also shows that the projected distance of \nuctwo\ from
the center of \n7727, $\dproj = 0.48$ kpc, is more than 12$\times$ smaller
than the projected distances of any of the four UCDs from their host galaxies.
In conjunction with the BTS, this small \dproj\ suggests that the tidal
stripping is currently strong because \nuctwo\ lies {\em physically
close} to the center of \n7727.
Therefore, \nuctwo\ has not only formed recently as a UCD, but it also
continues being stripped and, hence, evolving structurally and dynamically.

Yet, despite this tidal stripping, \nuctwo\ has remained very massive so far,
as judged by its high \sigvel, \mdyn, and \mstar\ (Table \ref{tab05}).
As already noted, it is on a par in mass with record-holding M59-UCD3 and
well ahead of (M87-)VUCD3, M59cO, and former record holder M60-UCD1.
An interesting question is whether, and for how long, \nuctwo\ will keep
its high mass.
Several studies of ``galaxy threshing'' suggest that once a galaxy has been
tidally whittled down to its nucleus, the resulting UCD manages to survive
for a prolonged period because of its high internal binding energy
\citep[e.g.,][]{bass94,bekk01,pfef13}.
Longevity may also be enhanced for UCDs that form in {\em minor}\, mergers
($m/M < 1/3$) and for those that harbor an SMBH
\citep[e.g.,][]{call11,vwas12}.

It is, therefore, of interest that \nuctwo\ appears to harbor an AGN,
as indicated by its central X-ray source and discussed in detail in
Section \ref{sec422} below.
This AGN is likely driven by an accreting SMBH, whose mass \mbh\ we estimate
to lie in the range $3\times 10^6\,\msun \la \mbh \la 3\times 10^8\,\msun$,
with perhaps a most likely range of (4--8)$\times 10^7\,\msun$.
While three of the other four UCDs listed in Table~\ref{tab05} also harbor
SMBHs, only the two most massive ones, M60-UCD1 and M59-UCD3, exhibit
possible AGN activity through weak central X-ray sources.
However, the X-ray source in M60-UCD1 \citep{stra13,hou16} is
$\sim$\,20$\times$ less luminous than that of \nuctwo, while the X-ray
source in M59-UCD3 \citep{hou16} is $\sim$\,$10^2\times$ less luminous.
 
As various possible formation scenarios for UCDs have been sorted out,
a broad consensus has emerged that extended star-formation histories or
the presence of tidal streams would be powerful, ``smoking gun''
signatures in support of at least some UCDs having formed as nuclei stripped
of their host galaxies; yet only a few such signatures have been found so far
\citep[e.g.,][]{pfef13,norr15,janz16,vogg16,hilk17}.
Interestingly, \nuctwo\ sports {\em both}\, of these smoking-gun signatures
in extreme forms.

Concerning an {\em extended star-formation history}, \nuctwo\ is one of
only a handful of UCDs with good evidence for it.
Another such UCD is M59cO, which features a bimodal age distribution for
its inner and outer components, with mean ages of 5.5 and 11.5 Gyr,
respectively (Table 5; from \citealt{ahn17}).
Perhaps the best case published so far is \n4546-UCD1: its main star
formation lasted from $\sim$\,12 Gyr to $\sim$\,2 Gyr ago, with some
evidence that it may have continued weakly until $\sim$\,0.6 Gyr ago
\citep[][esp.\ Fig.\ 3]{norr15}.
Yet, \nuctwo\ seems more extreme: it formed nearly $2/3$ of its stars in
the distant past ($\sim$\,9--13 Gyr ago) and then the rest ($\sim$\,38\%)
in a major starburst $\sim$\,1.2--1.5 Gyr ago, likely followed by a weak
one $\sim$\,500--650 Myr ago (Section \ref{sec35}).
About 1.5 Gyr ago is when its gas-rich host galaxy likely began the recent
strong interaction with \n7727, leading to its disintegration except for
the nucleus (Section \ref{sec43} below).

Concerning a {\em tidal stream}, the BTS associated with \nuctwo\ is
nearly unique among UCDs, both for its blue color and for its large extent.
With its $\vitotzero\approx 0.75$ it is $\sim$\,0.2 mag bluer than the
already-blue nucleus, and its measured extent to the NNW is $\sim$\,1.5 kpc,
while we suspect---but cannot prove---that it also extends $\sim$\,1.0 kpc
to the SE (Section \ref{sec323}).
Although a number of UCDs have been found to feature minor asymmetries
that may reflect tidal distortions, few have been found to be associated
with clearly extended tidal streams.
The best case among them so far is UCD-FORS2 (a.k.a. \n1399:78:12),
which sports a tidal stream extending $\sim$\,350 pc on either side of
its main body \citep{vogg16}.
Like \nuctwo, it is blue ($\vi\approx 0.96$) and shows strong Balmer
absorption lines in its spectrum \citep{rich05}.
However,  UCD-FORS2 lies at $\dproj\approx 37$ kpc from \n1399 and has a
luminosity of only $\mv=-11.3$, whence it may be an old globular cluster
or the nucleus of a former {\em dwarf}\, galaxy in the Fornax Cluster.
In either case, its blue color is likely due to low metallicity, whereas
the blue color of \nuctwo\ is clearly due to the recent starburst in
a former host galaxy of about solar metallicity.

In short, with its extended star-formation history, relative youth, and
long BTS, \nuctwo\ really {\em is}\, a ``smoking gun'' for UCD formation
and evolution!

\subsection{Activity in the Two Nuclei of \n7727}
\label{sec42}

Two questions of interest are as follows:
(1) Does \n7727 harbor any SMBHs in its two nuclei?
And (2), does either nucleus show signs of AGN activity or central star
formation?
The main observational surprise here is our finding that the one bright
X-ray point source in the central region of the galaxy, ``Source 5''
detected by \citet{bras07}, coincides with \nuctwo\ rather than with \nucone\
(Section \ref{sec31}), which may point toward some AGN activity in \nuctwo.
In the following we discuss any evidence, or the lack thereof, for the
presence and activity of SMBHs in both nuclei.
We begin with \nucone\ and conclude with the more intriguing \nuctwo.

\subsubsection{\nucone: A Weak LINER}
\label{sec421}
An approximate expected mass for the hypothetical SMBH at the center of
\n7727 can be estimated from the general fundamental-plane relations
derived by \citet{vdbo16} as follows.
From \n7727's absolute $K_s$-band magnitude of $\mk = -24.51$ \citep{jarr03}
we derive a luminosity of $\lk = 1.31\times 10^{11}\,\lksun$ and, via the
mean relation $\mstar/\lk = 0.10\,\sigveleff^{0.45}$ \citep{vdbo16} and 
the measured $\sigvelone = 190$ \kms\ (Section \ref{sec342}), a stellar mass
for the galaxy of $\mstar = 10^{11.14\pm 0.10} \msun$.
Inserting this stellar mass and \n7727's effective radius of
$\reffone = 3.7 \pm 0.2$~kpc (Section \ref{sec33}) into Equation (5) by
\citet{vdbo16} yields an expected SMBH mass of $\logmbh = 8.25\pm 0.30$.
Hence, \nucone\ is expected to host an SMBH in the mass range of
$\mbh \approx (1$--$4)\times 10^8\,\msun$.

In a search for this SMBH, \nucone\ was observed with the Space Telescope
Imaging Spectrograph (STIS) aboard \hst\, on 2001 September 1 for
$\sim\,$5 hrs (Program 9107, PI: D.\ Richstone), and a future paper describing
it and two other program galaxies was announced \citep{guel09}.
However, no mass determination for the expected SMBH has been published to
date, possibly because of interpretative difficulties or the lack of
supporting ground-based spectra.
Hence, there is currently no direct {\em stellar-dynamical\,} evidence for
the existence of an SMBH in \nucone.

%%%%%%%%%%%%%%%%%%%%%%%%%%%%%%%   Fig. 10   %%%%%%%%%%%%%%%%%%%%%%%%%%%%%%%%
\begin{figure*}
  \begin{center}
  \includegraphics[width=17.2cm]{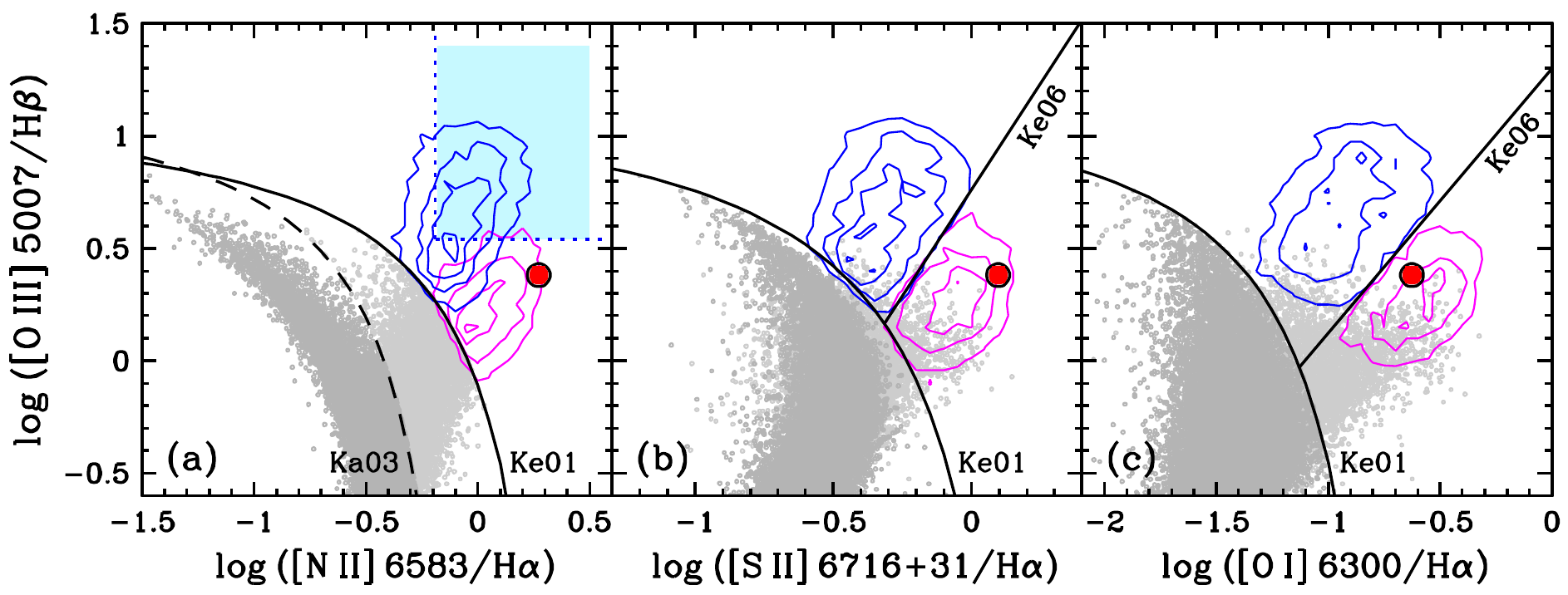}
    \caption{
Positions of \n7727's \nucone\ (red dots) and \nuctwo\ (light-blue shaded
area) in three BPT diagrams that plot the line ratio \oiiisev/\hbet\ versus
the ratios (a) \niithr/\halp, (b) \siiboth/\halp, and (c) \oizer/\halp.
The dashed curve in panel (a) marks the empirical upper boundary for purely
star-forming galaxies \citep{kauf03}, while the solid curves in each panel
mark theoretical upper boundaries for extreme starbursts \citep{kewl01}.
Straight lines in panels (b) and (c) show empirical boundaries between
LINERs (below) and Seyfert galaxies (above) \citep{kewl06}.
For reference, the line ratios of a subsample of SDSS emission-line galaxies
selected for high signal-to-noise line measurements are shown in gray for
star-forming (dark) and composite-spectrum (light) galaxies, and as contours
for LINERs (red) and Seyfert galaxies (AGNs; blue).
\nucone\ clearly lies in the domain of LINERs and is a weak LINER probably
excited by hot pAGB stars, LMXBs, and/or the central X-ray gas.
\nuctwo\ appears to lie in Seyfert territory and may be experiencing weak
AGN activity; for details, see text.
    \label{fig10}}
  \end{center}
\end{figure*}
%%%%%%%%%%%%%%%%%%%%%%%%%%%%%%%%%%%%%%%%%%%%%%%%%%%%%%%%%%%%%%%%%%%%%%%%%%%%

The optical spectrum of \nucone\ features several of the well-known emission
lines from ionized gas, of which the \niithr\ line is the most prominent
(see Figures~\ref{fig07} and \ref{fig08}).
Based on the relative strength of low-ionization lines from \nii, \oi, \oii,
and \sii, the nucleus was identified as a likely LINER already early on
\citep{keel83,daha85}.
To check on these early assessments, we measured emission-line fluxes and
widths from our MagE and LDSS-3 spectra as described in Appendix~\ref{appb},
where line fluxes and ratios are listed in Table~\ref{tab_b1}.
Figure~\ref{fig10} shows the location of \nucone\ in the three standard
BPT diagrams \citep{bald81}, based on our measured line ratios corrected
for both Milky Way foreground reddening ($\avmw = 0.098$) and internal
reddening in \n7727 ($\avint = 1.4\pm 0.2$).
As can be seen, \nucone\ falls into the LINER region in each of the three
diagrams and is, therefore, a LINER beyond any reasonable doubt.

The measured emission lines are relatively weak, and all appear to be narrow,
with---after correction for the instrumental FWHM of 58.0 \kms---a weighted
mean width of FWHM$= 294\pm 7$ \kms\ at rest wavelength.
This width corresponds to a central velocity dispersion of
$\sigma_{\rm g,1} = 125\pm 3$ \kms\ for the ionized gas of \nucone.
Note that this dispersion is only about 2/3 of the velocity dispersion
measured for the stars, $\sigvelone = 190\pm 5$ \kms\ (Section \ref{sec342}).
The difference could be due to a number of causes, including gaseous
dissipation and/or the measured ionized gas lying perhaps preferentially in
the foreground of the stellar nucleus.
The most likely cause seems to be gaseous dissipation, since the central
gas must be part of the counterrotating ionized- and molecular-gas disk
described in Section \ref{sec372}.

Note that even after subtraction of the stellar continuum we were unable to
detect any broad emission-line components, including in \halp\ and \hbet.
However, the wavy zero level and limited signal-to-noise ratio of the
spliced and continuum-subtracted MagE spectrum did not permit any reliable
multicomponent line decomposition.
With this caveat expressed, we conclude that there is no evidence for any
AGN activity and, hence, the presence of an SMBH in \nucone\ from the
ionized-gas kinematics either.

What, then, does the fact that \nucone\ is a LINER mean in terms of central
accretion and the expected presence of an SMBH?
The answer is---very likely---that the ionization of the central gas is
due entirely to the {\em exclusively old stellar population} of \nucone\
(Section \ref{sec35}), especially to its hot post-asymptotic-giant-branch
(pAGB) stars, as first proposed for elliptical galaxies by \citet{bine94}
and since demonstrated for E to Sa galaxies in many studies
\citep[e.g.,][]{stas08,sarz10,cidf11,yan12}.
There may also be significant ionization contributions from low-mass X-ray
binaries (LMXBs) and diffuse X-ray-emitting plasma \citep{lho08}.
As \citet{heck14} nicely review, {\em weak LINERs}---defined by equivalent
widths $EW(\halp)\la 1$ \AA\ \citep{bine94} or $EW(\oiiisev)\la 2.4$ \AA\
\citep{sarz10}---form the majority of all LINERs found in the Sloan Digital
Sky Survey (SDSS).
Their nuclear emission lines are excited neither by shocks nor by any
central AGN, but mainly by the above-mentioned old stellar populations.

With its measured properties, \nucone\ is clearly a weak LINER.
Its equivalent widths of 1.1 \AA\ for \halp\ and 0.7 \AA\ for \oiiisev\
emission (Table~\ref{tab_b1}) place it at about the upper \halp\ limit,
but well below the upper \oiiisev\ limit quoted above for weak LINERs.
Also, the spatial extent of its low-ionization emission lines over several
arcseconds in radius ($\sim\,$0.5--1 kpc) supports the notion that its
ionization sources are spatially extended themselves, whether they be
pAGB stars, LMXBs, or X-ray plasma.
And third, the \chandra\, observations by \citet{bras07} do not show
any X-ray point source coincident with \nucone\ (Section \ref{sec31}).
Therefore, we lack any evidence of central AGN activity, a central SMBH,
or central star formation (Section \ref{sec35}) in \nucone.
This, of course, does not in any way prove the absence of an SMBH there.
Such an SMBH may well exist, but may currently not be accreting or may be
heavily obscured.

\subsubsection{\nuctwo: Evidence for an AGN}
\label{sec422}

Although \nuctwo\ has been reduced to UCD status and is still being
tidally stripped, it shows at least three signs of recent activity:
a post-starburst spectrum, a high-excitation emission-line spectrum, and
a luminous X-ray source.

The marked post-starburst spectrum  of \nuctwo\ (Fig.~\ref{fig06}) points
to a relatively recent burst of star formation, with a light-weighted mean
age of $1.4\pm 0.1$ Gyr and having produced about 1/3 to 1/2 of the
present stellar mass (Section \ref{sec35}).
Although it remains unclear whether a second, much weaker burst of star
formation occurred 500--650 Myr ago, there is little question that the
entire starburst episode involved significant amounts of gas, at least the
several times $10^8\,\msun$ necessary to produce the currently observed
mass of $(7.5\pm 1.0)\times 10^7\,\msun$ in intermediate-age stars.
Presumably, this gas was funneled by tidal torques toward the center
of the infalling companion galaxy during its close approaches to \n7727,
a process well known to occur in strongly interacting galaxies
\citep[e.g.,][]{tt72,nogu88,bh96}.
Hence, it should not surprise if the same supply of gas also fed an SMBH
possibly residing at the center of the infalling companion
\citep[e.g.,][]{hopk10}.

There is some spectroscopic evidence that \nuctwo\ may indeed be
experiencing AGN activity.
Based on its measured emission lines (Table \ref{tab_b1}), Figure
\ref{fig10}(a) shows the area (shaded in light blue) where \nuctwo\
is likely to fall in the main BPT diagram that displays the excitation ratio
\oiiisev/\hbet\ versus the ratio \niithr/\halp\ \citep{bald81}.
Although we were unable to measure the flux of the \hbet\ emission line
itself, we determined an upper limit for it, leading to a {\em lower}\, 
limit of $\oiiisev/\hbet \ga 3.5$ (or $\ga 0.54$ logarithmically) for the
excitation ratio (Table \ref{tab_b1}).
This limit, shown as a horizontal dotted blue line in Figure \ref{fig10}(a),
strongly points to \nuctwo\ lying in Seyfert territory, which is outlined
by blue isopleths representing 5074 Seyfert galaxies with high S/N-ratio
spectra \citep{fust09}.

Since an emission-line flux for \halp\ could not be measured either,
we estimated an upper limit for it from the upper limit for \hbet\ and
a Balmer decrement of 3.1, assuming no internal extinction in
\nuctwo.\footnote{
Given the blue color of \nuctwo, $\vitotzero = 0.93$ (Table \ref{tab03}),
this assumption seems justified.}
This \halp\ limit yields a {\em lower}\, limit of $\niithr/\halp \ga 0.64$
(or $\ga -0.19$ logarithmically), marked by a {\em vertical}\, dotted blue
line in Figure \ref{fig10}(a).
Although this lower limit could move to the left (i.e., lower) if there
were significant internal extinction in \nuctwo, it would not affect our
conclusion that the elevated value of \oiiisev/\hbet\ places this nucleus
in Seyfert territory.
The inferred AGN nature of \nuctwo\ will have to be checked via
well-calibrated new spectra of higher S/N ratio, permitting better
modeling and subtraction of the underlying starlight.
The prediction is that such spectra should feature some broader lines (or
line components) characteristic of AGNs.

The strongest evidence for a weak AGN in \nuctwo\ may be the X-ray point
source named ``Source~5'' by \citet{bras07}, which coincides to within
$\sim$\,$0\farcs1$ with the optical center of the nucleus (Section
\ref{sec31}).
Although the intrinsic X-ray luminosity of this source is relatively low,
\lx(0.3-6.0 keV)\ $= (2.80\pm 0.34)\times 10^{39}\,\ergsec$ (for
$D=27.4$ Mpc), it does place Source~5 in the category of Ultraluminous
X-ray (ULX) sources, defined as having $\lx \ga 10^{39}\ \ergsec$ based
on assumed isotropic emission \citep[e.g.,][]{fabb89,fabb06,kaar17}.
Such sources exceed the Eddington limit for a 10\,\msun\ black hole and
have long been surmised to be associated with black holes of intermediate
mass (IMBHs; $10^2\,\msun \la M_{\rm IMBH} \la 10^5\,\msun$).
Yet recently some of them have been found to be associated with accreting
neutron stars \citep{bach14,kaar17} and others with UCDs
\citep[e.g.,][]{sori10,kim15,irwi16}.
Hence, their X-ray luminosities alone do not establish them unambiguously
as AGNs.

%%%%%%%%%%%%%%%%%%%%%%%%%%%%%%%   Fig. 11   %%%%%%%%%%%%%%%%%%%%%%%%%%%%%%%%
\begin{figure}
  \begin{center}
    \includegraphics[width=7.5cm]{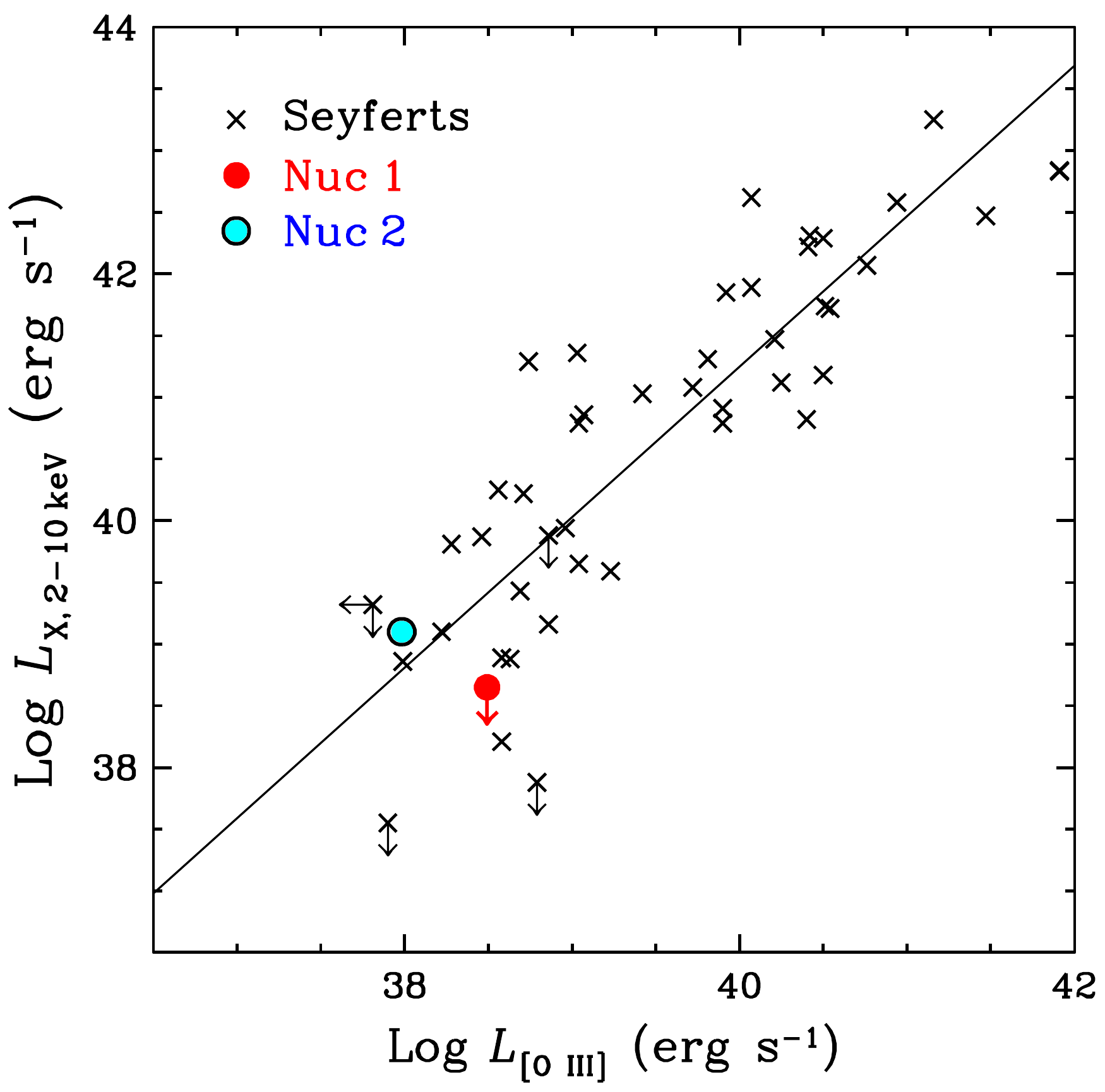}
    \caption{
Positions of \n7727's \nucone\ (red dot with downward-pointing arrow) and
\nuctwo\ (blue dot) on a correlation diagram showing X-ray luminosities of
45 local Seyfert galaxies plotted vs.\ their \oiiisev\ luminosities
\citep{pane06}.
While \nucone\ falls below the mean relation for the Seyferts (inclined
line), with only an upper limit available for its \lx, \nuctwo\ falls
close to the mean relation, yielding evidence in support of it hosting
a low-luminosity AGN.
    \label{fig11}}
  \end{center}
\end{figure}
%%%%%%%%%%%%%%%%%%%%%%%%%%%%%%%%%%%%%%%%%%%%%%%%%%%%%%%%%%%%%%%%%%%%%%%%%%%%

However, for a ULX that lies close to the center of a galactic nucleus or
UCD, one way to check on suspected AGN activity is via the well-known
correlation between the X-ray and \oiiisev\ luminosities of AGNs, which
holds over a range of at least
$10^{38}\,\ergsec \la \lx \la 10^{44}\,\ergsec$
\citep{mulc94,bass99,heck05,pane06}.
Figure \ref{fig11} shows that \nuctwo\ ({\em blue dot}) falls close to the
mean relation determined for 45 local Seyfert galaxies by \citet{pane06},
near the relation's low-luminosity end.
Since the X-ray luminosities of the Seyfert galaxies were measured in the
2--10 keV energy band, we converted the \lx(0.3--6.0 keV)
of \nuctwo\ \citep{bras07} to
\lx(2--10 keV) $= 1.25_{-0.36}^{+0.88}\times 10^{39}\,\ergsec$
via the photon index $\Gamma = 2.15_{-0.36}^{+0.37}$ measured by the same
authors.
To determine the nucleus's \oiiisev\ luminosity, we used the flux measured
with MagE (Table \ref{tab_b1}) and corrected for Milky-Way foreground
extinction, leading to $\loiiitwo = (9.6\pm 1.1)\times 10^{37}\,\ergsec$.
Clearly, this \loiii\ for \nuctwo\ agrees well with the \lx\ of
Source~5 and the expected relative X-ray and \oiii\ luminosities of
Seyfert galaxies.
This greatly strengthens the case for Source 5 being a low-luminosity AGN
at the center of \nuctwo, rather than an accreting neutron star or
stellar black hole.

For comparison, an upper limit of
\lxone(2--10 keV) $\la 4.5\times 10^{38}\,\ergsec$
determined from the \chandra\, data for \nucone\ places this primary
nucleus ({\em red dot}) clearly {\em below} the mean relation for local
Seyfert galaxies, in agreement with our conclusion above that it is
currently inactive (Section \ref{sec421}).

Returning to \nuctwo, note that with a photon index of
$\Gamma = 2.15_{-0.36}^{+0.37}$ \citep{bras07} its X-ray spectrum is
unusually soft, suggesting that the inferred AGN is in a high/soft state.
This state agrees with the absence of any detectable radio point source
at the location of \nuctwo\ in the 3-mm continuum map \citep{ueda14},
since AGNs in the high/soft state are well known to be radio quiet
\citep[e.g.,][and references therein]{svob17}.

The {\em mass of the SMBH}\, causing the observed low-level AGN activity
in \nuctwo\ is of great interest and can---despite the lack of spatially
resolved kinematic information---be bracketed approximately as follows.

First, assume that \nuctwo\ is a bulge-like stellar system in equilibrium,
with measured $\sigveltwo = 79\pm 4$ \kms\ (Table \ref{tab04}).
The \mbh--\sigvel\ relation then yields an estimated black hole mass of
$\mbhtwo = 1.8_{-0.5}^{+0.8}\times 10^6\,\msun$ when computed with
Equation (2) by \citet{vdbo16}, or one of
$\mbhtwo = 5.3_{-1.2}^{+1.4}\times 10^6\,\msun$ when computed with
Equation (7) by \citet{koho13}.
Since until recently \nuctwo\ was part of a more massive galaxy, we take
these two values to represent a likely {\em lower limit}\, in the range of
$\mbhtwo \ga$ (2--5)$\times 10^6\,\msun$.

Second, assume that the only reason why the dynamical and stellar masses
of \nuctwo\ differ is the presence of an SMBH.
Then the difference $\mdyntwo - \mstartwo$ sets an {\em upper limit}\, to
the mass of this SMBH,
$\mbhtwo \la \mdyntwo - \mstartwo = (2.5\pm 0.6)\times 10^8\,\msun$.

In short, taking the various uncertainties into account, we estimate that
\mbhtwo\ probably lies in the range
$3\times 10^6\,\msun \la \mbhtwo \la 3\times 10^8\,\msun$.
Although this range is large, it does suggest that the SMBH of \nuctwo\
is at least comparable in mass to that of our own Galaxy
($\mbhmw \approx 4.3\times 10^6\,\msun$), and possibly an order of magnitude
more massive.
Specifically, if \mbhtwo\ were to contain 10\%--20\% of the dynamical mass
of \nuctwo, as has been found for the SMBHs in two massive UCDs of the Virgo
Cluster \citep{ahn17}, its range would narrow to what we consider more
likely values of $4\times 10^7\,\msun \la \mbhtwo \la 8\times 10^7\,\msun$.

\subsection{\n7727: The Aftermath of Ingesting a Gas-Rich Disk Companion}
\label{sec43}

In trying to unravel \n7727's recent merger history, we rely on the
assumption that its complex structure resulted entirely from one (or
possibly two) merger event(s), and not from any tidal interaction with
one of the other two main members of this small group.
This assumption seems well justified.
Based on their relative \ks-band luminosities, \n7724 at $\dproj = 97$
kpc to the WNW has an \mstar\ of only $\sim$\,9\% that of \n7727 and itself
shows no tidal damage (Figure \ref{fig02}), while the 11$\times$ more
massive protagonist shows plenty of it.
And similarly, \n7723 at $\dproj = 338$ kpc to the SSW has an \mstar\
of $\sim$\,58\% that of \n7727, yet itself shows no tidal damage, even
on deep exposures.
Hence, the history of {\em recent}\, gravitational interactions must be
buried in \n7727 itself.

A helpful clue is the presence of various blue structures and point-like
sources in \n7727's main body in addition to the blue \nuctwo\ and BTS.
Foremost are the two major blue arcs described in Section \ref{sec321}
(\arcone\ and 2), several fainter bluish arcs and shells visible on
color-index maps \citep{lho11}, and a few dozen star clusters and
associations of intermediate age \citep{crab94,tran03}.
Whereas the compact star clusters are spread throughout the main body,
the associations seem to be located preferentially in blue arcs.
A combination of \hst\, $V\!I$ photometry and ground-based \ks\ photometry
shows that the compact clusters have ages of $0.9\pm 0.8$ Gyr and about
solar metallicities \citep{tran14}.
Together with \nuctwo\ and its BTS, these various forms of bluish detritus
in the otherwise reddish main body of \n7727 suggest strongly that we
are observing the {\it shredded remains of a gas-rich disk companion}\,
that fell into \n7727 within the past 2 Gyr.

Knowing the stellar mass \mhost\ of this former companion, host to \nuctwo,
is of interest both in absolute terms and relative to the mass of \n7727.
Yet any direct estimate of this mass, for example by summing up the various
bluish components listed above, is well beyond what our limited data permit.
Instead, we use the estimated mass range for the SMBH of \nuctwo,
$3\times 10^6\,\msun \la \mbhtwo \la 3\times 10^8\,\msun$ (Section
\ref{sec422}), to estimate the range of \mhost\ via the
\mbh--\mbulge\ relation \citep[][Equation (10)]{koho13}.
The resulting stellar mass range for the {\em bulge}\, of the former companion
is\,\ $2\times 10^9\,\msun \la \mbulge \la 6\times 10^{10}\,\msun$.
For comparison, the stellar mass of \n7727, estimated from the galaxy's
luminosity $\lk = 1.31\times 10^{11}\,\lksun$ and $\sigvel = 190\pm 5$ \kms,
is $\mseven = (1.35\pm 0.10)\times 10^{11}\,\msun$.
Hence, the stellar-mass ratio between the bulge of the former companion and
\n7727 likely lies in the range\,\ $0.015 \la \mbulge/\mseven \la 0.45$.
Note, however, that since the companion clearly possessed a stellar disk
of significant mass in addition to its bulge, a more likely range of mass
ratios for the entire companion is\,\ $0.05 \la \mhost/\mseven \la 1.4$
(assuming a disk twice as massive as the bulge).
By itself, this range suggests that the merger involving the companion
hosting \nuctwo\ may have been a minor merger of at least $\sim$\,5\% the
mass of \n7727 up to---possibly---a major merger ($m/M \ga 1/3$).

If, instead of using the full estimated mass range for the SMBH of \nuctwo\
(as done above), we use the more restricted range of its {\em likely}\,
mass, $4\times 10^7\,\msun \la \mbhtwo \la 8\times 10^7\,\msun$ (Section
\ref{sec422}), the estimated bulge mass of the former companion becomes
$1.1\times 10^{10}\,\msun \la \mbulge \la 2.1\times 10^{10}\,\msun$, and
the corresponding mass ratio for the entire companion becomes
$0.25 \la \mhost/\mseven \la 0.47$ (again for D/B~$\approx 2$).
In that case, the merger involving the host galaxy of \nuctwo\ lies in
the transition region between minor and major mergers.

This conclusion agrees with the impression one gets from visual inspection
of \n7727 images that this messy merger involved {\em two}\, rather massive
disk galaxies.
Specifically, it agrees with the W Tail appearing to be about 2--3$\times$
less luminous than the E Tail, yet of similar color, which might imply a
companion 2--3$\times$ less massive than the main participant galaxy.
Yet, since we cannot directly link the W Tail to the blue arcs and bluish
detritus, we cannot currently exclude the possibility that a third galaxy
was involved---as an early player---in a triple merger.
Nevertheless, the solar metallicity of \nuctwo\ (Section \ref{sec36}) and
the plentiful detritus make a strong case for the host galaxy of this nucleus
having been a relatively massive and gas-rich disk companion, and certainly
{\em not}\, a nucleated dwarf elliptical!

The current {\em spatial position}\, and {\em orbit}\, of \nuctwo\ relative to
\nucone\ are crucial parameters for any successful future simulation of this
late-stage merger system.
Some limited information is currently available.
Because of their blue color, \nuctwo\ and the BTS very likely lie between
us and the sky plane containing \nucone, with the surmised SSE leg of the
BTS cutting right across the center of \n7727.
Interestingly, the position angle of the BTS to the NNE,
P.A.(BTS)~$= 331\degr\pm 2\degr$, agrees closely with that between the two
optical nuclei, P.A.(2$-$1)~$= 331\fdg9\pm 0\fdg1$.
Since models of tidal stripping show that tidal streams following and
preceding dense star clusters tend to align with the projected orbit rather
closely \citep{oden03,lux13}, this suggests that we may be seeing the
orbital plane of \nuctwo\ nearly edge-on and at P.A.~$\approx 332\degr$.
Given the measured small $\dvlos = +18.8\pm 3.1$ \kms\ between \nuctwo\
and \nucone\ (Section \ref{sec342}), our inference that \nuctwo\ probably
lies physically close to \nucone\ (Section \ref{sec41}), and the nearly
edge-on view of the orbit, we tentatively conclude that (1) \nuctwo\ lies
probably close to its orbit's present pericenter, and (2) its tangential
velocity probably exceeds 100 \kms.

Given \nuctwo's current near-pericentric position and its past one or
two starbursts (Section \ref{sec35})---each presumably triggered by
a pericentric passage---the present {\em orbital period}\, is probably
either $\sim$\,1.4 Gyr, if there was a single starburst, or
$\sim$\,0.6 Gyr, if there were two.
In either case, the present osculating orbit is likely to be highly
eccentric, if our guess is correct that \nuctwo\ lies physically close to
\nucone\ (say, $\la 3$ kpc).\footnote{
For comparison, in our Galaxy the period of an object in circular orbit at
the sun's distance of 8.2 kpc from the center is a mere 0.21 Gyr (for
$v_{\rm c} = 238$ \kms; see \citealt{blan16}).}
The true physical separation between the two nuclei and the direction of
motion of \nuctwo\ (to the SSE versus NNW) are perhaps the two most
important orbital parameters still missing, but needed to reconstruct the
recent history of this late-stage merger.

In short, the one firm conclusion is that until about 2~Gyr ago the
\n7727 group contained at least four major disk galaxies, of which one
harbored \nuctwo.
This gas-rich companion was likely 1/20\,--\,1/2 as massive as then-\n7727
itself and experienced strong gas flows into its center as it interacted
gravitationally with \n7727. 
By now most of its disk and bulge has been tidally shredded, leaving its
nucleus as a freshly formed and still evolving UCD within \n7727.

\section{SUMMARY}
\label{sec5}

We have described imaging, photometric, and spectroscopic observations of
\n7727 (Arp 222), a late-stage merger and dominant member in a nearby
small group of galaxies.
The observations include images taken with \hst/WFPC2 and with the du Pont
2.5 m telescope at Las Campanas Observatory, as well as long-slit spectra
obtained there with the LDSS-3 and MagE spectrographs on the Clay 6.5 m
telescope.
Although hitherto known mainly as an abnormally gas-poor \sapec\ galaxy,
\n7727 features within its complex disturbed body young massive star clusters
and a bright second nucleus that has often been mistaken for a Milky Way
foreground star.

The main results of our observations and analysis are as follows:

1. \n7727 is relatively luminous ($\mv = -21.7$ for $H_0 = 73$ \kms\ 
Mpc$^{-1}$) and located at a distance of 27.4 Mpc ($\czhel = 1863\pm 9$ \kms)
in Aquarius ($\delta = -12\degr$).
Besides a disturbed main body rich in fine structure, it features one
prominent tidal tail of  projected length $\ga$\,60 kpc, a second, shorter 
tidal tail, various bluish arcs and luminous star clusters, and {\em two
bright nuclei}\, separated by 483 pc in projection near its center, where a
small molecular- and ionized-gas disk counterrotates to the stars.
Its stellar mass is $\mstar\approx 1.4\times 10^{11}\,\msun$.
Except for the presence of a second nucleus, it shares many properties with
recent merger remnants.

2. The two nuclei of \n7727 differ in radial velocity by only
$\dvlos = +18.8\pm 3.1$ \kms, yet are strikingly different.
\nucone\ appears to fit smoothly into the central luminosity profile of the
galaxy, has spectral features characteristic of a weak LINER in an old
stellar population, and shows---so far---no signs of AGN activity at X-ray,
optical, and radio wavelengths, appearing instead as ``red and dead'' as
can be.
In contrast, \nuctwo---which appears clearly offset from the center---is 
very compact ($\reff = 29$ pc), has a tidal-cutoff radius of 103~pc, and
shows at least three signs of recent activity: a post-starburst spectrum,
an \oiiisev\ emission line indicative of high excitation, and a luminous
X-ray source of \lx(0.3--6.0 keV)\ $= 2.8\times 10^{39}\,\ergsec$.
Its spectral line ratios place it among Seyfert nuclei, although direct
evidence in the form of broad emission lines is currently lacking.
The stellar velocity dispersions of the two nuclei are $\sigvelone = 190\pm
5$ \kms\ and $\sigveltwo = 79\pm 4$ \kms, respectively, with \sigveltwo\
significantly higher than expected for \nuctwo, given the nucleus's
luminosity of $\mv = -15.5$ and stellar mass of
$\sim$\,$1.7\times 10^8\,\msun$.

3. A comparison of \nuctwo\ with well-studied UCDs suggests that it may
be the best case yet for a massive UCD having formed recently through
tidal stripping of a gas-rich disk galaxy.
Evidence for this comes from its extended star-formation history,
$\ga$\,1.5~kpc long blue tidal stream (BTS), and elevated ratio of dynamical
to stellar mass, $\mdyntwo/\mstartwo = 2.5\pm 0.6$, a characteristic of
massive UCDs.
With $\mdyntwo = (4.2\pm 0.6)\times 10^8\,\msun$, \nuctwo\ is slightly more
massive than M59-UCD3, the most massive UCD known so far.
While the majority of its stars formed in the distant past ($\ga$\,10 Gyr),
$\sim$\,38\%--50\% of them by mass formed during one or two starbursts in
the past 2 Gyr.
Its weak AGN activity is likely driven by an SMBH of mass
$3\times 10^6\,\msun \la \mbhtwo \la 3\times 10^8\,\msun$, with a most
likely range of perhaps (4--8)$\times 10^7\,\msun$.

4. The combined morphological and kinematic evidence suggests that \n7727
experienced one or perhaps two mergers during the past few Gyr.
The most recent merger involved a gas-rich disk companion and began about
2 Gyr ago, leading to strong gas inflows and one or two starbursts at the
center of the companion, as recorded in the post-starburst spectrum of
\nuctwo.
The initial mass of this accreted companion was probably between 1/20th and
1/2 that of then-\n7727, making for a likely minor merger or---possibly---one
near the  minor/major merger boundary ($1/4 \la m/M \la 1/2$).
By now most of this former companion has been shredded, leaving behind
various forms of blue detritus (BTS, arcs, young clusters) and \nuctwo\
as a freshly minted UCD. 
This UCD likely moves on a highly eccentric orbit of period $\sim$\,0.6 Gyr
(or 1.4 Gyr) and may currently be close to pericenter.
The persistence of \nuctwo\ makes \n7727 a late-stage merger rather than
a merger remnant.

5. Before its most recent merger, \n7727 was likely an S0 galaxy.
It is the blue detritus of the infallen companion that earned this
galaxy an \sapec\ classification from the classifiers.
Clearly, a mild rejuvenation of the stellar populations took place.
A mystery remains why \n7727 is so gas-poor and \nucone\ appears so
inactive.
To improve our understanding of the assembly history of this galaxy,
kinematic mapping of the stars and gas will be essential.
Note that even at only five times its distance, say $z\approx 0.03$,
figuring out the details of this likely minor, but significant, merger
would have been extraordinarily difficult.

\acknowledgments

We thank Herman Olivares, Hern\'{a}n Nu\~nez, and Mauricio Martinez for
expert assistance at the telescopes;
Oscar Duhalde, Mauricio Navarrete, and Vincent Suc for technical help with
the instruments;
Michael Fitzpatrick from the former IRAF Help Desk at NOAO for his steady
support;
Roberto Cid Fernandes for making his software package STARLIGHT generally
available;
Werner Probst for his kind permission to reproduce his deep wide-field
image of \n7727 in Figure~\ref{fig02}(f);
Scott Burles, George Jacoby, Steve Shectman, and Ian Thompson for technical
advice;
and Barry Madore, Bryan Miller, John Mulchaey, Yue Shen, Alar Toomre, and
Gelys Trancho for helpful discussions.
The UK Schmidt plate shown in part in Figure~\ref{fig02}(e) is copyright (c)
the Royal Observatory Edinburgh and the Anglo-Australian Observatory, and the
scan of it that we used to produce the figure was made at the Space Telescope
Science Institute (STScI) under US Government grant NAG W-2166.
Support for \hst\,\ Program GO-7468 was provided by NASA through a grant from
STScI, which is operated by AURA, Inc., under NASA contract No.\ NAS5-26555.
This work is based in part on observations made with \hst\, and obtained
from the Hubble Legacy Archive, which is a collaboration between STScI/NASA,
the Space Telescope European Coordinating Facility (ST-ECF/ESA), and the
Canadian Astronomy Data Centre (CADC/NRC/CSA).
Our research has greatly benefited from the use of
(1) NASA's Astrophysics Data System (ADS) Bibliographic Services, operated
by the Smithsonian Astrophysical Observatory, and
(2) the NASA/IPAC Extragalactic Database (NED), which is operated by the
Jet Propulsion Laboratory, California Institute of Technology, under
contract with NASA.

\appendix

\section{A. DETAILS ABOUT MASKED IMAGES}
\label{appa}

Three of the images illustrating the fine structure of \n7727 in the present
paper---Figures~\ref{fig02}(a), \ref{fig02}(b), and \ref{fig03}(c)---have
been digitally masked to enhance the visibility of faint small-scale
features relative to the strongly varying surface brightness of the
galaxy.\notetoeditor{
It seems to be both appendices could be in two-column format, which I do not
know how to generate with pdfTeX and emulateapj.  Please feel free to change
to two-column format.}
This appendix provides some details about the adopted algorithms that---at
least in part---imitate the ``unsharp masking'' technique often used for
photographic reproductions in the past \citep[e.g.,][]{sand64,mees66,mali77}.
The main goal of unsharp masking was to diminish large-scale background
variations by superposing a photographic negative with an out-of-focus
(``unsharp'') {\em positive} copy of it and making a photographic print
from the negative\,+\,positive sandwich placed in an enlarger.
This process can be imitated digitally in various ways
\citep[e.g.,][]{schw85,schw88,berr05} or can be replaced by the digital
subtraction of---or division by---a model light distribution derived from
the galaxy image (e.g., \citealt{fort86,jedr87,prie88}; and many since).

Figures~\ref{fig02}(a) and \ref{fig02}(b) were both produced from the same
composite $R$-band image of \n7727 obtained with the du Pont 2.5~m telescope
(Section \ref{sec212}) and shown directly (i.e., unmasked) at two different
contrasts in Figures~\ref{fig02}(c) and \ref{fig02}(d).
To produce a mask for them, the $R$-band image was first cleaned of
bright stars and \nuctwo\ by local interpolation with the IRAF task
{\em imedit}.
The ``mask'' was then computed by running a circular median filter of
10\arcsec\ radius across the precleaned image with task
{\em rmedian}.\footnote{
Various filter sizes and shapes were tried, with a filled circular median
filter of 10\arcsec\ radius yielding the best displays as judged visually.}
The resulting smoothed mask image was combined with the original $R$\,
image in two different ways.
In Figure~\ref{fig02}(a) the original image was {\em divided}\, by the
mask image raised to the power of 0.95, while in Figure~\ref{fig02}(b)
the mask image was first multiplied by 0.90 and then {\em subtracted}\,
from the original image.
Both techniques accomplish partial masking of the original image and aim
at preserving a small fraction of the original brightness gradient in
the galaxy image, while enhancing the visibility of fine-structure details.
Whereas the division-masked Figure~\ref{fig02}(a) shows the inner fine
structure better than the subtraction-masked Figure~\ref{fig02}(b), the
latter shows the outer fine structure at higher contrast.
Nevertheless, these two masked versions of the original $R$\, image of
\n7727 correspond to each other in considerable detail, emphasizing that
the depicted fine structure is real and relatively independent of details
in the adopted masking algorithms.

Figure~\ref{fig03}(c) was produced from the PC $V$ image of \n7727 obtained
with \hst/WFPC2 (Section \ref{sec211}).
This image is implicitly shown in Figure~\ref{fig03}(a), which is a
color rendition based on the PC $V$ and $I$\, images.
To produce a mask for it, the original $V$ image was first cleaned of a
bright star by linear interpolation with IRAF task {\em imedit} and then
smoothed by running a circular median filter of $2\farcs0$ radius across
it with {\em rmedian}.
The chosen 2\arcsec\ radius was sufficiently large to remove not only some
point-like sources (globular clusters in \n7727 and faint foreground stars),
but also the entire \nuctwo, which is very compact (see Section \ref{sec33}).
Since the resulting smoothed image still showed some pixelated noise,
it was slightly smoothed further by convolving it with a 2D Gaussian
function of $\sigma = 0\farcs25$ and limiting radius $3\,\sigma$ to form
the final mask.
The original $V$ image was then divided by this mask (this time without
exponentiation) to produce the ``unsharply masked'' $V$ image shown in
Figure~\ref{fig03}(c).

\section{B. EMISSION-LINE SPECTRA OF NUCLEI}
\label{appb}

In support of our discussion of possible AGN activity in the nuclei
(Section \ref{sec42}), we measured emission-line widths, equivalent widths,
and fluxes for each nucleus from our spectra.
The main challenge was---in both nuclei---the weakness of the ionized-gas
emission lines relative to the underlying bright absorption-line spectrum
of the stars.

For \nucone, the broad wavelength coverage of the MagE spectrum permitted
measuring the \oiisev\ doublet and the five \halp, \nii, and \sii\ lines,
none of which are covered by the LDSS-3 spectrum.
However, since for technical reasons we could not reliably flux-calibrate
the MagE spectrum, we had to measure emission-line equivalent widths from
its continuum-normalized version (Fig.~\ref{fig07}) and then convert these
widths to line fluxes via the flux-calibrated continuum of our LDSS-3
spectrum (Fig.~\ref{fig08}).
We did this using slight extrapolations at both ends of the LDSS-3 spectrum,
helped by a published spectrum of \n7727 \citep{liu95b}.
A crucial step to reveal and measure the fainter emission lines was the
subtraction of the stellar background spectrum.
After some experimentation, and lacking an early-type galaxy spectrum
taken with MagE, we subtracted a continuum-normalized spectrum of the
G8\,III star HD\,165760 obtained with MagE on the same night
(Fig.~\ref{fig07}).
To achieve proper subtraction, this spectrum was first convolved with the
stellar velocity dispersion measured for \nucone, $\sigvelone = 190$ \kms,
before being subtracted.\footnote{
The spectrum of a K2\,III star might have produced a slightly better match,
but was not obtained with MagE.}

For \nuctwo, we followed a similar procedure.
However, we were unable to subtract the stellar continuum from the
normalized MagE spectrum, which is seriously noisy owing to the preceding
subtraction of the \n7727 background (Section \ref{sec34}).
As a result, we were able to determine only an upper limit for the flux of
the \hbet\ emission line, which is buried in the noisy \hbet\ absorption
trough of the post-starburst spectrum (Figs.~\ref{fig06} and \ref{fig07}).
The one reliably measured emission feature in \nuctwo\ was the \oiiisev\
line, whose equivalent width and flux strongly peak at the center.
For completeness we also measured the \niithr\ line, although its flux
appears not particularly peaked toward the center of \nuctwo\ and may be
contaminated by emission from the ionized-gas disk surrounding \nucone\
(Section \ref{sec372}).

%% ##############################  Table B1  ################################
\begin{deluxetable*}{lcccccccc}
\tablecolumns{9}
%\tablewidth{0pt}
\tablewidth{14.5truecm}
%\tabletypesize{}
%\tablenum{B1}				% = Table B (called B1 in paper)
					% Switched to letting it be "Table 6"
\tablecaption{Parameters of Emission Lines in Nuclei of \n7727\label{tab_b1}}
\tablehead{
   \colhead{Ion$\phn$} &
   \colhead{Line}      &
   \colhead{}          &
   \colhead{$-EW$\tablenotemark{a}\ }        &
   \colhead{Flux $F$\tablenotemark{b}\ }     &
   \colhead{\lineratzeromw\tablenotemark{c}} &
   \colhead{\lineratzero\tablenotemark{d}\ } &
   \colhead{}          &
   \colhead{FWHM\tablenotemark{e}\ }	       \\
   \colhead{}          &
   \colhead{}          &
   \colhead{}          &
   \colhead{(\AA)}     &
   \colhead{($10^{-16}$ cgs)} &
   \colhead{}          &
   \colhead{}          &
   \colhead{}          &
   \colhead{(\kms)}
}
\startdata
Nucl.\,1: &       &&                 &                 &                    &                    &&            \\
\ \ \oii  &  3727 &&  $7.56\pm 0.15$ & $29.20\pm 2.80\phn$& $10.70\pm 1.12\phn$& $16.29\pm 1.71\phn$&& \nodata \\
\ \ \hbet &  4861 &&  $0.26\pm 0.01$ &  $2.81\pm 0.11$ &     $1.00\pm 0.00$ &     $1.00\pm 0.00$ &&      282   \\
\ \ \oiii &  5007 &&  $0.68\pm 0.02$ &  $7.23\pm 0.42$ &     $2.56\pm 0.15$ &     $2.41\pm 0.17$ &&      326   \\
\ \ \oi   &  6300 &&  $0.26\pm 0.01$ &  $3.20\pm 0.15$ &     $1.10\pm 0.07$ &     $0.72\pm 0.05$ &&      298   \\
\ \ \nii  &  6548 &&  $0.50\pm 0.02$ &  $6.37\pm 0.30$ &     $2.19\pm 0.14$ &     $1.37\pm 0.09$ &&      271   \\
\ \ \halp &  6563 &&  $1.11\pm 0.02$ & $14.25\pm 0.56\phn$&  $4.90\pm 0.28$ &     $3.04\pm 0.17$ &&      298   \\
\ \ \nii  &  6583 &&  $2.08\pm 0.04$ & $26.70\pm 1.10\phn$&  $9.17\pm 0.52$ &     $5.70\pm 0.32$ &&      300   \\
\ \ \sii  &  6716 &&  $0.75\pm 0.02$ &  $9.55\pm 0.43$ &     $3.28\pm 0.20$ &     $1.98\pm 0.12$ &&      270   \\
\ \ \sii  &  6731 &&  $0.69\pm 0.02$ &  $8.77\pm 0.39$ &     $3.01\pm 0.18$ &     $1.82\pm 0.11$ &&      283   \\
Nucl.\,2: &       &&                 &                 &                    &                    &&            \\
\ \ \hbet &  4861 &&    $\la 0.50  $ &    $\la 2.77$   &     $1.00\pm 0.00$ &         \nodata    &&    \nodata \\
\ \ \oiii &  5007 &&  $1.89\pm 0.20$ &  $9.64\pm 1.06$ &        $\ga 3.5  $ &         \nodata    &&      226   \\
\ \ \nii  &  6583 &&  $1.23\pm 0.15$ &  $5.68\pm 0.74$ &        $\ga 2.0  $ &         \nodata    &&      194
\enddata
\tablecomments{Errors are measuring errors and their propagations. Systematic flux
	       errors are larger, but do not affect flux ratios (see text).}
\tablenotetext{a}{Equivalent width measured from the normalized MagE spectrum through an
		  aperture of $0\farcs82\times 0\farcs70$ ($109\times 93$ pc).}
\tablenotetext{b}{Observed flux computed from measured MagE equivalent width and LDSS-3
		  continuum, in units of $10^{-16}$ \ergscm.}
\tablenotetext{c}{Ratio of fluxes corrected for Milky Way foreground extinction of
		  $\avmw = 0.098$.}
\tablenotetext{d}{Ratio of fluxes corrected for total (Milky Way plus internal)
		  extinction \av.  Available only for \nucone, where $\av = 1.5 \pm 0.2$.}
\tablenotetext{e}{{L}ine widths determined from Gaussian fits and corrected for
		  instrumental broadening; typical errors are $\pm 6$\%.}
\end{deluxetable*}
%% ##########################  End of Table B1  #############################

Table \ref{tab_b1} presents our measured equivalent widths and computed
line fluxes and flux ratios to \hbet.
For \nucone, the flux ratio corrected for Milky Way foreground reddening,
$$F_{0,{\rm MW}}(\halp)/F_{0,{\rm MW}}(\hbet) = 4.90\pm 0.28,$$
yielded---when compared to the ratio of 3.1 for a Case B spectrum for
$T = 9000$\,K and low density \citep{of06}---an {\em internal} visual
extinction of $\avint = 1.4\pm 0.2$ within \n7727.
Although relatively large, this value is not surprising given the many
strong dust lanes and patches visible near the nucleus (Figure \ref{fig03}).
The penultimate column of Table \ref{tab_b1} gives the fully
reddening-corrected ($\avmw + \avint$), intrinsic line ratios relative to
\hbet, \lineratzero.
For \nuctwo, our inability to detect and measure \halp\ and \hbet\ line
emission deprives us of a measure of its total extinction, whence we cannot
derive intrinsic line ratios.
Finally, the last column of Table \ref{tab_b1} gives line widths measured
at half-maximum and corrected for instrumental broadening.

%%%%%%%%%%%%%%%%%%%%%%%%%%%%%%%  REFERENCES  %%%%%%%%%%%%%%%%%%%%%%%%%%%%%%%%

\notetoeditor{
Three references include an arXiv:ID because they are otherwise difficult
and/or costly to access.}

\end{document}